\documentclass[oneside]{jfm}

\usepackage{graphicx}
\usepackage{newtxtext}
\usepackage{newtxmath}
\usepackage{natbib}
\usepackage{hyperref}
\hypersetup{
    colorlinks = true,
    urlcolor   = blue,
    citecolor  = black,
}

\newcommand{\RomanNumeralCaps}[1]

\usepackage{subcaption}
\usepackage{placeins}
\usepackage{multirow}

\usepackage{xurl}

\usepackage{xcolor}

\newcommand{\ii}[0]{\mathrm{i}}

\newcommand{\ee}[0]{\mathrm{e}}

\shorttitle{Jet-noise reduction via streak generation in the nozzle boundary layer}
\shortauthor{F.R. Amaral, P.A.S. Nogueira, I.A. Maia, A.V.G. Cavalieri and P. Jordan}

\title{Jet-noise reduction via streak generation in the nozzle boundary layer}

\author{Filipe R. do Amaral\aff{1,3}\corresp{\email{filipefra@ita.br}},
	Petrônio A. S. Nogueira\aff{2},
	Igor A. Maia\aff{3},
	André V. G. Cavalieri\aff{3}
	\and Peter Jordan\aff{1}}

\affiliation{\aff{1}Institut Pprime, CNRS–Université de Poitiers–ENSMA, 86000 Poitiers, France
\aff{2}Monash University, Melbourne, Victoria 3800, Australia
\aff{3}Instituto Tecnológico de Aeronáutica, São José dos Campos, SP 12228-900, Brazil}

\begin{document}
\maketitle


\begin{abstract}

We study the hydrodynamic and acoustic fields of turbulent jets issuing from nozzles modified by the addition of cylindrical tabs on the inner surface, one diameter upstream of the exit.
The tabs are designed to promote significant growth of steady streaks in the nozzle turbulent boundary layer.
A baseline smooth nozzle is also studied for comparison.
Acoustic measurements are made using an azimuthal array for Mach numbers in the range 0.4 $\leq M_j \leq$ 0.9.
The tabs are found to reduce the emitted sound levels by up to 3 dB/St.
In terms of overall sound pressure levels (OASPL), reductions of up to 3 dB are observed at all measured polar angles in the range 20 deg $\leq \theta \leq$ 90 deg.
Time-resolved particle image velocimetry (TR-PIV) experiments are conducted to measure the three components of velocity for a series of cross-stream planes at $M_j =$ 0.7.
A Floquet-based Fourier decomposition is applied for the azimuthally periodic flow field, and spectral proper orthogonal decomposition (SPOD) is then employed to extract coherent structures.
Comparison of the structures obtained for nozzles with and without tabs shows an enhancement of the streaky structures by the tabs and a damping of Kelvin--Helmholtz (K--H) wavepackets.
A linear model based on the one-way Navier--Stokes equations (OWNS) is employed to explore the underlying amplification mechanisms and how these are impacted by the tabs.
The model reproduces the growth--attenuation mechanism observed in the data, showing that the changes in the mean flow induced by the streaks work to reduce the amplification of the noise-generating coherent structures associated with linear spatial growth mechanisms.

\end{abstract}


\begin{keywords}
Authors should not enter keywords on the manuscript, as these must be chosen by the author during the online submission process and will then be added during the typesetting process (see \href{https://www.cambridge.org/core/journals/journal-of-fluid-mechanics/information/list-of-keywords}{Keyword PDF} for the full list).  Other classifications will be added at the same time.
\end{keywords}


\section{Introduction}
\label{sec:introduction}

Great advances have been made in the field of jet noise since the theoretical development of~\citet{lighthill1952sound}, especially regarding the understanding of physical mechanisms associated with jet noise and the development of technologies to lower the noise emission by turbulent jets.
Modern aircraft make use of numerous noise reduction strategies, including the use of ultra-high-bypass-ratio turbofans~\citep{saiyed2003acoustics,huff2007noise,casalino2008aircraft}, sound absorbers (liners) on the nacelle~\citep{envia2002fan}, chevrons~\citep{bridges2004parametric,callender2005far,alkislar2007effect,zaman2011evolution}, among others.
Chevrons have been shown to produce moderate noise reductions at low frequencies, a high-frequency noise increase~\citep{saiyed1999tabs,tam2000subsonic,bridges2004parametric,callender2005far,alkislar2007effect} and a penalty in thrust~\citep{krothapalli1993role,saiyed2003acoustics,zaman2011evolution}.

It is often argued that chevrons applied to jet nozzle exits reduce jet noise primarily by generating streamwise vortices that enhance shear layer mixing.
These vortices disrupt the formation and evolution of large-scale coherent structures--the dominant sources of low-frequency noise in turbulent jets--by promoting earlier and more vigorous mixing of the high-speed jet core with the ambient air.
As the vortices interact with the shear layer, they induce increased entrainment and velocity gradients, thereby accelerating the breakdown of coherent structures and shortening the potential core length~\citep{alkislar2007effect, wernet2021characterization}.
Specifically, chevrons tend to reduce low-frequency noise by weakening large eddies, although this often comes at the cost of increased high-frequency noise due to elevated turbulence levels near the nozzle exit~\citep{saiyed1999tabs,tam2000subsonic,bridges2004parametric,callender2005far,alkislar2007effect}.
The balance between noise reduction and high-frequency penalty depends on chevron geometry parameters such as count, penetration depth, and asymmetry, which influence the strength and persistence of the induced vortices~\citep{bridges2004parametric,zaman2011evolution}.
Overall, the development of chevrons builds upon earlier studies with tabs and notches, in which streamwise vortex generation was identified as a key mechanism for both mixing enhancement and noise suppression.

Turbulent jets contain large-scale coherent structures, frequently referred to as wavepackets, which are known to underpin jet noise (see reviews by~\citet{jordan2013wave} and~\citet{cavalieri2019wave} and references therein).
Wavepackets are characterised by axially extended waves which, despite their low-energy content, can be effective in the generation of sound thanks to their space--time organisation~\citep{mollo1967jet,crow1971orderly,crighton1975basic}.
The most acoustically efficient sound-generating structures are those with lowest azimuthal wavenumber, and these are underpinned by the modal Kelvin--Helmholtz (K--H)~\citep{michalke1964inviscid,michalke1965spatially} and non-modal Orr~\citep{tissot2017wave} amplification mechanisms.
Other coherent structures recently identified in unbounded shear flows such as jets and mixing layers, are rolls and streaks~\citep{jimenezgonzalez2017transient,nogueira2019large,pickering2020lift,maia2023effect,maia2024effect}.
In wall-bounded flows these are generated by the lift-up effect~\citep{ellingsen1975stability,landahl1980note,brandt2014lift}, a non-modal mechanism characterised by longitudinal vorticity that transports momentum from low- to high-speed flow regions and vice versa~\citep{marusic2017scaling}.
This leads to low- and high-speed streaks in the streamwise direction with slow time scales.
These structures have been shown to be important in the turbulent dynamics of a range of free and wall-bounded flows~\citep{gustavsson1991energy,hultgren1981algebraic,smits2011high,hutchins2007evidence}.

Stability studies~\citep{sinha2016parabolized,marant2018influence,lajus2019spatial,wang2021effect} show that streaky structures can interact with K--H modes, decreasing their growth rates. Similar phenomena occur in wall-bounded flows, such as boundary-layers, where optimal (stable) streaky structures were found to stabilise the Tollmien--Schlichting instability mechanism ~\citep{fransson2004experimental,fransson2005experimental}, even delaying transition to turbulence~\citep{fransson2006delaying}.
Studies in turbulent flows also demonstrated the potential of using designed roughness elements to generate coherent streaks for control purposes~\citep{pujals2010forcing}, leading to lower drag in bluff bodies, flow separation control~\citep{pujals2010drag} and reduction of tonal noise in aerofoils at low Reynolds number~\citep{alva2023reduction}.

Coherent structures can be empirically extracted from experimental or numerical flow databases with data post-processing techniques such as spectral proper orthogonal decomposition (SPOD)~\citep{lumley1967structure,lumley1981coherent,picard2000pressure,towne2018spectral,schmidt2020guide}, which identifies structures that are optimal according to an energy metric, e.g. the turbulent kinetic energy (TKE).
On the other hand, simplified models based on linearisation of the Navier--Stokes equations about a mean flow~\citep{cavalieri2013wavepackets,rodriguez2015study, nogueira2019large,lesshafft2019resolvent,pickering2021resolvent,towne2022efficient}, can provide further insight on the physical mechanisms that underpin coherent structures in turbulent jets.  
Recently, one-way Navier–Stokes (OWNS) equations~\citep{towne2015one,zhu2023recursive} have been used to educe wavepackets from turbulent round~\citep{zhu2023recursive}, elliptical~\citep{nogueira2023prediction} and chevron~\citep{nogueira2024wavepackets} jets.
Such spatial marching methods enable computation of the full spatial structure of wavepackets at a fraction of the computational cost needed to perform global linear analysis while retaining all downstream-travelling modes, unlike parabolic stability equations (PSE) which captures only a single local mode.
OWNS is thus a well-suited tool for flows that contain multimodal effects, as is the case for chevron nozzles and other flows containing streak instability.

In this paper we use azimuthally-periodic tab elements (in the shape of small cylinders) inside a round nozzle to generate steady streaks in the nozzle boundary-layer, in contrast to devices such as chevrons and serrated extensions that are applied directly to the nozzle exit.
In this study, the tab design is informed by nozzle boundary layer measurements, with the specific objective of activating the lift-up mechanism within the nozzle boundary layer and optimising streak growth between the tabs and the nozzle exit plane~\citep{fransson2005experimental, pujals2010forcing}.
Motivated by previous studies that show how streaks can produce a passive control effect~\citep{fransson2004experimental,fransson2005experimental,fransson2006delaying,pujals2010forcing,pujals2010drag} and the attenuation of the growth of K--H wavepackets~\citep{sinha2016parabolized,marant2018influence,lajus2019spatial,wang2021effect}, we explore the possibility of obtaining noise reduction in turbulent jets through the same mechanism.
The tab design is based on the nozzle boundary-layer thickness for Mach number $M_j = 0.4$ and aims to produce significant growth of steady streaks between the tab position and the nozzle exit plane.
We investigate the effect of tabs on the acoustic and turbulent fields of turbulent jets at different Mach numbers.
The acoustic field is characterised using an azimuthal microphone array, such that the effect of tabs on the azimuthal modes of the sound field can be assessed.
The turbulent velocity field is characterised through time-resolved particle image velocimetry (TR-PIV) measurements in cross-flow planes at a number of streamwise positions (enabling Floquet-Fourier decomposition of the flow field) at $M_j = 0.7$.
The snapshot version of spectral proper orthogonal decomposition (SPOD)~\citep{towne2018spectral,schmidt2020guide} is used to extract structures from the flow field.
The Floquet-based decomposition is motivated by the $L$-fold rotational symmetry of the flow produced by tabs~\citep{sinha2016parabolized,lajus2019spatial,rigas2019streaks}.
Wavepacket modelling is achieved using a 3D recursive OWNS code, and the solutions are compared to the structures educed from the PIV data~\citep{nogueira2023prediction,nogueira2024wavepackets}.

The paper is organised as follows.
The tabs design strategy is addressed in \S~\ref{sec:nozzle_design} and the acoustic measurements, including the experimental setup and results, are provided in \S~\ref{sec:acoustic}.
The TR-PIV database is presented in \S~\ref{sec:piv}, which contains the experimental setup and a brief description of the post-processing procedures (Floquet decomposition and SPOD), as well as the main results.
Wavepacket modelling, with a brief description of the OWNS methodology, is shown in \S~\ref{sec:wavepacket}.
A summary of the conclusions is given in \S~\ref{sec:conclusions}.

\section{Nozzle design}
\label{sec:nozzle_design}

\begin{figure}
	\centering
	\includegraphics[width=\textwidth]{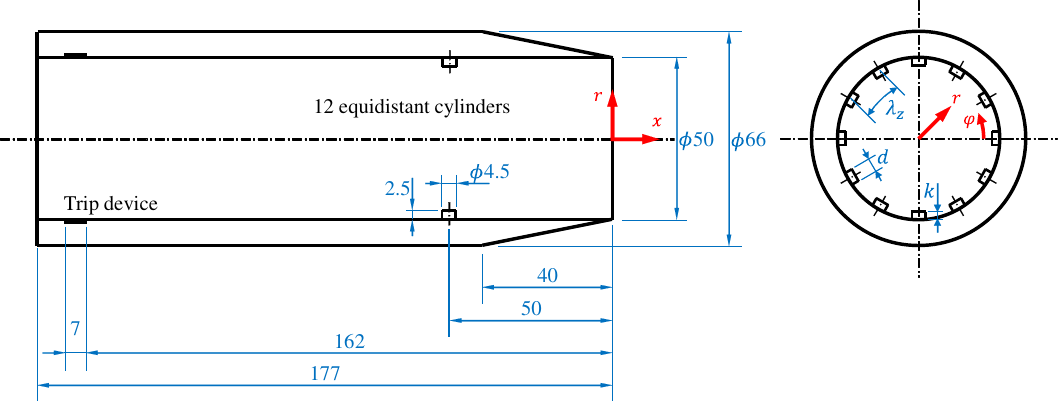}
	\caption{Sketch of the nozzle with tabs and coordinate system (in red colour). Left frame: longitudinal section view (flow is directed from left to right). Right frame: frontal section view (flow is directed towards the reader, leaving the page). Dimensions in millimetres.}
	\label{fig:sketch_nozzle}
\end{figure}

Two different nozzles are employed in this study.
The first is a baseline geometry without tabs.
The second is a tabbed geometry, containing 12 small cylinders equally distributed in the azimuthal direction on the nozzle internal surface, one diameter upstream of the exit.
Figure~\ref{fig:sketch_nozzle} shows a sketch of the nozzle with tabs, indicating all relevant parameters and dimensions.
The baseline nozzle has the same dimensions and characteristics as the nozzle with tabs, except for the cylindrical elements.
Indications of the coordinate system are also provided in the figure in red colour, with $r$, $x$ and $\varphi$ as the radial, streamwise and azimuthal directions, respectively.

Both nozzles have an internal diameter of $D = 50$ mm and are 177-mm long.
Experiments were performed with the nozzles containing a boundary-layer trip, i.e., a transition element made of carborundum.
The boundary-layer trip is located 162 mm upstream the nozzle exit ($x/D = -3.24$), is 7-mm long ($l/D = 0.14$, where $l$ is the length of the trip) and was designed following practices established by~\citet{cavalieri2013wavepackets}.
The carborundum strip is designed as a separate ring piece, which can be removed in both baseline and tabbed nozzles. 

A key difference between the present work and earlier studies of jets in tabs (see review by~\citet{zaman2011evolution} and references therein) is the design of tabs based on the growth of streaks in the turbulent boundary layer, such that one obtains large-amplitude streaks at the nozzle exit with small roughness elements.
The design of the tab elements (small cylinders) followed the works by~\citet{fransson2005experimental} and~\citet{pujals2010forcing}, aiming to generate boundary-layer streaks with significant downstream development.
The boundary layer thickness ($\delta$) upstream the nozzle exit at $x/D = -1$ and $M_j = 0.4$ was measured by~\citet{kaplan2021nozzle}, and is used in the present calculations.
The design follows the parameters of case B in~\citet{pujals2010forcing}, for which the cylinder height is set at $k = 0.8 \delta$ and the ratio between the cylinder height and diameter is $k/d = 0.597$.
Based on the boundary-layer thickness provided by~\citet{kaplan2021nozzle}, this gives $k = 2.8$ mm and $d = 4.1875$ mm.
For machining reasons, these parameters are approximated as $k = 2.5$ mm and $d = 4.5$ mm.

The cylinder spacing is set at $\lambda_z/d \approx 3$, such that one expects significant streak amplitudes at a downstream distance $L_x = 4\lambda_z \approx 48$ mm; hence, tabs are placed 50 mm upstream of the nozzle exit.
The cylinder spacing $\lambda_z/d \approx 3$ is slightly different from the optimal parameter $\lambda_z/d = 4$ used by~\citet{pujals2010forcing}.
Nevertheless, we will show in the following that it also leads to the development of strong streaks in the jet.
Moreover, the number of tabs is given as $L = \pi D/\lambda_z = 12$, i.e. the nozzle perimeter divided by the cylinder spacing.
The left frame of figure~\ref{fig:sketch_nozzle} shows a sketch of the tabs together with the design parameters, including the cylinder height ($k$), diameter ($d$) and spacing ($\lambda_z$).

Note that the tab geometry employed in this study is based on nozzle boundary-layer measurements at $M_j = 0.4$~\citep{kaplan2021nozzle}.
However, the nozzle was tested at Mach numbers up to 0.9, hence operating off-design.
Nevertheless, as will be seen in the next sections, even for Mach numbers different from 0.4, the tabbed geometry was able to generate large-amplitude streaks at the nozzle exit and to affect both the jet noise signature and the hydrodynamic field.
To obtain the largest-amplitude streaks at each Mach number, an adaptive geometry would be required, since the tab dimensions depend on the boundary-layer thickness~\citep{pujals2010forcing}, which varies with Mach number.

\begin{figure}
	\centering
	\includegraphics[width=0.50\textwidth]{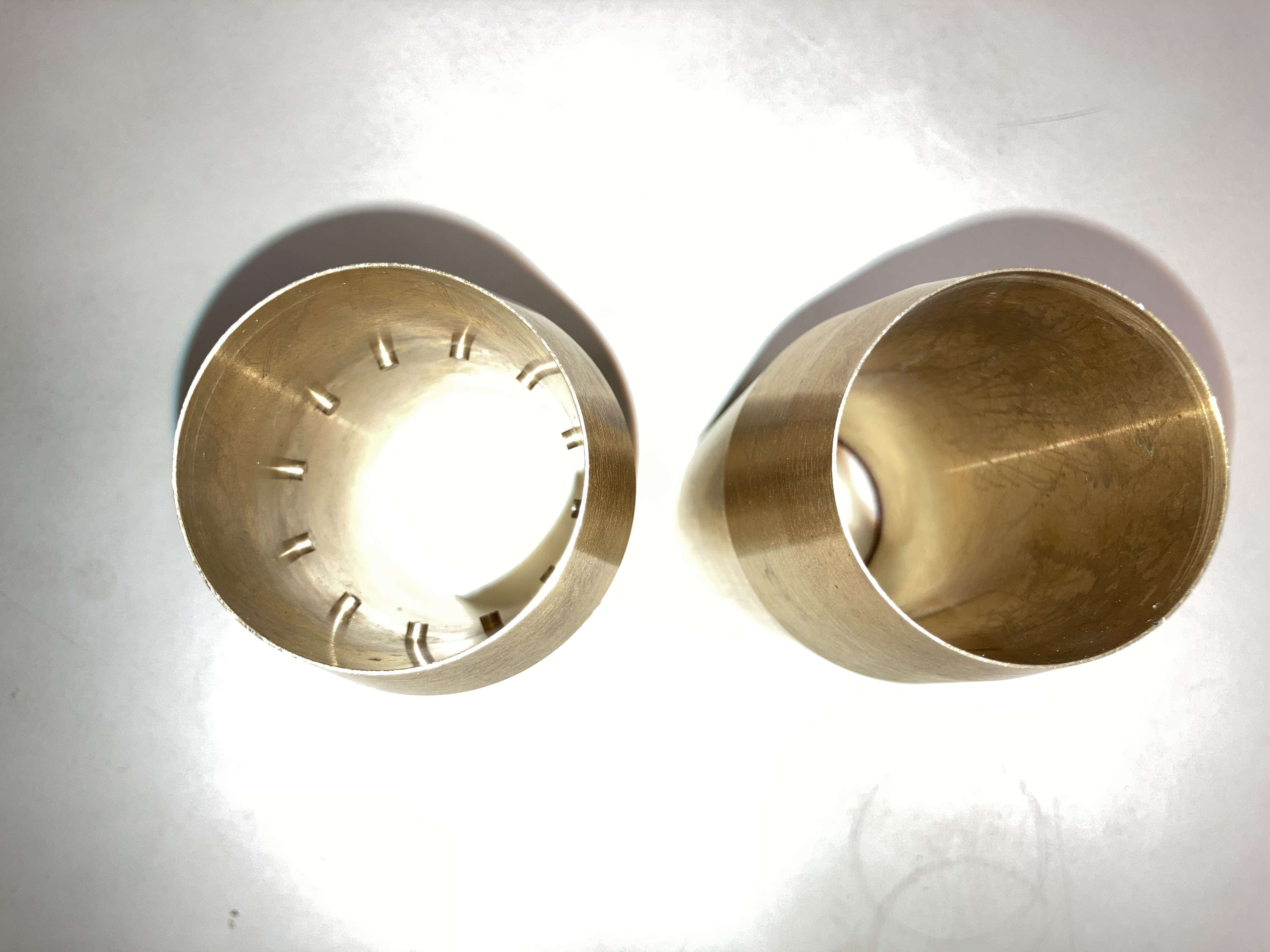}
	\caption{Picture of the two studied nozzles. Left: nozzle with the cylindrical tabs. Right: baseline nozzle.}
	\label{fig:picture_nozzle}
\end{figure}

\begin{figure}
	\centering
	\begin{subfigure}{0.48\textwidth}
		\includegraphics[width=\textwidth]{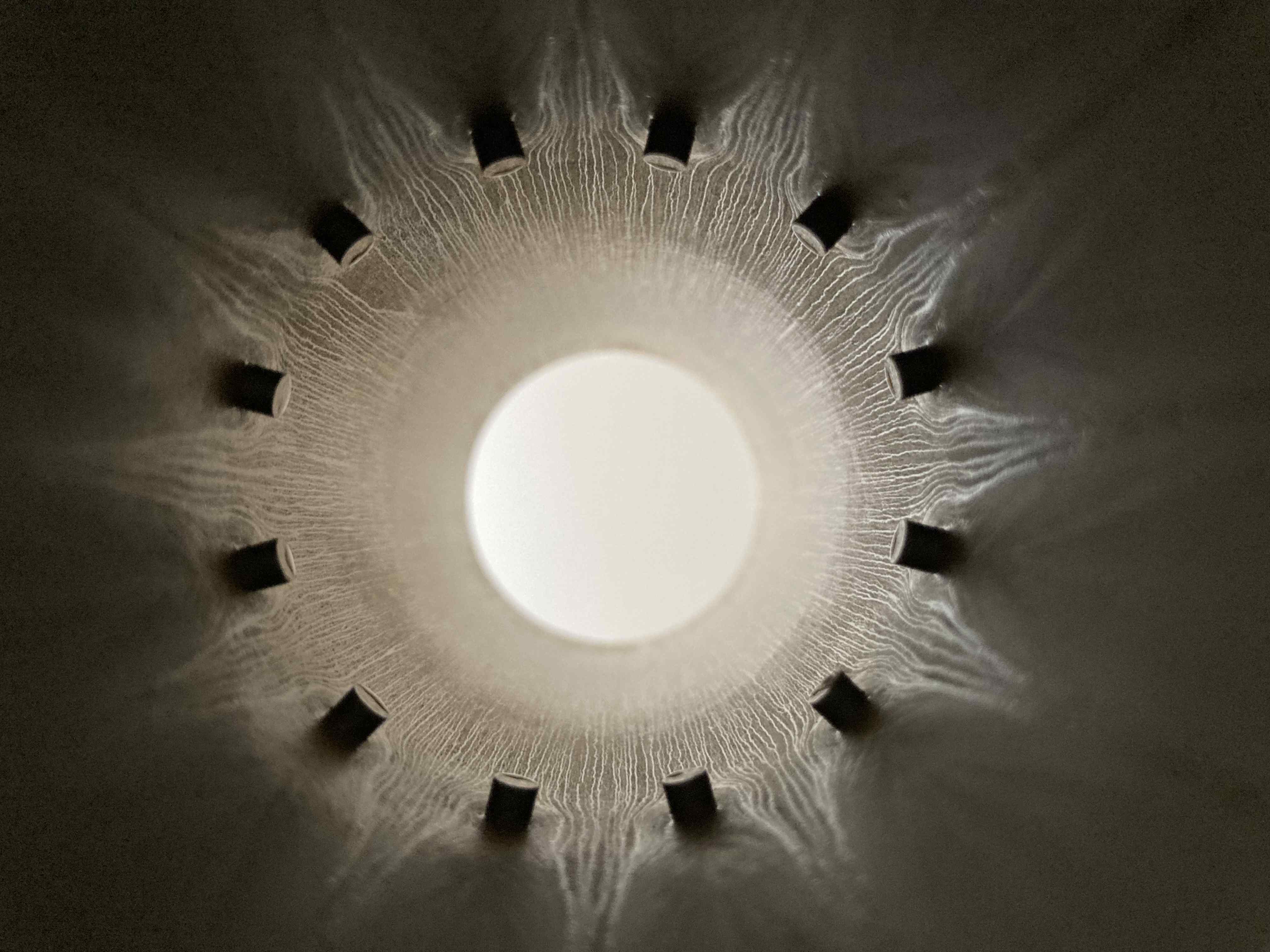}
		\caption{Front view}
		\label{fig:tabbed_nozzle_streaks_front}
	\end{subfigure}
	\begin{subfigure}{0.48\textwidth}
		\includegraphics[width=\textwidth]{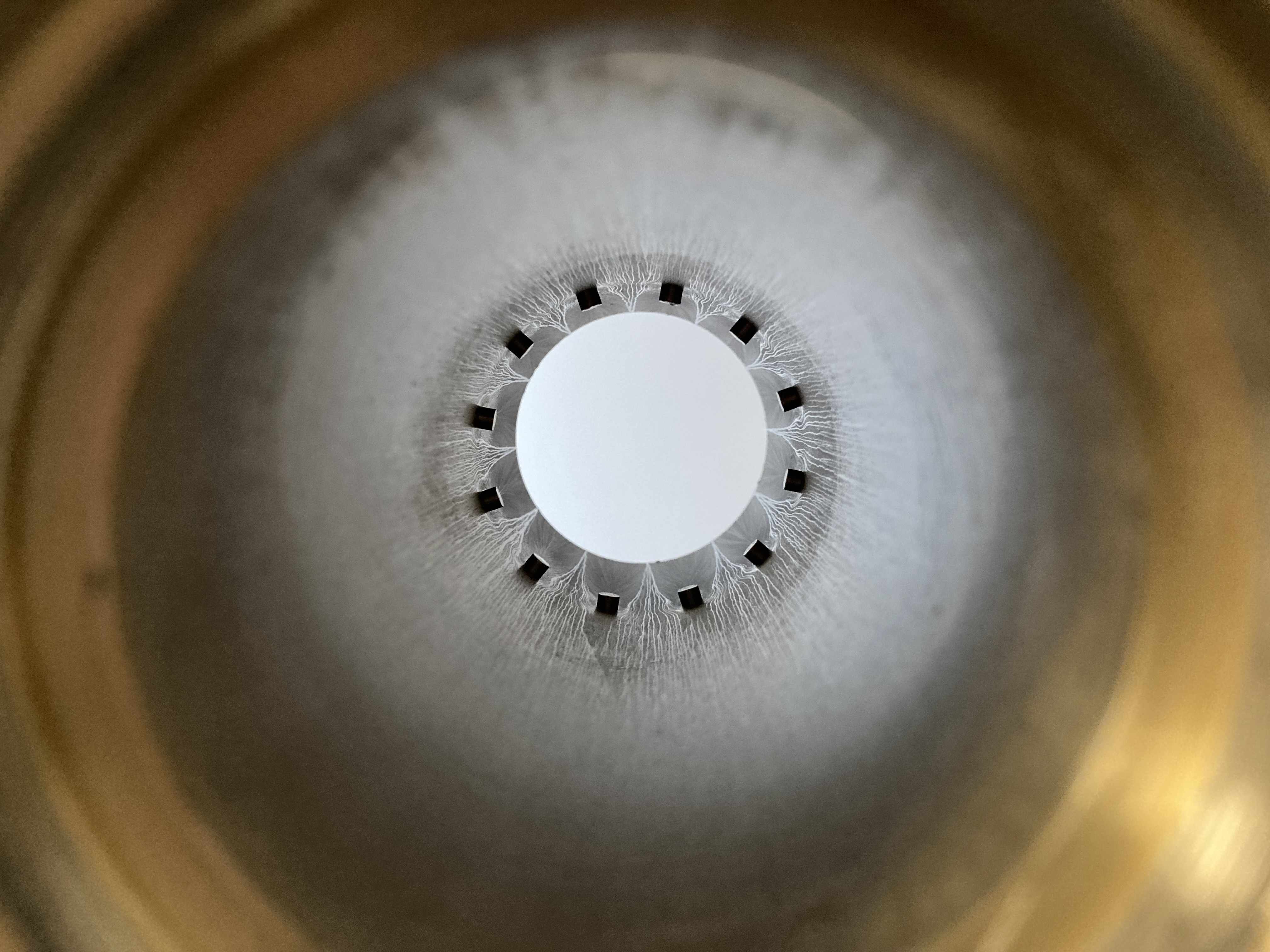}
		\caption{Rear view}
		\label{fig:tabbed_nozzle_streaks_rear}
	\end{subfigure}
	\caption{Pictures of the internal walls tabbed nozzle after the stereo PIV experiments, where it is possible to verify the flow footprints left on the nozzle walls by the seeding particles.}
	\label{fig:tabbed_nozzle_streaks}
\end{figure}

Figure \ref{fig:picture_nozzle} displays a picture of the nozzles employed in this study side by side.
The left nozzle contains the cylindrical tabs, whereas the right one regards the baseline configuration.
Figure \ref{fig:tabbed_nozzle_streaks} shows the internal walls of the tabbed nozzle with the footprints left by the TR-PIV seeding particles.
The pictures were taken from the nozzle front (nozzle exit) and rear (nozzle entry) views, and it is possible to see the flow signature, which are directed from the nozzle entry to exit.
The flow signature shows lines that keep straight and the footprints are very strong up to the cylindrical tabs.
At the tabs wake, the footprints are no longer visible at the walls.

\section{Acoustic measurements}
\label{sec:acoustic}

\subsection{Experimental setup}
\label{sec:acoustic_setup}

The experiments were performed at the \emph{Bruit \& Vent} facility, located at the \emph{Prometée plateforme}, in Chasseneuil-du-Poitou (France).
The facility is composed of an anechoic room, measuring $12.6 \times 10.6 \times 7.85$ m$^3$ and a cut-off frequency of 200 Hz.
The facility allows jet flows up to Mach number 1.4.
Acoustic measurements were conducted with a circular microphone array with radius $R/D = 14.3$, containing 18 B\&K type 4944-B 1/4-inch pressure-field microphones, equally distributed in the azimuthal direction, with $\Delta \varphi = 20$ deg azimuthal angle between neighbouring microphones~\citep{amaral2023nozzlestreaks}.

\begin{figure}
	\centering
	\includegraphics[width=\textwidth]{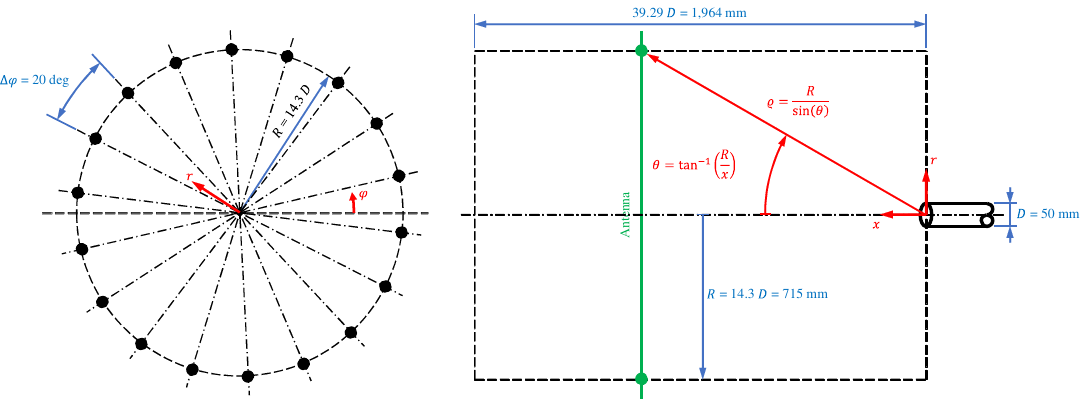}
	\caption{Sketch representing the microphones distribution over the circular antenna (right frame) and the cylindrical surface covered by the acoustic measurements (left frame). Coordinate system in red colour.}
	\label{fig:sketch_antenna}
\end{figure}

The right frame of figure~\ref{fig:sketch_antenna} shows a sketch of the microphones distribution on the circular antenna.
The coordinate system, i.e., the radial and azimuthal directions ($r$ and $\varphi$, respectively), is shown in red colour.
The streamwise direction ($x$), not shown in the figure, points out of the page toward the reader.

A side view of the experimental apparatus is shown in the left frame of figure~\ref{fig:sketch_antenna}.
The coordinate system, shown in red, has origin at the nozzle exit centre, i.e. $(x/D,r/D) = (0,0)$.
Moreover, $\theta = \tan^{-1}(R/x) = \tan^{-1}(14.3D/x)$ indicates the polar angle and $\varrho = R/\sin(\theta)$ is the polar radius.
With the aid of a traverse system, the antenna, represented in green, can be moved in the streamwise direction~\citep{hasparyk2023two,hasparyk2024shape,amaral2023elliptical,amaral2023installed}.
Hence, in the current setup, it is possible to measure the sound in a cylindrical surface of $R = 14.3 D = 715$ mm radius in the $0 \leq x/D \leq 39.29$ streamwise range, corresponding to a polar range of 20 deg $\leq \theta \leq$ 90 deg.
The current measurements were discretised using a 5-degree step in polar angle ($\Delta \theta =$ 5 deg).
A projection of the cylindrical surface used in the acoustic experiments is denoted by the dashed rectangle in the figure, which contains relevant dimensions.

Acoustic signals were sampled at 200 kHz, with 10 seconds of acquisition time.
Regarding jet Mach numbers, measurements were conducted in the range $0.4 \leq M_j \leq 0.9$ with increments of $\Delta M_j = 0.1$.
The flow was set at isothermal condition for all Mach numbers studied herein.~\citet{amaral2023installed} describes with more details the control strategies employed to maintain the jet in isothermal condition and constant Mach number during the experiments.
Table \ref{tab:Mach_Re_U} displays the Reynolds number $Re_D = \frac{U_j D}{\nu}$, with $\nu$ the kinematic viscosity, and $U_j$ the jet exit speed for the Mach number of interest.

\begin{table}
  \begin{center}
		\def~{\hphantom{0}}
		\begin{tabular}{c | c c c c c c}
			$M_j$				& 0.4								& 0.5								& 0.6								& 0.7								& 0.8								& 0.9								\\
			$Re_D$			& $4.5 \times 10^5$	& $5.6 \times 10^5$	& $6.7 \times 10^5$	& $7.8 \times 10^5$	& $8.9 \times 10^5$	& $1.0 \times 10^6$	\\
			$U_j$ [m/s]	& 135								& 170								& 205								& 240								& 275								& 310								\\
		\end{tabular}
		\caption{Values of Reynolds number ($Re_D$) and jet flow speed at the nozzle exit ($U_j$) as a function of the jet Mach number ($M_j$).}
		\label{tab:Mach_Re_U}
  \end{center}
\end{table}

\subsection{Data post-processing}
\label{sec:acoustic_processing}

The first step to post-process the data began with the application of a Gaussian filter over the time series prior to downsampling the data from 200 kHz to 100 kHz.
The pressure measurements $\boldsymbol{p}$ were then Fourier decomposed in the azimuthal direction as
\begin{equation}
	\boldsymbol{p}\left(x,r=R/D,\theta,t\right) = \sum_{m} \boldsymbol{p}_m\left(x,r=R/D,t\right) e^{i m \theta} + \mbox{c.c.} \mbox{,}
	\label{eq:p_fourier_azimuth}
\end{equation}
\noindent where $i = \sqrt{-1}$, $t$ denotes time, $m$ denotes the azimuthal mode and $\mbox{c.c.}$ indicates the complex conjugate.
Although the tabbed nozzle is not axisymmetric, we chose to show the acoustic results in terms of azimuthal modes in order to compare the tabbed nozzle results with the reference sound radiation by the baseline jet, whose azimuthal mode decomposition of the sound field is well documented~\citep{jordan2013wave}.

The noise spectra were evaluated using Welch's method~\citep{welch1967fft}, by segmenting the pressure time series into blocks of $N_b = 2,048$ size, with 75\% overlap.
Each block was windowed using a Hann function prior to computing the fast Fourier transform (FFT).
The resulting spectra were obtained by averaging the power spectra of each segment.
Correction factors associated with the windowing function energy loss as well as with the mirroring of spectra to negative frequencies owing to the FFT real-valued signals.
The spectral representation of each azimuthal component of the data is given by
\begin{equation}
	\hat{\boldsymbol{p}}_m\left(x,r=R/D,\omega\right) = \sum_\omega \boldsymbol{p}_m\left(x,r=R/D\right) e^{i \omega t} + \mbox{c.c.} \mbox{,}
	\label{eq:p_fourier_time}
\end{equation}
\noindent where $\omega = 2 \pi f$ denotes the frequency and hat indicates component transformed from time to frequency domain.

The dimensionless power spectral density (PSD, in dB/St) was obtained for each azimuthal mode and Strouhal number following~\citet{bres2018importance}, 
\begin{equation}
	\mathit{PSD} = 10 \log_{10}\left[\frac{\text{diag}\left(\mathsfbi{\hat{P}}\right)}{{p_\mathit{ref}}^2}\frac{U_j}{D}\frac{1}{{M_j}^4}\right] \mbox{,}
	\label{eq:psd}
\end{equation}
\noindent where $\mathsfbi{\hat{P}} = \left\langle \hat{\boldsymbol{p}}_m \hat{\boldsymbol{p}}_m^{\dagger} \right\rangle$ indicates the pressure cross-spectrum density (CSD), in Pa$^2$/St, $\text{diag}\left(\cdots\right)$ denotes only the diagonal values are gathered (i.e., the pressure auto-spectrum) and $p_\mathit{ref} = 2 \times 10^{-5}$ Pa is the reference acoustic pressure level.
All the data will be presented as function of Strouhal number, which is given by
\begin{equation}
	St = \frac{f D}{U_j} \mbox{.}
	\label{eq:St}
\end{equation}

\subsection{Results}
\label{sec:acoustic_results}

\begin{figure}
	\centering
	\includegraphics[width=\textwidth]{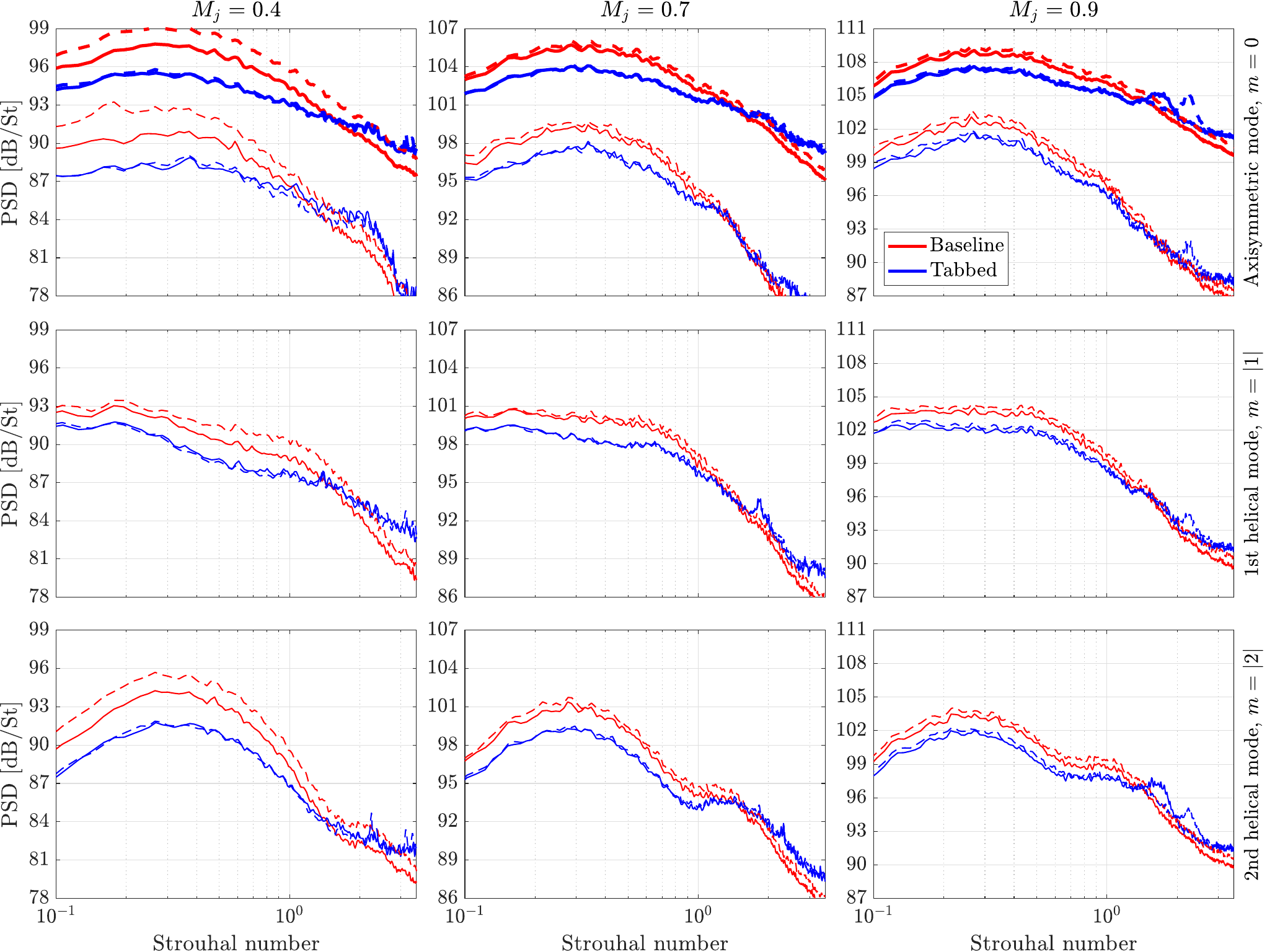}
	\caption{Acoustic spectra for polar angle $\theta =$ 90 deg (nozzle exit, sideline) for both tripped ($-$) and untripped ($--$) configurations. Frames, from top to bottom: axisymmetric mode ($m =$ 0), 1st helical mode ($|m| =$ 1) and 2nd helical mode ($|m| =$ 2). Frames, from left to right: Mach 0.4, 0.7 and 0.9. The top frames display the full sound radiation without azimuthal decomposition, shown as thick curves.} 
	\label{fig:spectra_az_theta90}
\end{figure}

\begin{figure}
	\centering
	\includegraphics[width=\textwidth]{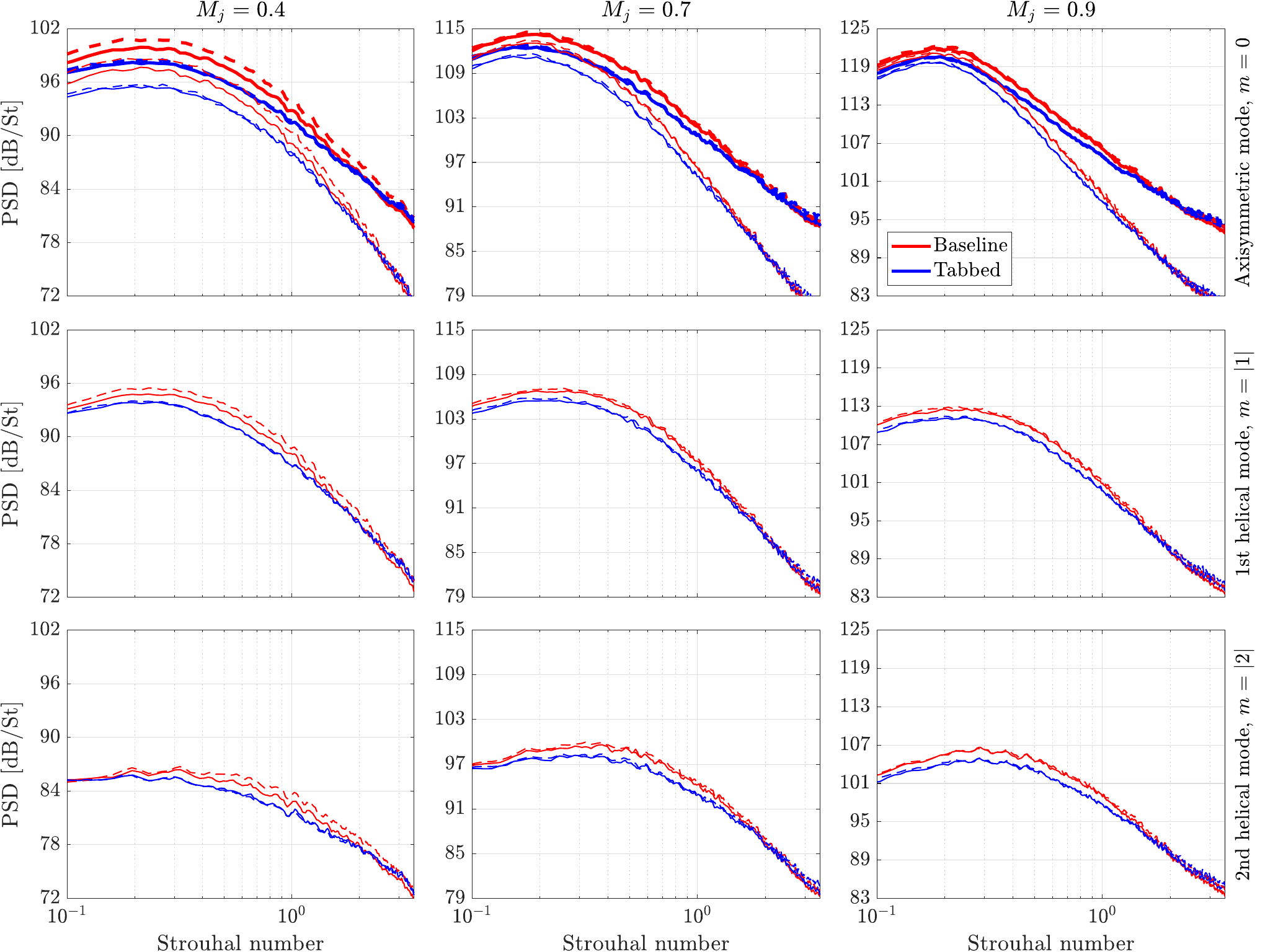}
	\caption{Acoustic spectra for polar angle $\theta =$ 30 deg (downstream). See the comments in the caption of figure~\ref{fig:spectra_az_theta90}.}
	\label{fig:spectra_az_theta30}
\end{figure}

Figures~\ref{fig:spectra_az_theta90} and \ref{fig:spectra_az_theta30} show noise spectra at 90 deg (nozzle exit plane, sideline) and 30 deg (downstream) polar angles, respectively, for the baseline (red curves) and tabbed (blue curves) nozzles.
The figures display the spectra for the axisymmetric mode, $m = 0$, the first two helical modes, $|m| = 1$ and $|m| = 2$ (where $|\cdots|$ meaning both the positive and negative counterparts of the helical modes are taken into account), and Mach numbers $M_j =$ 0.4, 0.7 and 0.9.
Results are shown for the nozzles with a boundary-layer trip (continuous lines) and without it (dashed lines), with the full sound radiation shown as thick curves.
Results for $M_j =$ 0.5, 0.6 and 0.8 follow the same trend, and are included in the appendix \ref{app:acoustic_extra}.
The acoustic spectra are presented in terms of azimuthal decomposition, following verification of far-field homogeneity in the present measurements—an observation that is also supported by previous studies~\citep{tam2000subsonic}.
Appendix~\ref{app:acoustic_extra} provides an example of the acoustic spectra prior to azimuthal decomposition, where the homogeneity of the acoustic field can be verified.
 
Figure~\ref{fig:spectra_az_theta90} shows the sideline results.
Starting at the lowest Mach number studied ($M_j=0.4$), one can observe a noise reduction of up to 3 dB/St in the axisymmetric component of the tripped jet, for a wide range of Strouhal numbers, up to $St \approx$ 1, with larger reductions at lower frequencies.
The reduction is even more pronounced for the untripped configurations, with the tabbed nozzle presenting a noise reduction of up to 6 dB/St for $M_j =$ 0.4.
Differences between tripped and untripped round jets are expected, since the resulting noise is dependent on the state of the boundary layer inside the nozzle~\citep{bres2018importance}.
It should be noted that one can only ensure that the boundary layer has transitioned to turbulence upstream of the nozzle exit when the trip ring is employed, especially at lower Reynolds numbers.
Interestingly, the presence of the boundary layer trip has virtually no effect on the noise generated by the tabbed jets, indicating that the boundary layer dynamics in this case might be dominated by the tab-generated streaks, potentially causing the flow to transition even without the presence of the trip.
Increasing the Mach number reduces the gap between the tripped and untripped spectra, diminishing the noise reduction provided by the tabs to approximately 2 dB/St for $M_j =$ 0.7 and 1.5 dB/St for $M_j =$ 0.9.
It is expected that the tabs will be more effective at $M_j =$ 0.4, since their design was based on boundary layer thickness obtained at this Mach number (\S~\ref{sec:nozzle_design}).
Similar trends are observed for the first two helical modes. 

In most cases, for the sideline results, a small increase in the noise generated is observed at higher frequencies, around the roll-off region of the spectrum.
However, the sound produced at these frequencies has lower amplitudes, which mitigates the impact of this amplification. In some cases ($M_j=0.9$, $m=0$ for instance), small peaks are observed in the spectrum at high frequencies in the tabbed jets, which could be related to a secondary instability of the strong streaks generated by the tabs. 

At the polar angle $\theta =$ 30 deg, the noise reduction is not as pronounced, as shown in figure~\ref{fig:spectra_az_theta30}.
However, reductions are still observed in a wide range of frequencies, with benefits of up to 2 dB/St for the tripped nozzle and 3 dB/St for the untripped one (for the $M_j =$ 0.4, $m =$ 0 case).
As before, increasing the Mach number reduces the trip and tab effects, but reductions of up to 1.5 dB/St are still attained for $M_j =$ 0.9.
As before, the boundary layer trip is more effective for the baseline configuration, and has virtually no effect on the tabbed nozzles.
As observed in~\citet{cavalieri2012axisymmetric}, the axisymmetric mode shows higher intensity levels at lower frequencies and also dominates the spectra for lower polar angles (e.g. at $\theta=$ 30 deg).
The helical modes, on the other hand, tend to dominate the spectra at higher polar angles (e.g. at $\theta=$ 90 deg).

\begin{figure}
	\centering
	\includegraphics[width=\textwidth]{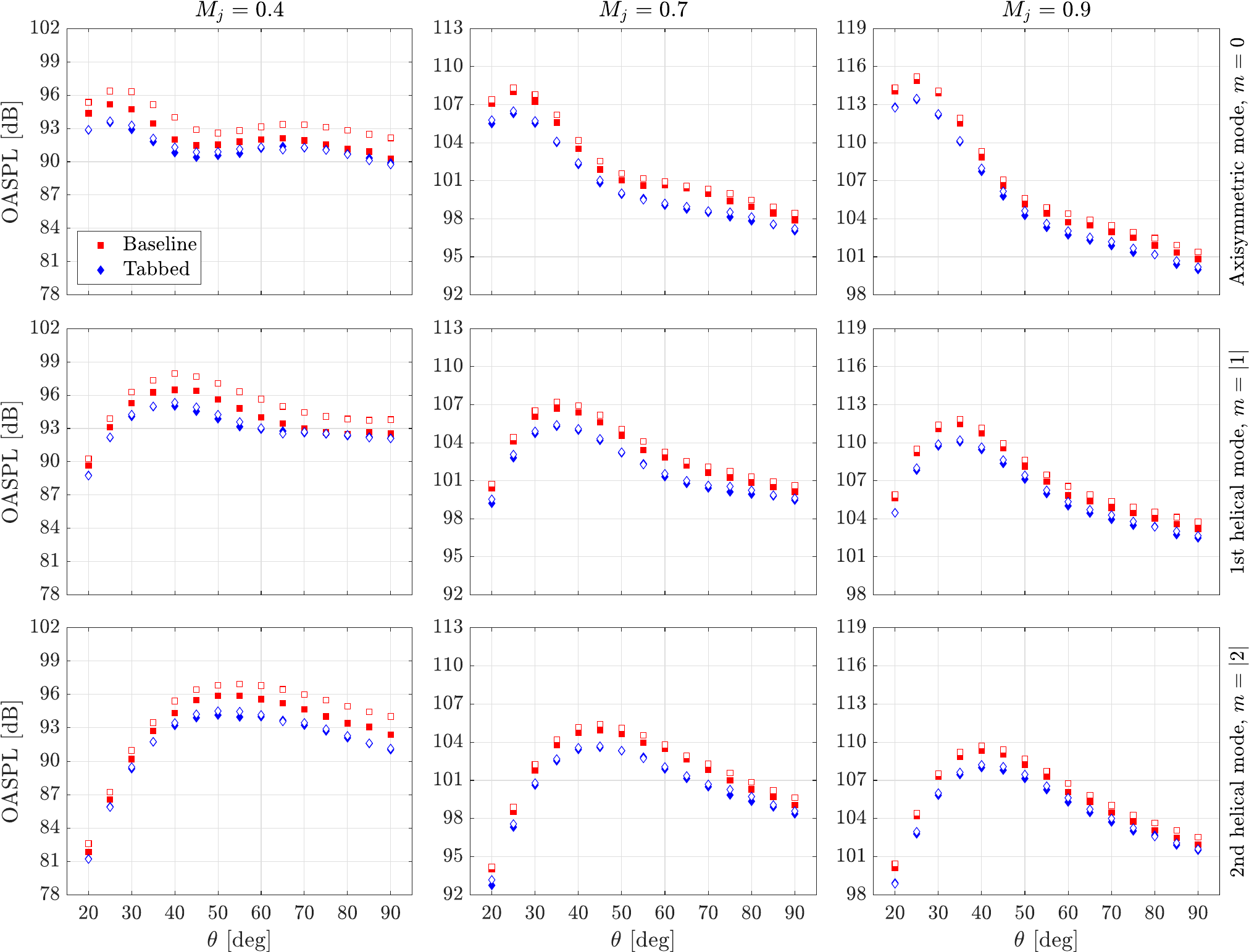}
	\caption{Overall sound pressure level for polar angles in the $\theta =$ 20 deg to 90 deg interval for tripped (closed symbols) and untripped (open symbols) configurations. Power levels evaluated for 0.1 $\leq St \leq$ 3.5. Frames, from top to bottom: axisymmetric mode ($m =$ 0), 1st helical mode ($|m| =$ 1) and 2nd helical mode ($|m| =$ 2). Frames, from left to right: Mach 0.4, 0.7 and 0.9.} 
	\label{fig:oaspl_az}
\end{figure}

The overall sound pressure level (OASPL) as a function of the polar angle for the same cases shown in figures~\ref{fig:spectra_az_theta90} and \ref{fig:spectra_az_theta30} is shown in figure~\ref{fig:oaspl_az}.
To evaluate the OASPL as a function of polar angle, the microphone traverse was moved downstream to obtain measurements at the desired angular increment $\delta \theta$, and the were spectra adjusted using a far-field assumption~\citep{cavalieri2012axisymmetric}.
The noise spectra were integrated over the 0.1 $\leq St \leq$ 3.5 range.
Noise reductions here follow roughly the same trends observed in the previous spectra; for instance, for the axisymmetric mode, noise reductions of up to 2 and 3 dB are observed when applying the tabs for both the tripped and untripped configurations, respectively, especially up to $\theta \approx$ 60 deg.
For the tripped case with $M_j=0.4$ and $m=0$, the noise benefit reduces for higher polar angles.
Nonetheless, consistent reductions in OASPL are observed in all analysed cases and across all polar angles, indicating that the reduction trends observed in the spectra are consistent.
For instance, for higher azimuthal modes, untripped configuration and low Mach numbers, the tabbed nozzle provides noise reductions of up to 4 dB in these conditions.
As in~\citet{cavalieri2012axisymmetric}, the axisymmetric mode shows a marked directivity towards low polar angles, whereas the helical modes are primarily directed in the 35 to 55 deg polar angle range.
The axisymmetric mode exhibits higher levels at low polar angles, whereas the helical modes dominate for polar angles higher than 30 degrees.
Further comparisons for other Mach numbers also exhibit the same behaviour, and can be found in appendix \ref{app:acoustic_extra}.

Overall, the acoustic data show that the application of tabs consistently reduces the acoustic radiation of all the jets studied for a wide range of subsonic Mach numbers, even though tabs were only designed to induce significant streak growth at the lowest Mach number studied. 
Reductions are observed at all angles, and for all azimuthal wavenumbers, without the usual high-frequency penalty observed in chevron jets~\citep{callender2005far,zaman2011evolution}.
Considering the link between coherent structures and noise radiation~\citep{mollo1967jet,crighton1975basic,jordan2013wave,cavalieri2019wave}, a physical explanation for these reductions requires detailed analysis of the flow hydrodynamics and the changes in coherent structures induced by the presence of streak-generating tabs.
This is addressed in the following.

\section{Time-resolved velocity measurements}
\label{sec:piv}

\subsection{Experimental setup}
\label{sec:piv_setup}

Time-resolved particle image velocimetry (TR-PIV) measurements were based on LaVision hardware, using two Phantom 2640 CMOS cameras and a Continuum Mesa solid Nd-YAG laser.
The optical setup was designed to obtain a laser sheet parallel to the nozzle plane and the seeding particles made of atomized glycerin and heavy fog fluid.
The laser sheet thickness and the mean size of the seeding particles are approximately 3 mm and 1.5 $\mu$m, respectively.
As in the acoustic experiments, the flow was set at isothermal condition, and the Mach number was chosen as $M_j = 0.7$ (see table \ref{tab:Mach_Re_U}).
LaVision hardware and software (Davis) were employed to perform the TR-PIV measurements, with an acquisition frequency of 12.5 kHz (which yields a Nyquist Strouhal number of approximately $\mathit{St}_\mathit{Nyquist} \approx 1.3$), and the time between two laser exposures was set at 2 $\mu$s.
Measurements were performed in cross-flow planes for streamwise stations from near the nozzle exit up to $x/D = 10$, with the positions specified in table \ref{tab:PIV_stations}.
These were chosen so as to capture the initial growth and saturation of the K--H mode, the end of the potential core, and to evaluate the recircularisation of the jet further downstream.
The three velocity components were measured for each cross-flow plane.
After processing the raw data, the snapshots contain $(N_y,~N_z) = (194,~142)$ grid points in Cartesian coordinates and the field of view has approximately $250 \times 200$ mm$^2$ ($5 D \times 4 D$).

\begin{table}
  \begin{center}
		\def~{\hphantom{0}}
		\begin{tabular}{c | c c c c c c}
			$x$ [mm]	& 1.5		& 5		& 50	& 150	& 290	& 500	\\
			$x/D$			& 0.03	& 0.1	& 1		& 3		& 5.8	& 10	\\
		\end{tabular}
		\caption{Streamwise positions measured in the TR-PIV campaign.}
		\label{tab:PIV_stations}
  \end{center}
\end{table}

\begin{figure}
	\centering
	\begin{subfigure}{0.55\textwidth}
		\includegraphics[width=\textwidth]{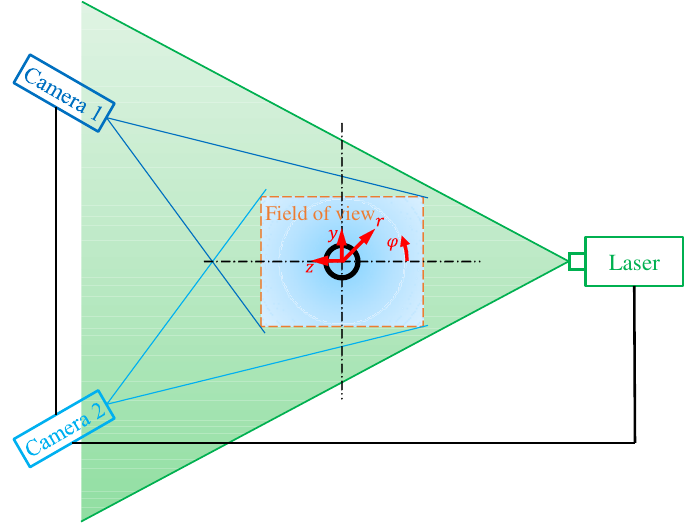}
		\caption{Front view}
		\label{fig:sketch_PIV_front}
	\end{subfigure}
	\begin{subfigure}{0.43\textwidth}
		\includegraphics[width=\textwidth]{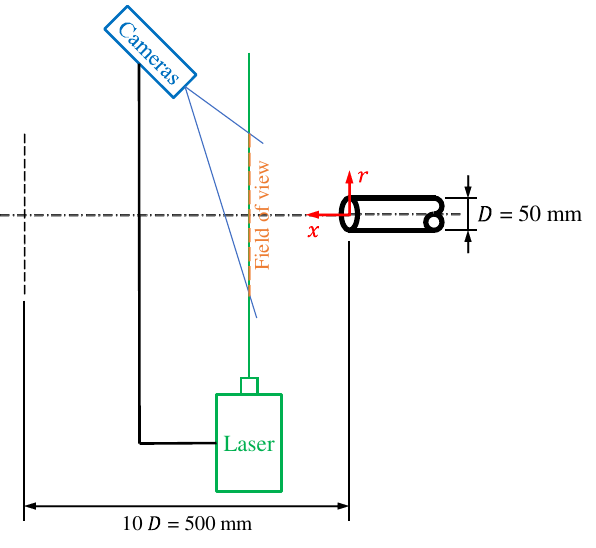}
		\caption{Top view}
		\label{fig:sketch_PIV_top}
	\end{subfigure}
	\caption{Sketch representing the TR-PIV setup, containing the nozzle, cameras, laser, field of view, coordinate system and all the relevant dimensions.}
	\label{fig:sketch_PIV}
\end{figure}

Figure~\ref{fig:sketch_PIV} shows a sketch of the stereo PIV setup, which includes the coordinate system used in this study.
The streamwise direction is denoted by $x$, whereas $(y,~z)$ indicate the cross-flow directions in Cartesian coordinates and $(r,~\varphi)$ are the radial and azimuthal directions in polar coordinates, i.e. $r = \sqrt{y^2+z^2}$, $y = r \cos(\varphi)$ and $z = -r \sin(\varphi)$.
A traverse system enables the movement of the cameras and laser in the streamwise direction keeping the cameras and laser relative position fixed for all measurements.

\subsection{Data post-processing}
\label{sec:piv_processing}

The first step of the data post-processing is the interpolation from the Cartesian PIV grid to a polar coordinate system.
Prior to this, it is required to find the jet axis $(y_c,~z_c)$ for each measurement plane ($x$).
The mean streamwise velocity component ($u$) was used to determine the geometric centre of the jet axis, following the procedures by~\citet{gudmundsson2011instability}, i.e.
\begin{subeqnarray}
	y_c &=& \frac{\sum_{n=1}^{N_y} \bar{u}_n y_n}{\sum_{n=1}^{N_y} \bar{u}_n} \\
	z_c &=& \frac{\sum_{n=1}^{N_z} \bar{u}_n z_n}{\sum_{n=1}^{N_z} \bar{u}_n} \mbox{,}
	\label{eq:center}
\end{subeqnarray}
\noindent where $n$ denotes a grid point index.
A bi-cubic interpolation scheme was employed to obtain the snapshots in polar coordinates~\citep{jaunet2017two, lesshafft2019resolvent, nogueira2019large}, using a grid composed of $(N_\varphi,~N_r) = (144,~75)$ points in the azimuthal and radial directions, respectively.

The jet flow issuing from the tabbed nozzle exhibits a discrete rotational symmetry due to the presence of $L = 12$ tabs, i.e. the flow is invariant under rotations of $2\pi/L$.
Consequently, the flow is not axisymmetric, but periodic in the azimuthal direction with period $2\pi/L$.
This periodicity allows the flow to be decomposed using a Floquet expansion in the azimuthal direction.
In this decomposition, the flow state can be represented as a superposition of azimuthal harmonics grouped according to their Floquet coefficient $\mu \in (-L/2,L/2]$, which defines a symmetry group of the solution.
Each Floquet coefficient corresponds to a family of azimuthal wavenumbers $m$, given by the relation
\begin{equation}
	m = \mu - jL \mbox{,}
	\label{eq:m_azimuth}
\end{equation}
where $j$ is an integer indexing the repetitions of the pattern every $L$ harmonics.
Thus, the flow field can be expressed as~\citep{sinha2016parabolized,lajus2019spatial,rigas2019streaks}
\begin{equation}
	\boldsymbol{q}(x,r,\varphi,t) = \sum_{\omega} \sum_{\mu} \sum_{j} \boldsymbol{\tilde{q}}(x,r) e^{i \omega t + i(\mu - jL)\varphi} + \mbox{c.c.} \mbox{,}
	\label{eq:q_Floquet}
\end{equation}
\noindent where $\boldsymbol{q}$ is the physical space solution and $\boldsymbol{\tilde{q}}$ represents the spatial structure of the mode associated with the Floquet coefficient $\mu$, frequency $\omega = 2 \pi f$, and azimuthal wavenumber $m = \mu - jL$.

This formulation makes explicit the fact that in the presence of azimuthal periodicity, the wavenumbers are grouped: each Floquet mode $\mu$ contains contributions from an infinite set of azimuthal harmonics spaced by $L$.
For instance, for $L = 12$ and an azimuthal resolution of $N_\varphi = 144$ points, one obtains $j \in (-N_\varphi/2L,N_\varphi/2L] = (-6,6]$, resulting in twelve harmonics per Floquet group.

For clarity, the correspondence between the Floquet coefficient $\mu$ and the set of azimuthal wavenumbers $m$ is summarized in table~\ref{tab:Floquet}.
In this work, both the baseline (axisymmetric) and tabbed nozzle configurations are decomposed using this Floquet-based approach to ensure a consistent comparison, even though for the baseline case a standard Fourier transform would suffice.

\begin{table}
  \begin{center}
		\def~{\hphantom{0}}
		\begin{tabular}{c | c c c c c c c}
			\multicolumn{8}{c}{$m = \mu - Lj$}																																											\\
			\hline
			\multirow{2}{*}{$\mu$}	& \multicolumn{7}{c}{$j$}																																				\\
			\cline{2-8}
			\\
														& -$N_\varphi/2L+1$	& $\cdots$	& -1				& 0					& 1					& $\cdots$	& $N_\varphi/2L$	\\
			\hline
			-5										& -65								& $\cdots$	& -17				& -5				& 7					& $\cdots$	& 67							\\
			-4										& -64								& $\cdots$	& -16				& -4				& 8					& $\cdots$	& 68							\\
			-3										& -63								& $\cdots$	& -15				& -3				& 9					& $\cdots$	& 69							\\
			-2										& -62								& $\cdots$	& -14				& -2				& 10				& $\cdots$	& 70							\\
			-1										& -61								& $\cdots$	& -13				& -1				& 11				& $\cdots$	& 71							\\
			0											& -60								& $\cdots$	& -12				& 0					& 12				& $\cdots$	& 72							\\
			1											& -71								& $\cdots$	& -11				& 1					& 13				& $\cdots$	& 61							\\
			2											& -70								& $\cdots$	& -10				& 2					& 14				& $\cdots$	& 62							\\
			3											& -69								& $\cdots$	& -9				& 3					& 15				& $\cdots$	& 63							\\
			4											& -68								& $\cdots$	& -8				& 4					& 16				& $\cdots$	& 64							\\
			5											& -67								& $\cdots$	& -7				& 5					& 17				& $\cdots$	& 65							\\
			6											& -66								& $\cdots$	& -6				& 6					& 18				& $\cdots$	& 66							\\
		\end{tabular}
		\caption{Floquet coefficients $\mu$ and the associated azimuthal wavenumbers $m$ for $L =$ 12 and $N_\varphi =$ 144.}
		\label{tab:Floquet}
  \end{center}
\end{table}

Spectral proper orthogonal decomposition (SPOD) of the velocity data is performed extract the most energetic, coherent structures.
SPOD is a technique that decomposes the cross-spectral density (CSD) of the flow state components into an orthonormal basis that optimally represent the data ensemble in terms of energy for a given inner product~\citep{taira2017modal}.
For the present study, the three velocity components are used to build the CSD and the inner product corresponds to the turbulent kinetic energy (TKE) norm, accounting for numerical quadrature weights on the discrete grid.
Hence, for each frequency, the SPOD modes are evaluated after the eigenvalue decomposition of the CSD i.e.
\begin{equation}
	\mathsfbi{C} \mathsfbi{W} \boldsymbol{\Phi} = \boldsymbol{\Phi} \boldsymbol{\Lambda} \mbox{,}
	\label{eq:spod}
\end{equation}
\noindent where $\mathsfbi{W}$ accounts for both the TKE norm and the numerical quadrature weights, $\boldsymbol{\Phi}$ denotes the SPOD modes, $\boldsymbol{\Lambda}$ is a diagonal matrix containing real eigenvalues.
$\mathsfbi{C}$ is the CSD of the velocity field, with
\begin{equation}
	\mathsfbi{C} = \left\langle \boldsymbol{\tilde{q}} \boldsymbol{\tilde{q}}^{\dagger} \right\rangle \mbox{,}
	\label{eq:state_CSD}
\end{equation}
\noindent where $\left\langle \dots \right\rangle$ denotes the expected-value operator and $\dagger$ indicates the trans-conjugate (hermitian) operator.
The CSD is computed frequency-by-frequency for each Floquet coefficient $\mu$.
The contribution of each SPOD mode to the CSD representation is characterised by its associated eigenvalue, which is ranked according to the energetic content from the most to the least energetic, i.e. $\lambda_1 \geq \lambda_2 \geq \cdots \lambda_{n}$.
In this work, we employ the snapshot version of the SPOD algorithm~\citep{towne2018spectral,schmidt2020guide} to extract the modes and the CSD is evaluated using Welch's method~\citep{welch1967fft}, blocks of $N_\mathit{fft} = 128$ size and 50\% overlap, yielding a total of 421 blocks.
To mitigate spectral leakage, a Hamming tapering window was applied over all blocks.
These parameters, together with the jet diameter and the jet flow speed, provide a Strouhal number resolution of $\Delta St \approx 0.02$.

\subsection{Mean fields and fluctuation statistics}
\label{sec:piv_mean_rms}

\begin{figure} 
	\centering
	\includegraphics[width=\textwidth]{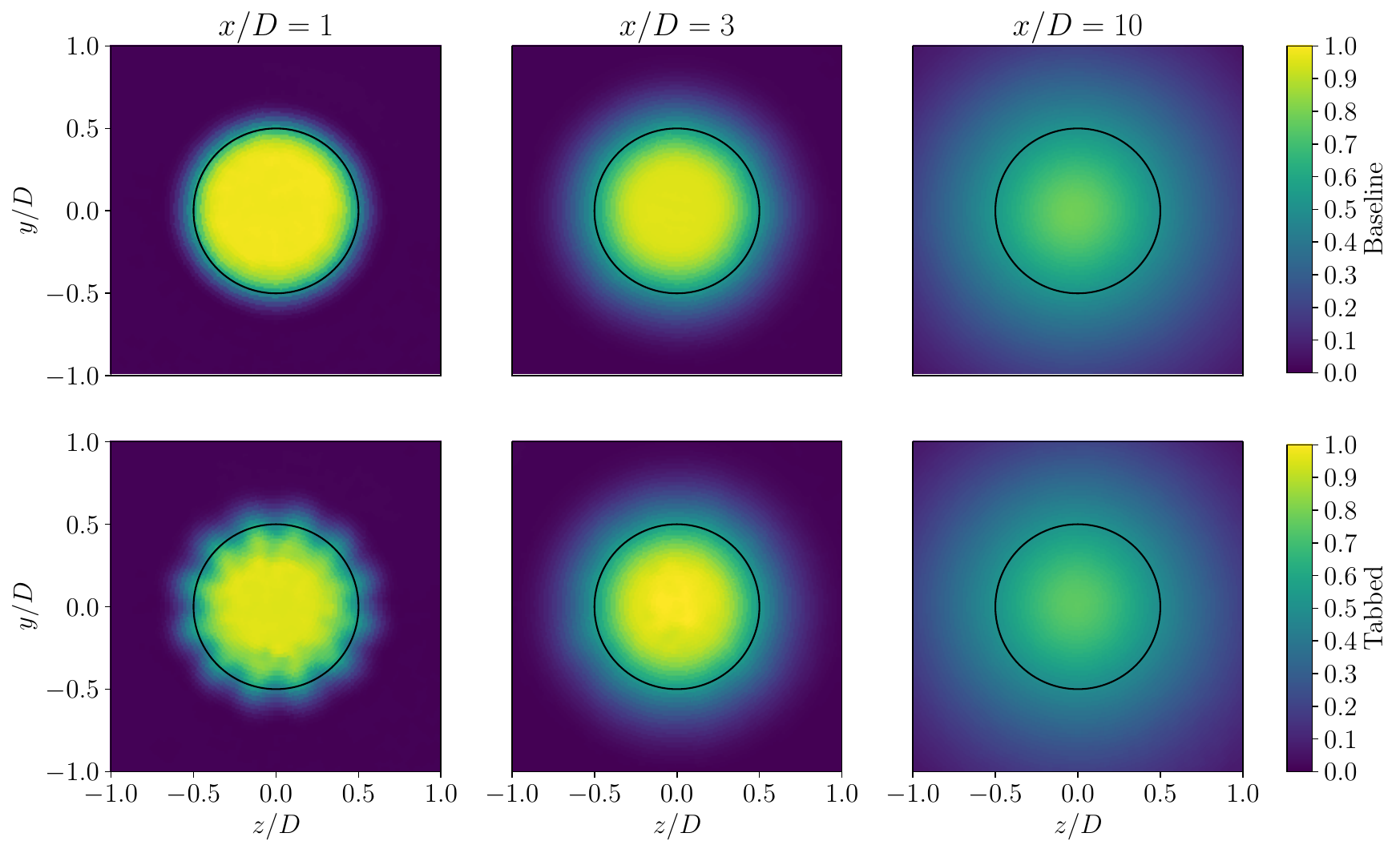}
	\caption{Mean streamwise velocity component ($\bar{u}/U_j$). The lipline is indicated by continuous circle. Top frames: baseline nozzle. Bottom frames: tabbed nozzle. Frames, from left to right: $x/D =$ 1, 3 and 10, respectively.}
	\label{fig:mean_flow_xD1_3_10}
\end{figure}

The mean streamwise velocity component, interpolated in the polar grid, is displayed in figure~\ref{fig:mean_flow_xD1_3_10} for the baseline (top frames) and tabbed (bottom frames) nozzles.
The lipline is shown in the figure as a continuous line circle.
Near the nozzle exit ($x/D = 1$, right frame), the influence of the tabs on the flow field is made clear by the presence of regions of low and high-flow velocity appearing periodically in azimuth.
In this region, the velocity distribution exhibits twelve vertices (equivalent to the number of tabs) in the region around the jet shear-layer.
Further downstream, the jet flows from the baseline and tabbed nozzles are almost indistinguishable, especially far from the nozzle exit ($x/D = 10$, right frames).
Note that all results presented in \S~\ref{sec:piv_mean_rms} were obtained before the application of Floquet decomposition.

\begin{figure}
	\centering
	\includegraphics[width=\textwidth]{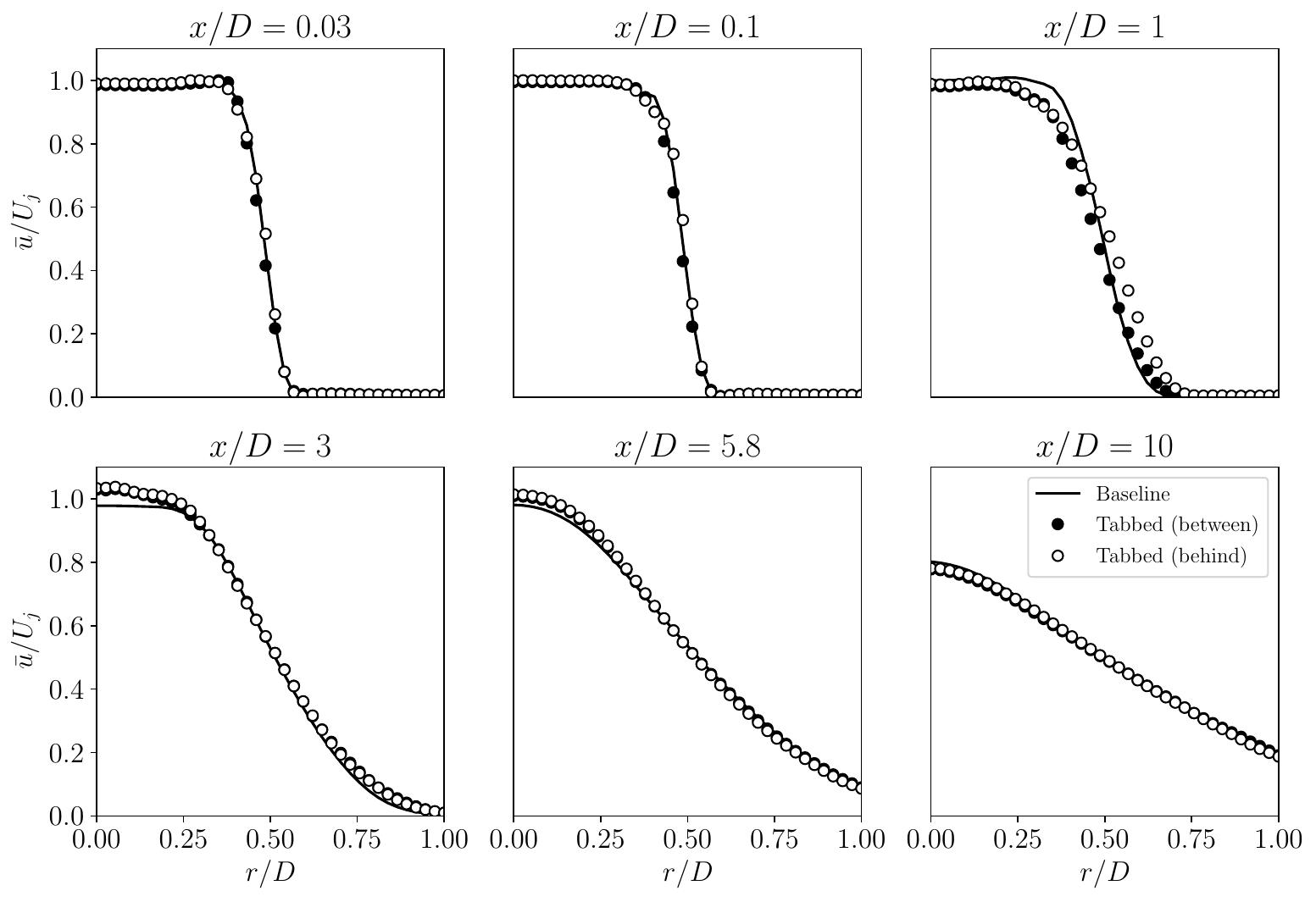}
	\caption{Radial profiles at different streamwise positions. Lines denote baseline nozzle and symbols tabbed nozzle. Closed symbols denote profiles taken in between two tabs and open symbols indicate profiles taken behind a tab.}
	\label{fig:radial_profiles}
\end{figure}

The streamwise velocity profiles as a function of radius for the baseline and tabbed nozzles for several streamwise positions are shown in figure~\ref{fig:radial_profiles}. 
For the tabbed nozzle, the profiles are shown for two positions: between two tabs (closed symbols) and directly behind a tab (open symbols).
Overall, the jet development has the expected trends, with the development of the shear layer in the downstream direction leading to the formation of the potential core.
Clear differences between the shear layers of the baseline and the tabbed case are observed in the region close to the nozzle, with the tabs tending to thicken the shear layer.
For instance, at $x/D = 1$, the shear layer is a strongly modulated by well-developed streaks, leading to distinct profiles behind and between tabs.
The differences are progressively less prominent in positions further downstream, possibly due to the small characteristic length of the streaks for this high value of $L$~\citep{nogueira2019large}.
This difference diminishes as the streaks decay, and the profiles become virtually the identical beyond $x/D = 3$, indicating a re-circularisation of the jet.
After the end of the potential core (which is around $x/D = 5.8$ in the present case), the profile assumes a Gaussian-like profile.
At $x/D = 10$, the profiles are only slightly different, and the induced streaks seem to have little effect on the mean flow in that region.

\begin{figure}
	\centering
	\includegraphics[width=\textwidth]{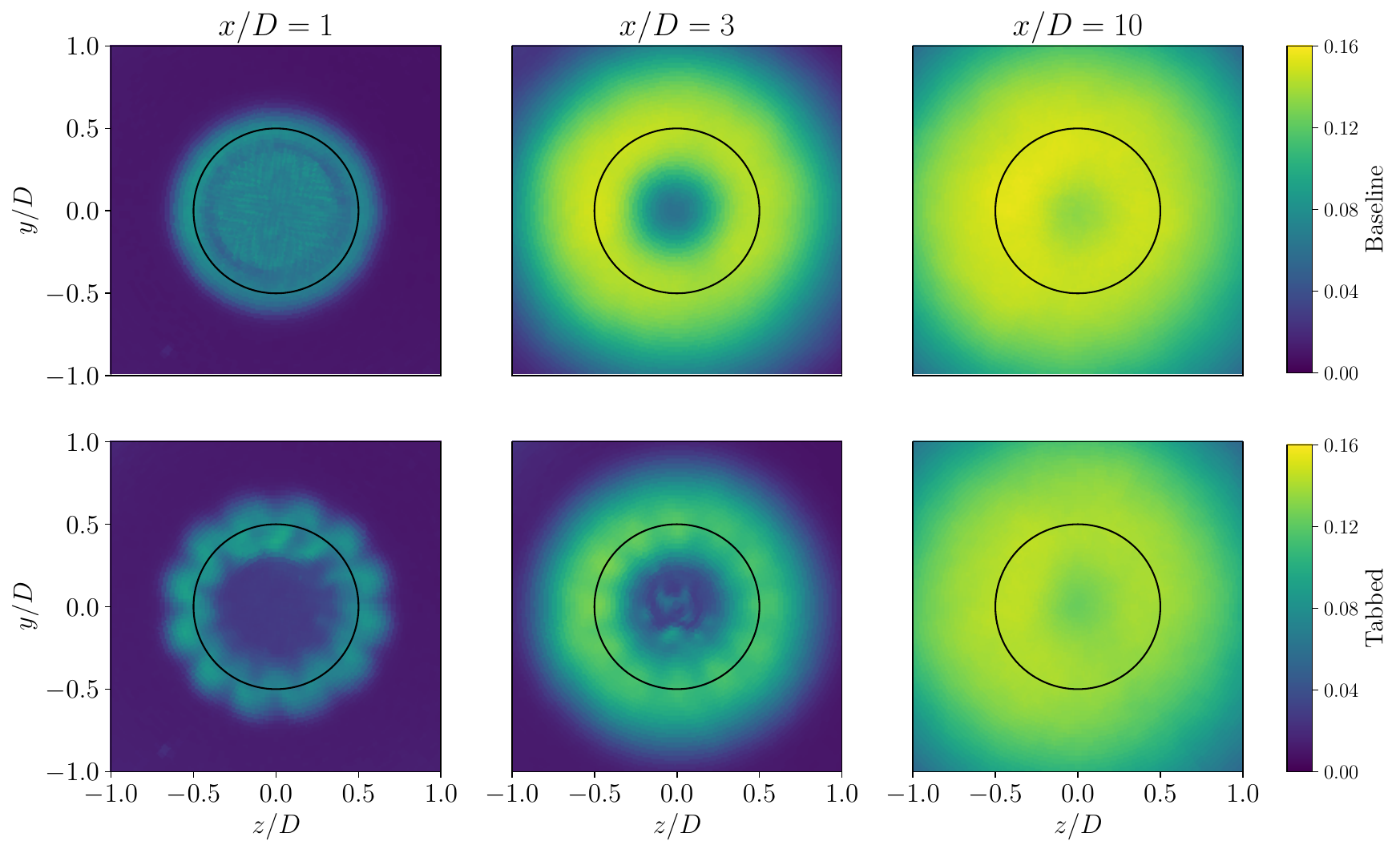}
	\caption{Root-mean square of the streamwise velocity component ($u_{rms}/U_j$). The lipline is indicated by continuous circle. See the comments in the caption of figure~\ref{fig:mean_flow_xD1_3_10}.}
	\label{fig:rms_xD1_3_10}
\end{figure}

The streamwise velocity root-mean-square (rms) distributions for the baseline and tabbed nozzles are shown in figure~\ref{fig:rms_xD1_3_10}.
As expected, the peak rms values for these jets start at the centre of the shear layer and spread radially; this holds for both baseline and tabbed nozzles.
The tabbed jet, however, displays a series of spots of high- and low-intensity rms in the jet shear-layer region, creating a corrugated pattern in the distribution.
This is in agreement with what is expected for flows containing streaky structures, where the development of the K--H is modulated by the streaks.
Such behaviour is observed at least up to $x/D = 3$, and is in contrast with the baseline case, which shows an approximately homogeneous ring structure in the shear-layer region.
Interestingly, the rms intensity is higher for the tabbed case near the nozzle, but decays more rapidly downstream than the baseline nozzle. 
This suggests that, although coherent structures modulated by streaks grow more rapidly near the nozzle, they stabilize more quickly than those in the baseline case.

\begin{figure}
	\centering
	\includegraphics[width=0.5\textwidth]{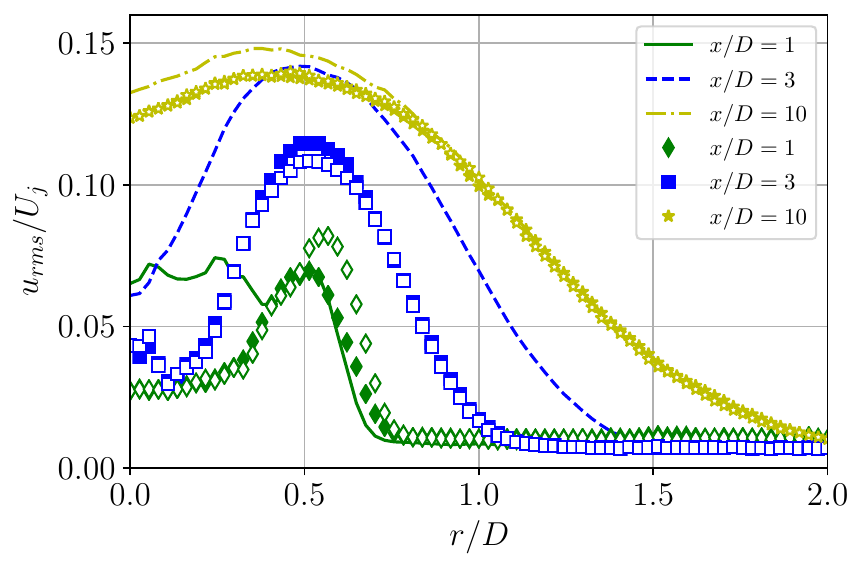}
	\caption{RMS profiles. See the comments in the caption of figure~\ref{fig:radial_profiles}.}
	\label{fig:rms_profiles_xD1_3_10}
\end{figure}

For completeness, figure~\ref{fig:rms_profiles_xD1_3_10} shows the rms profiles.
The reductions in rms due to the tabs is significant.
At $x/D = 3$ and 10, the rms peak values are noticeably higher for the baseline configuration in comparison with the tabbed configuration.
Closer to the nozzle, the peak rms magnitudes are similar, further supporting that the main changes in the flow from the application of tabs are in fact associated with the development of coherent structures further downstream.
This aspect will be investigated in the following.

Note that the rms profile for the baseline configuration at $x/D = 1$ (top-right frame in figure~\ref{fig:rms_xD1_3_10}, green curves in figure~\ref{fig:rms_profiles_xD1_3_10}) is noisy.
However, this will not hamper the coherent structures analysis, as the SPOD also acts as filter, as will be seen in the following section.

Since thrust cannot be measured directly in our facility, Appendix~\ref{app:momentum_estimates} presents momentum flux estimates derived from the TR-PIV measurements.
Within the margin of error, the impact of the tabs appears to be small; however, it should be emphasized that an accurate assessment of thrust penalty would require a dedicated measurement instrument.

\subsection{Coherent structures}
\label{sec:piv_spod}

The signature coherent structures present in the turbulent flow field were obtained through SPOD of the velocity field for each Floquet exponent.
Modal energy maps of the leading SPOD modes are exhibited in figures~\ref{fig:modal_energy_xD1} to \ref{fig:modal_energy_xD10} as a function of the Floquet coefficient $\mu$.
Only positive frequencies and Floquet coefficients are shown.
Fourier decomposition symmetry rules apply for negative frequencies and Floquet coefficients.
Most of the current analysis will be focused on the $\mu=0$ component, which is expected to contain the coherent structures most relevant to downstream-sound radiation.
The convergence of the SPOD modes is discussed in appendix~\ref{app:SPOD_convergence}.

\subsubsection{Modal energy maps}

\begin{figure}
	\centering
	\includegraphics[width=\textwidth]{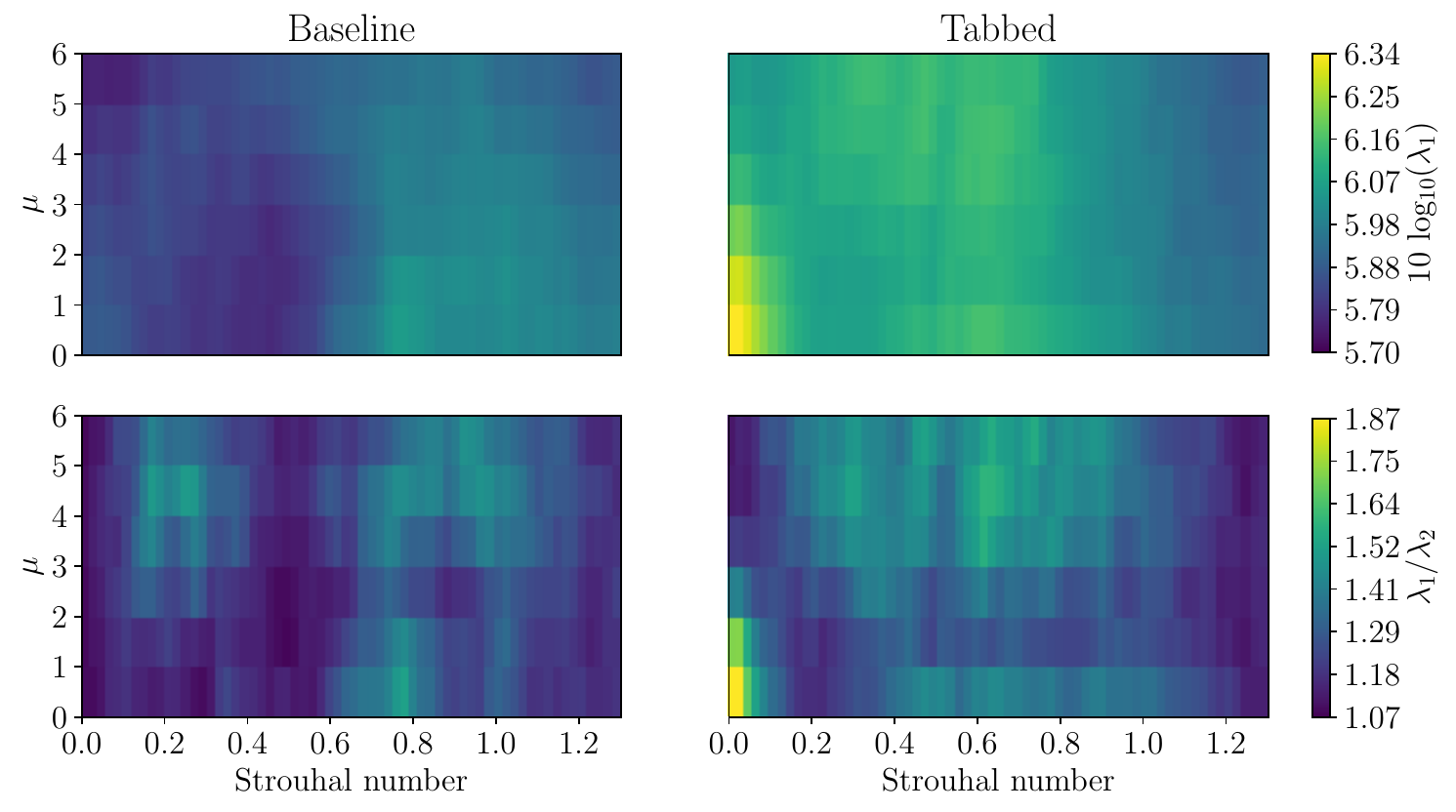}
	\caption{Modal energy maps for $x/D =$ 1. $\mu$ stands for Floquet coefficient. Left frames: baseline nozzle. Right frames: tabbed nozzle. Top frames: leading eigenvalue. Bottom frames: ratio between the leading and the first suboptimal eigenvalues.}
	\label{fig:modal_energy_xD1}
\end{figure}

Near the nozzle exit (figure~\ref{fig:modal_energy_xD1}), the spectrum for the first SPOD mode of the tabbed nozzle displays a clear peak for $\mathit{St} \rightarrow 0$.
This is the signature of the tab-generated streaks, which are expected to be dominant at $m=12$, and should thus be included in the SPOD modes for $\mu=0$.
Apart from the peak, energy amplifications occur over a broad region of the $\mu$-St spectrum with respect to the baseline case. 
The region exhibiting low-rank behaviour (indicated by high amplitudes in the $\lambda_1/\lambda_2$ maps) is also broader for the tabbed nozzle. 

\begin{figure}
	\centering
	\includegraphics[width=\textwidth]{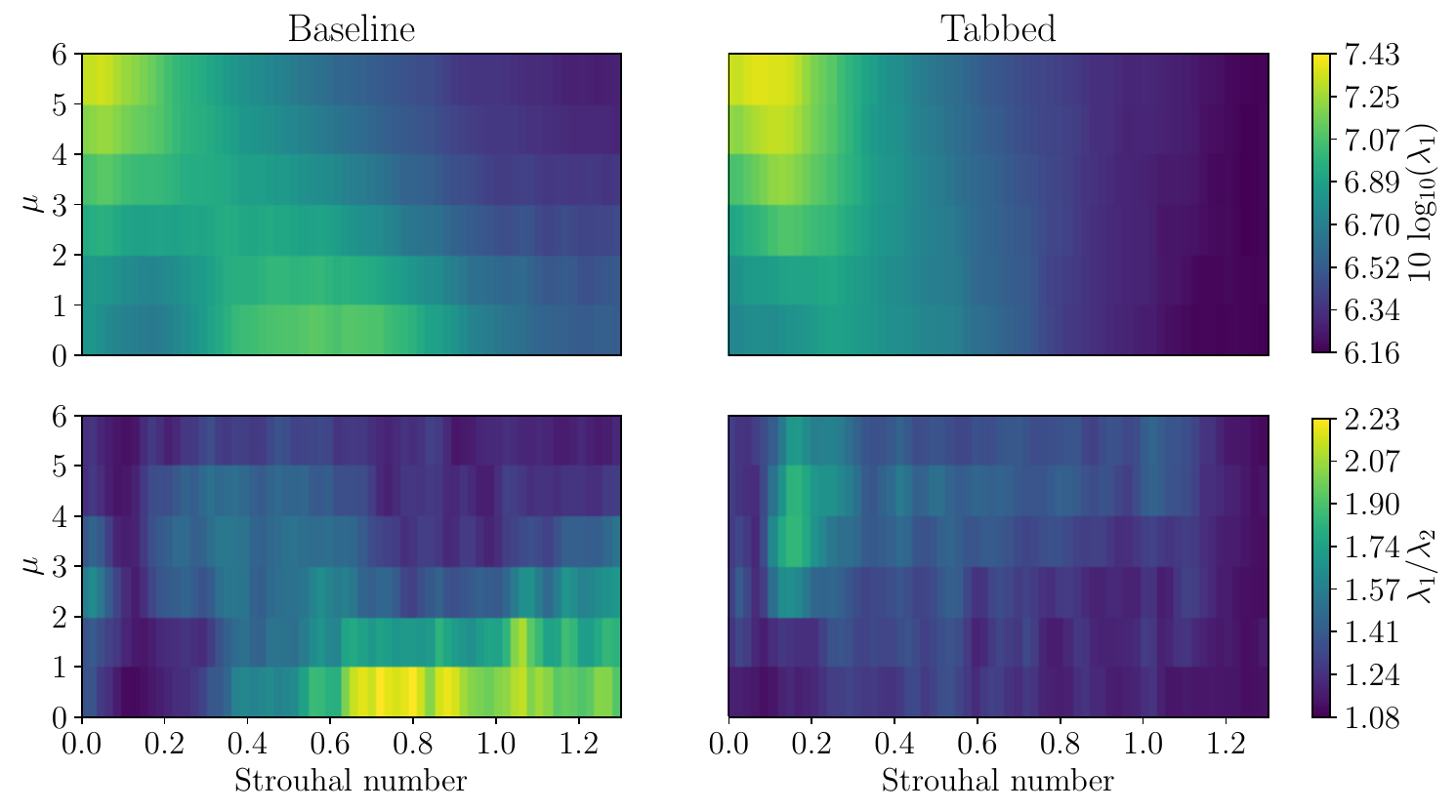}
	\caption{Modal energy maps for $x/D =$ 3. See the comments in the caption of figure~\ref{fig:modal_energy_xD1}.}
	\label{fig:modal_energy_xD3}
\end{figure}

Results for $x/D = 3$ (further downstream, but before the end of the potential core) are shown in figure~\ref{fig:modal_energy_xD3}.
In this region, coherent structures generated by the jet shear are expected to be more developed, especially the K--H mode.
At this location, the baseline spectrum exhibits a double-peak feature, concentrating energy at high $\mu$ and low $St$, and at very low $\mu$ and moderate $St$ ($0.4 \leq St \leq 1$).
As shown in a series of previous works~\citep{nogueira2019large,rigas2019streaks,pickering2020lift,nogueira2021dynamics,maia2023effect}, the first peak is associated with the growth of streaks via the lift-up effect in this free-shear flow, while the second is associated with the presence of the K--H wavepacket.
The K--H-associated peak is significantly attenuated by the tabbed nozzle, whereas the streak-related peak is accentuated and broadened.
This is consistent with the expected damping of K--H wavepackets by induced streaks.
Furthermore, the decrease in eigenvalue separation observed in the tabbed nozzle in the low-$\mu$ zone indicates a deterioration of the low-rank behaviour of the jet~\citep{schmidt2018spectral,lesshafft2019resolvent} in this spectral region.
This is also consistent with the weakening of the K--H mechanism.


\begin{figure}
	\centering
	\includegraphics[width=\textwidth]{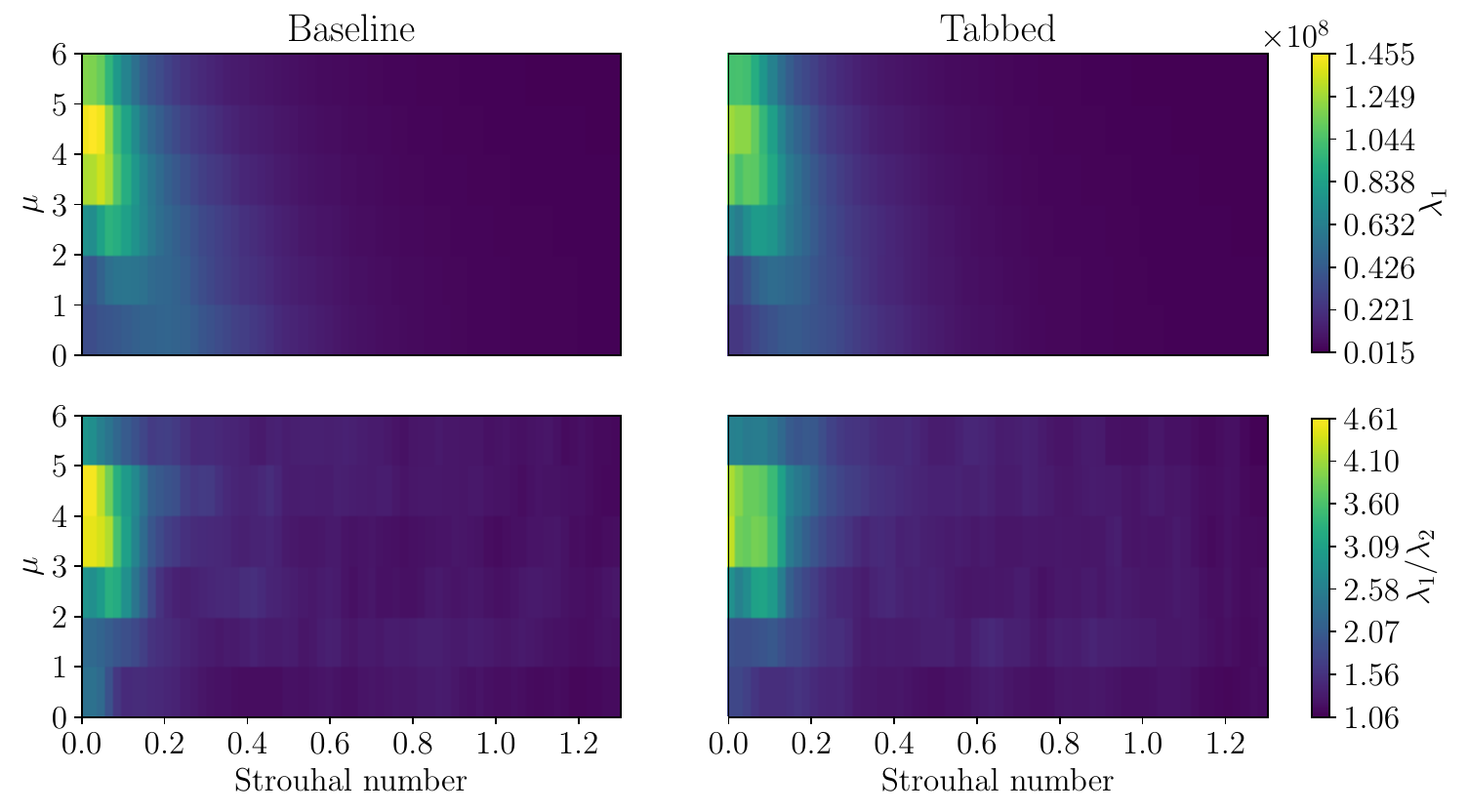}
	\caption{Modal energy maps for $x/D =$ 5.8. See the comments in the caption of figure~\ref{fig:modal_energy_xD1}.}
	\label{fig:modal_energy_xD6}
\end{figure}

Figure~\ref{fig:modal_energy_xD6} shows the modal energy maps at $x/D = 5.8$.
In this case, the energy is displayed on a linear scale rather than a logarithmic one, to facilitate direct comparison of energy levels between the baseline and tabbed nozzles.
At this streamwise location, the energy levels begin to approach the same order of magnitude.
Additionally, higher energy levels and greater gain separation are observed around $\mu = 4$ and as $\mathit{St} \rightarrow 0$.

\begin{figure}
	\centering
	\includegraphics[width=\textwidth]{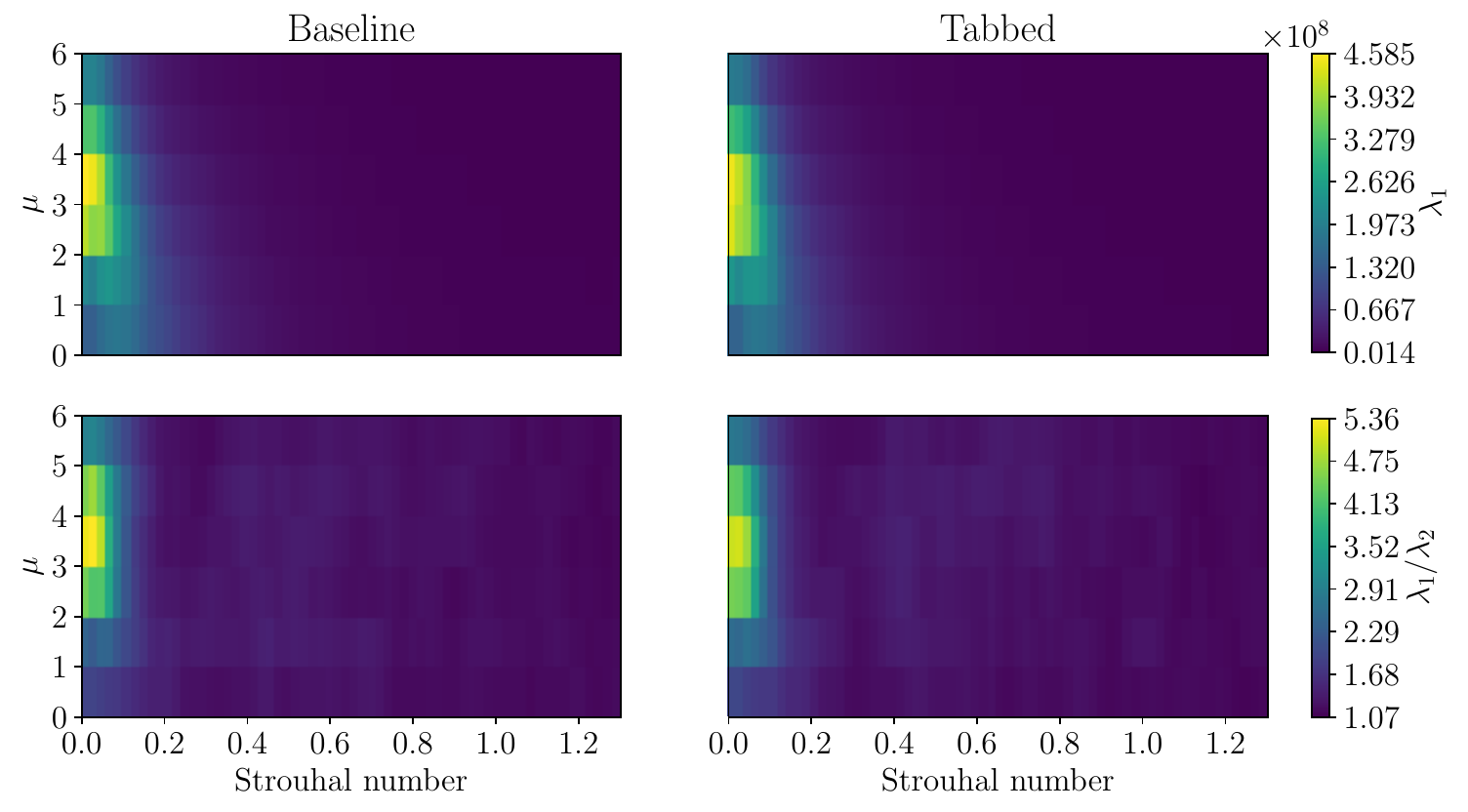}
	\caption{Modal energy maps for $x/D =$ 10. See the comments in the caption of figure~\ref{fig:modal_energy_xD1}.}
	\label{fig:modal_energy_xD10}
\end{figure}

Finally, far away from the nozzle exit, at $x/D = 10$ (figure~\ref{fig:modal_energy_xD10}), the energy maps of both nozzles are remarkably similar, although a slight attenuation can still be observed for the tabbed nozzle, with higher energy levels and eigenvalue separation around $\mu = 3$ and $\mathit{St} \rightarrow 0$.
This behaviour is consistent with previous literature, which shows the presence of large-scale streaks at lower $\mu$~\citep{nogueira2019large,pickering2020lift}.
The similarities between the two spectra are also consistent with the resemblance between the rms plots, indicating that the tabs have a minor influence on the turbulence dynamics in downstream stations.

\subsubsection{Eigenfunctions}

\begin{figure}
	\centering
	\includegraphics[width=\textwidth]{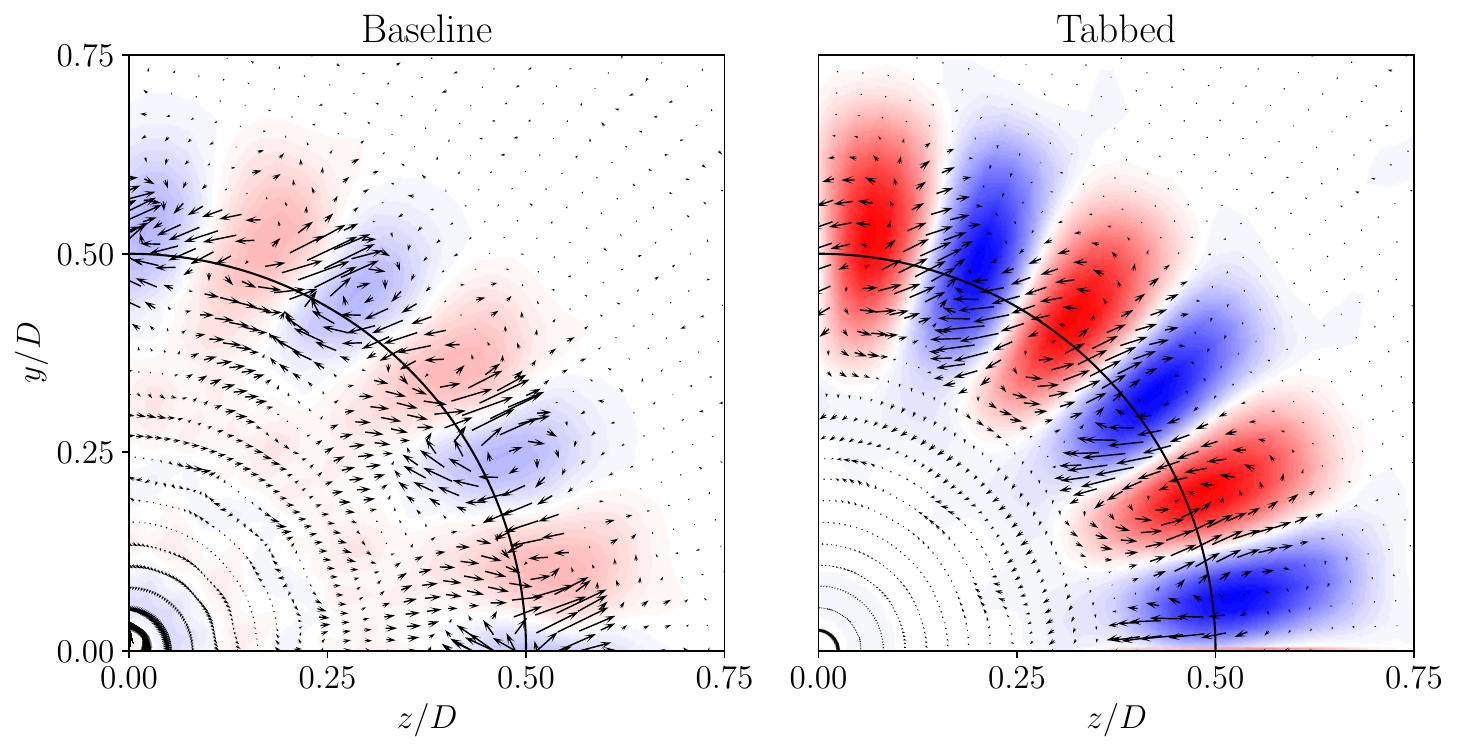}
	\caption{First SPOD mode for $x/D =$ 1, $\mu =$ 0 and $\mathit{St} \rightarrow$ 0. Contours, weighted by the eigenvalues, denote the real value of the streamwise velocity component, with the red to blue colourmap denoting positive to negative values. The arrows indicate the cross-flow velocity components. Only one-fourth of the domain is shown.}
	\label{fig:spod_xD1_M0_St0_mode}
\end{figure}

We now proceed to the analysis of selected eigenfunctions for $\mu = 0$ at several stations of the flow.
Starting at $x/D = 1$, as expected from the SPOD spectra shown in figure~\ref{fig:modal_energy_xD1}, the first mode of the tabbed nozzle has much more energy in the low Strouhal region, compared to the baseline case. 
The mode associated with this high-energy region is shown in figure~\ref{fig:spod_xD1_M0_St0_mode}, which is compared to the first mode at the same frequency for the baseline nozzle.
One-fourth of the domain is exhibited in the figure, with the streamwise velocity component represented by the contours (red-to-blue colourmaps) and the cross-flow velocity components given by the arrows.
In this figure, the modes are weighted by the corresponding eigenvalue to give a better sense of the difference in amplitude between the baseline and the tabbed nozzle--
the contour levels are kept constant between the baseline and tabbed cases, enabling direct comparison.
As expected, the mode in the tabbed nozzle displays streak-like structures similar to~\citet{nogueira2019large}--note that the $\mu = 0$ Floquet coefficient includes the wavenumbers $m = [0,~12,~24,~...]$. 

Past studies have demonstrated the importance of the lift-up mechanism~in shear-flow \citep{ellingsen1975stability, landahl1980note, brandt2014lift} and in the dynamics and energy spectrum of turbulent jets~\citep{nogueira2019large, pickering2020lift, maia2024effect}.
The lift-up mechanism is activated by streamwise vorticity which generates flow structures characterized by alternating regions of high and low streamwise velocity (streaks).
Characteristic features of lift-up, thus include streamwise vorticity both in the modes.
The fact that distinctive streamwise vortex and streak features are observed in both the baseline (where there is clearly no wake dynamics) and tabbed configurations suggests that the tabs have indeed enhanced the lift-up mechanism already present in the baseline case.
This conjecture is supported by Figure~\ref{fig:spod_xD1_M0_St0_mode}.

\begin{figure}
	\centering
	\includegraphics[width=\textwidth]{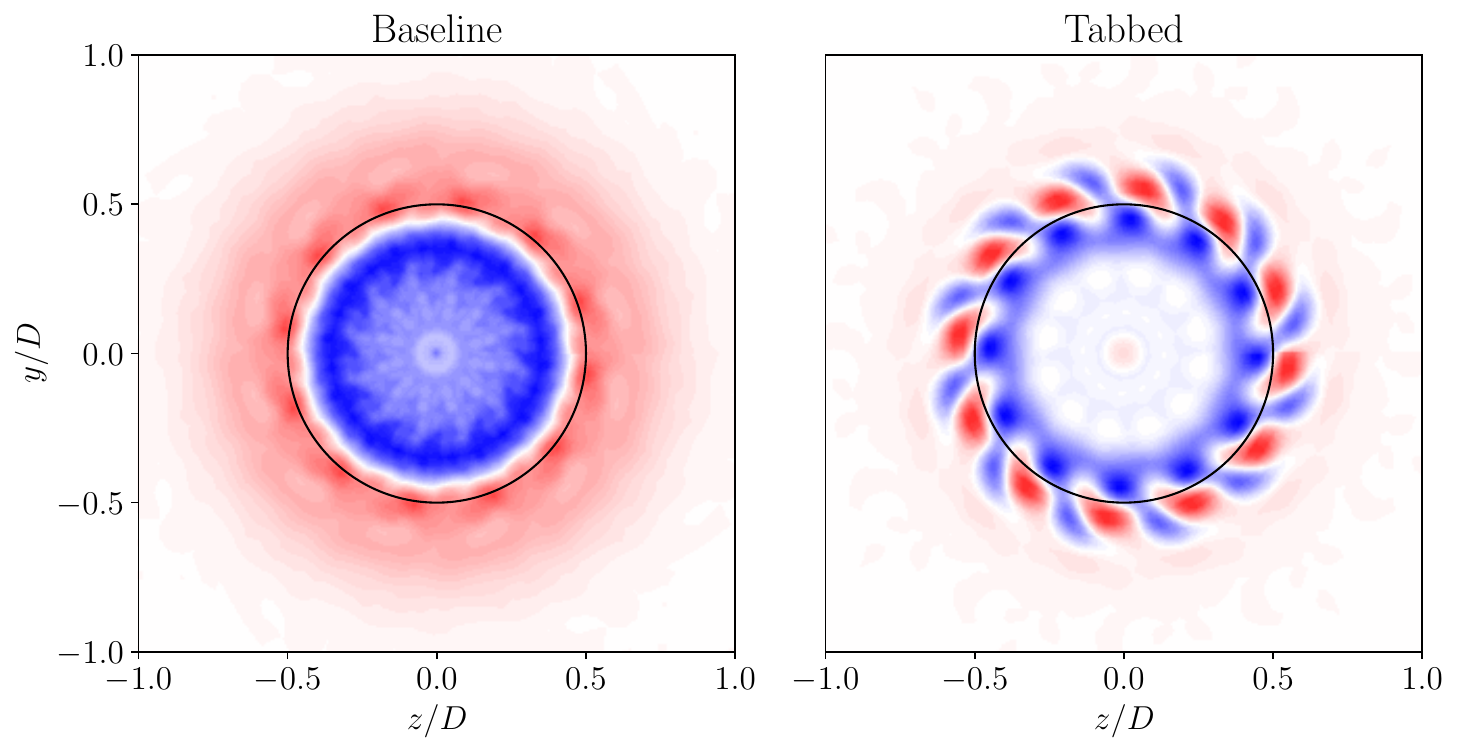}
	\caption{First SPOD mode for $x/D =$ 1, $\mu =$ 0 and $\mathit{St} \approx$ 0.8. Contours denote the real value of the streamwise velocity component, with the red to blue colourmap denoting positive to negative values.}
	\label{fig:spod_xD1_M0_St0.8_mode}
\end{figure}

Figure~\ref{fig:spod_xD1_M0_St0.8_mode} displays the first SPOD mode associated with $\mathit{St} \approx 0.8$.
Differently from figure~\ref{fig:spod_xD1_M0_St0_mode}, now only the streamwise velocity component mode is shown and for the full domain.
The baseline configuration clearly exhibits an axisymmetric K--H mode, whereas the tabbed case has a much more complex structure.
The dominant mode for the tabbed nozzle may be associated either with an azimuthal modulation of the original axisymmetric mode, or to modes of higher dominant azimuthal wavenumber ($m = 12$, in this case), as shown in~\citet{lajus2019spatial}.
At this position, it is unclear if this mode is directly associated with the sound-generating K--H wavepacket.

Now proceeding to $x/D = 3$ and Floquet coefficient $\mu = 0$, the first mode of the baseline case is more energetic than the tabbed case (see figure~\ref{fig:modal_energy_xD3}), suggesting a damping of the dominant physical mechanism generating the coherent structures (in this case, the K--H instability), in line with previous results~\citep{sinha2016parabolized,marant2018influence,lajus2019spatial,wang2021effect}.
A sharp decay in energy is observed at lower Strouhal number for the tabbed case, leading to a shift in the peak frequency for this case. 
This shift leads to strong reductions in the energy of the first mode around $0.3 \leq \mathit{St} \leq 1$.

\begin{figure}
	\centering
	\includegraphics[width=\textwidth]{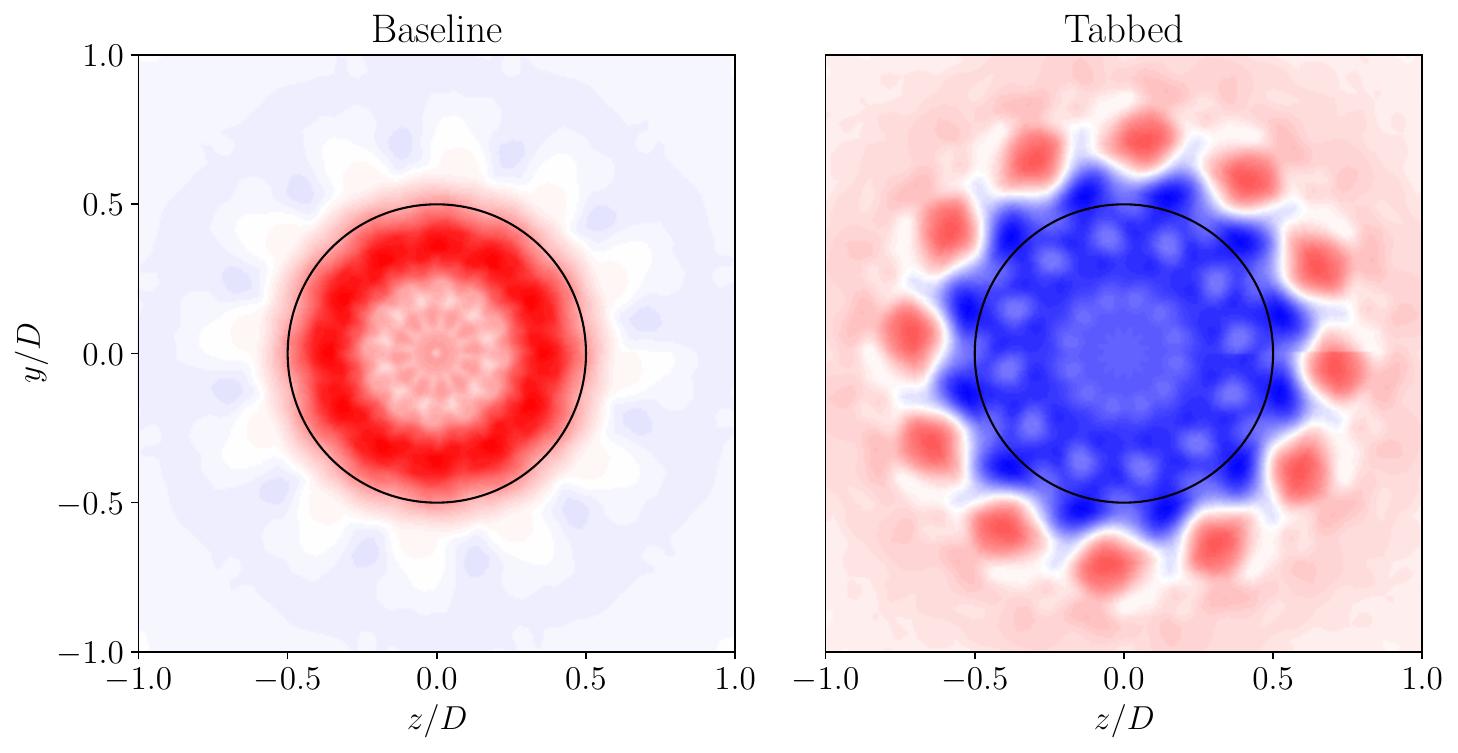}
	\caption{First SPOD mode for $x/D =$ 3, $\mu =$ 0 and $\mathit{St} \approx$ 0.3. See the comments in the caption of figure~\ref{fig:spod_xD1_M0_St0.8_mode}.}
	\label{fig:spod_xD3_M0_St0.3_mode}
\end{figure}

\begin{figure}
	\centering
	\includegraphics[width=\textwidth]{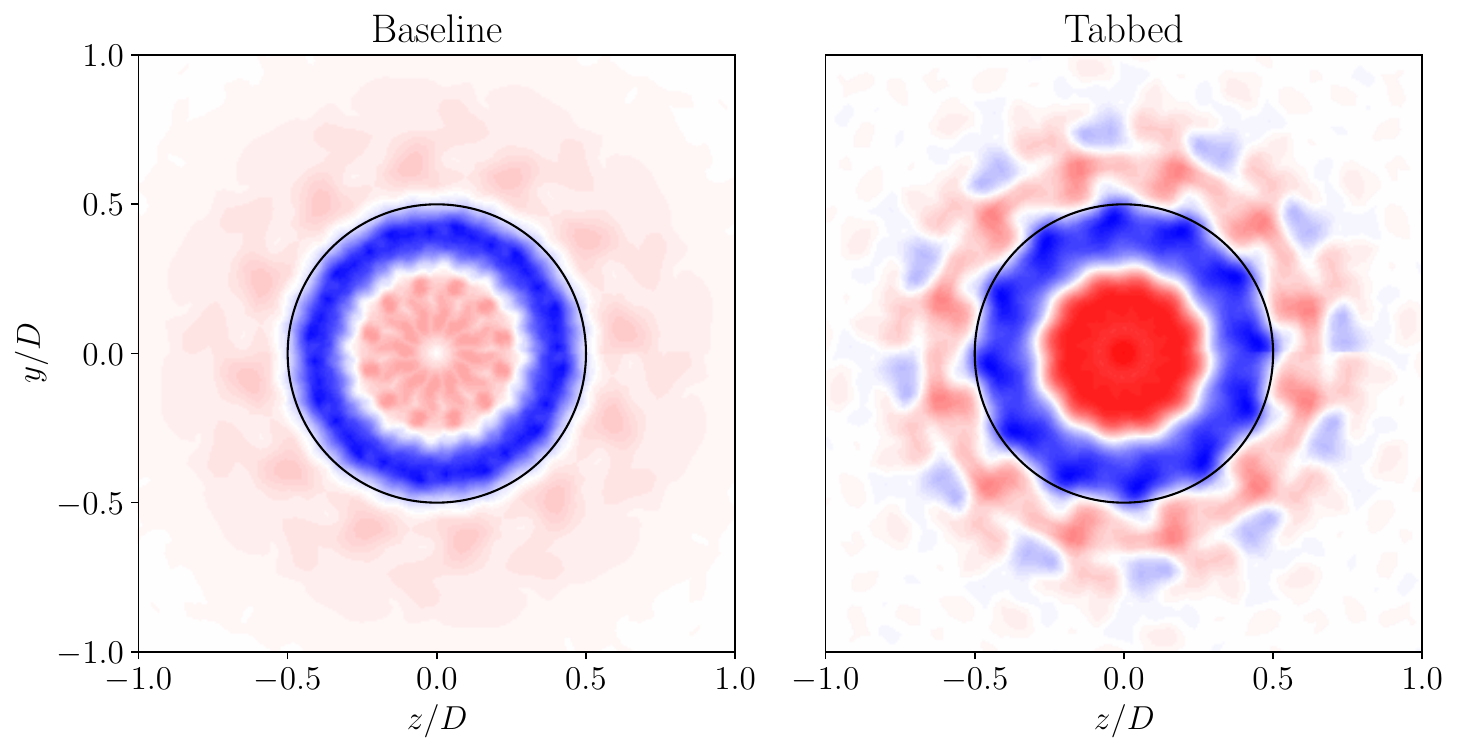}
	\caption{First SPOD mode for $x/D =$ 3, $\mu =$ 0 and $\mathit{St} \approx$ 0.5. See the comments in the caption of figure~\ref{fig:spod_xD1_M0_St0.8_mode}.}
	\label{fig:spod_xD3_M0_St0.5_mode}
\end{figure}

\begin{figure}
	\centering
	\includegraphics[width=\textwidth]{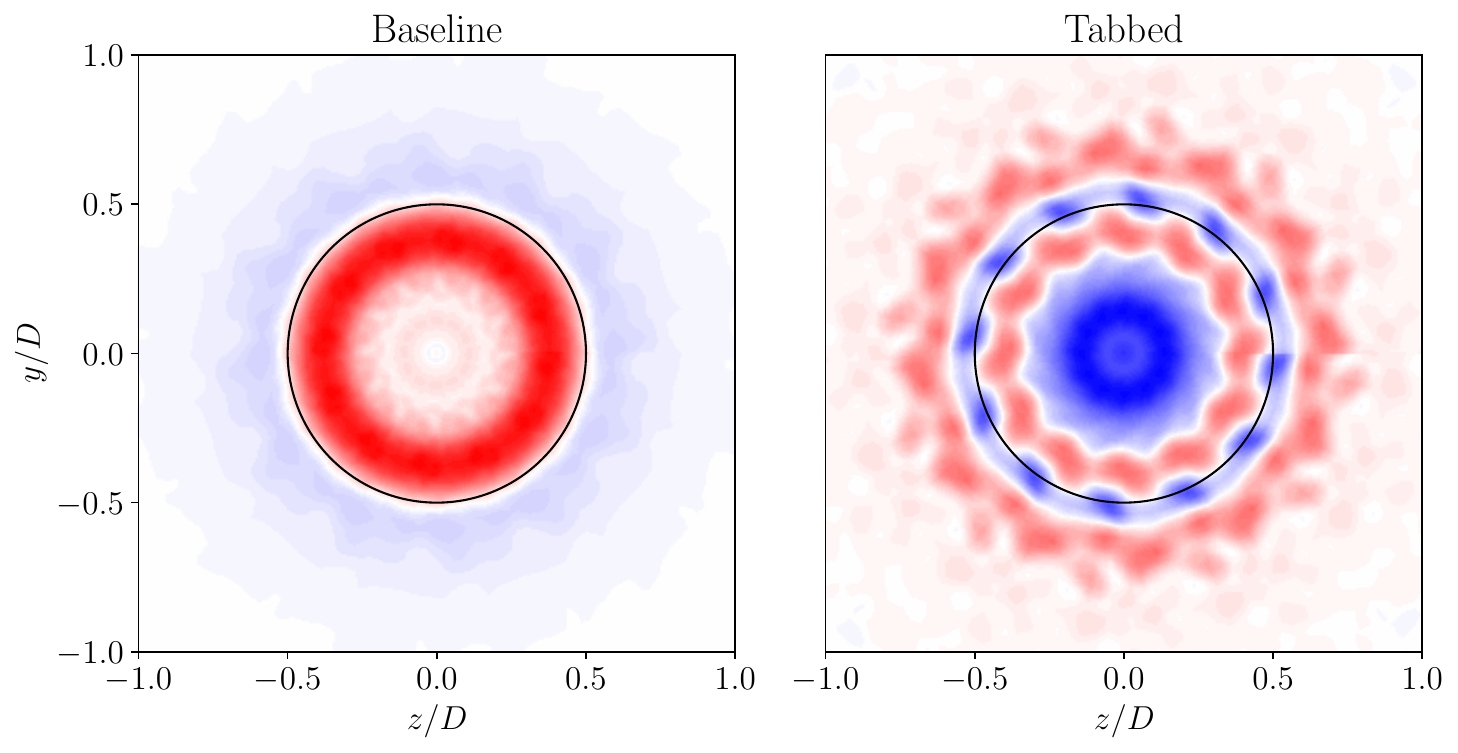}
	\caption{First SPOD mode for $x/D =$ 3, $\mu =$ 0 and $\mathit{St} \approx$ 0.7. See the comments in the caption of figure~\ref{fig:spod_xD1_M0_St0.8_mode}.}
	\label{fig:spod_xD3_M0_St0.7_mode}
\end{figure}

Shapes of the resulting eigenfunctions are also compared in figures~\ref{fig:spod_xD3_M0_St0.3_mode}, \ref{fig:spod_xD3_M0_St0.5_mode} and \ref{fig:spod_xD3_M0_St0.7_mode}, for $\mathit{St} \approx 0.3$, 0.5 and 0.7, respectively.
The baseline case has the expected K--H shape, peaking close to the centre of the shear layer, and with the usual phase shift across it. 
Very small modulations following the L-fold coupling are also observed throughout the domain.
This may result from SPOD's slow convergence in order to separate independent azimuthal modes $m$ under these conditions, as the flow is expected to be axisymmetric.
Regarding the tabbed nozzle, a strong modulation following the azimuthal wavenumber of the streaks ($m = 12$) is observed throughout the domain, being more clearly identified at higher radial positions.
This result is in line with previous linear-stability results~\citep{sinha2016parabolized,lajus2019spatial,wang2021effect}, which indicated that the presence of corrugations around the shear layer (in this case, induced by the tab-generated streaks).
This would change the shape of the K--H mode.

\begin{figure}
	\centering
	\includegraphics[width=\textwidth]{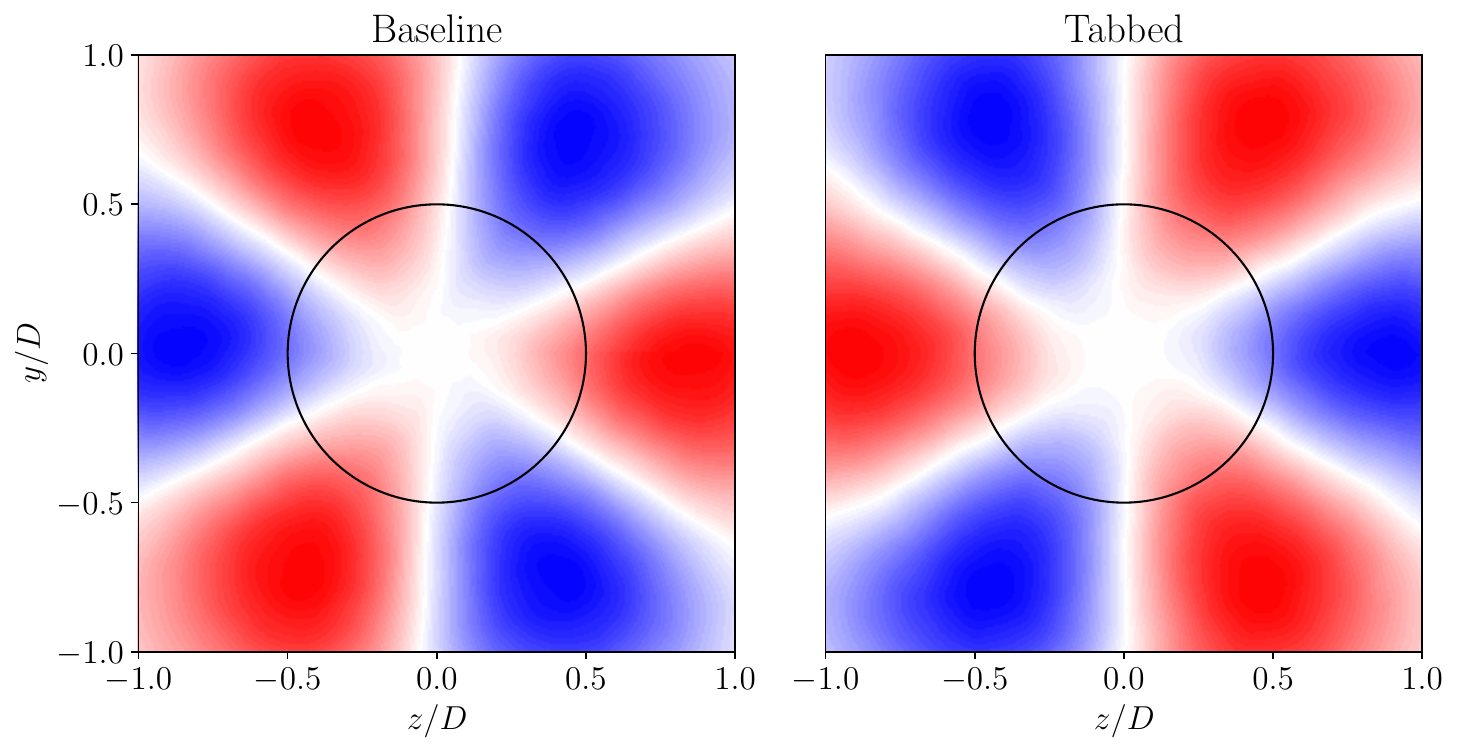}
	\caption{First SPOD mode for $x/D =$ 10, $\mu =$ 3 and $\mathit{St} \rightarrow$ 0. See the comments in the caption of figure~\ref{fig:spod_xD1_M0_St0.8_mode}.}
	\label{fig:spod_xD10_M3_St0_mode}
\end{figure}

Finally, let us analyse the structures for $x/D = 10$ and $\mu =3 $, which is the most energetic region of the modal energy map at this location (figure~\ref{fig:modal_energy_xD10}).
The spectra for both baseline and tabbed nozzles are very similar.
The peak in the spectrum is, once again, found at $\mathit{St} \rightarrow 0$, and associated with large-scale streaky structures, as seen in~\citet{nogueira2019large,pickering2020lift,maia2023effect}.
Mode shapes at this peak frequency for both baseline and tabbed nozzles are virtually identical and associated with streaks of dominant azimuthal wavenumber $m = 3$, as seen in figure~\ref{fig:spod_xD10_M3_St0_mode}.
As shown in~\citet{nogueira2019large}, low-wavenumber streaks extend further in the streamwise direction, peaking further downstream in the domain.

\section{Wavepacket modelling}
\label{sec:wavepacket}

Results in the previous sections highlight a clear physical mechanism that explains the noise reductions observed in the acoustic field: the presence of forced streaks in the mean flow affects the K--H mechanism, damping the associated wavepackets.
Since wavepackets are the noise-generating structures in these flows, this phenomenon is therefore expected to lead to reductions in the acoustic radiation.
However, the current experimental setup lacks information about the streamwise correlation, and the resulting 3D shape of the coherent structures is absent from the analysis.
Here, we attempt to overcome this issue by computing the solution of the linearised Navier--Stokes equations, which provides a good approximation of the wavepacket behaviour for this case (see~\citet{sinha2016parabolized}).
It should be noted that, since the computed solution is linear, the method does not provide direct information about the damping mechanism identified previously without some assumption regarding the initial amplitudes of the structures.
These assumptions will be inferred from the experimental data, and focus will be given on the associated growth of the wavepackets.
Furthermore, the approach is useful in analysing other changes induced in the wavepackets that could affect sound radiation.

\subsection{One-way Navier--Stokes}
\label{sec:wavepacket_owns}

The current analysis is based on the linearisation of the compressible viscous Navier--Stokes equations, written in a cylindrical polar coordinate system and normalised by the ambient specific volume, the jet diameter, and the ambient sound speed~\citep{nogueira2023prediction,nogueira2024wavepackets}.
Similar to previous studies, the Reynolds number selected for the analysis was $Re=3\times10^4$, which is lower than the experimental value to account for the effects of turbulent viscosity in the structures~\citep{pickering2021optimal}; no optimisation of this parameter was attempted in this study, but it is well known that the effect of this viscosity on the overall shape of the K--H instability waves is usually minor for sufficiently high values of Reynolds~\citep{edgington2021waves}.
After linearisation, the vector of flow disturbances $\boldsymbol{\hat{q}}$ is represented using the Floquet ansatz in azimuth~\citep{lajus2019spatial,nogueira2021investigation}, and in the frequency domain as
\begin{equation}
    \boldsymbol{\hat{q}}(x,r,\phi,\omega) = \boldsymbol{q}(x,r,\phi) \ee^{-\ii\omega t + \ii\mu\phi},
    \label{eqn:FloquetPres1}
\end{equation}
\noindent where $\omega=2\pi St M_j$ is the angular frequency, and $\mu$ is the Floquet coefficient.
Note that this approach takes advantage of the symmetry of the problem in the same way the post-processing of the experimental data did, which will facilitate the comparison between SPOD and OWNS results.
The linearised system of equations may then be written in operator form as
\begin{equation}
    -\ii \omega \boldsymbol{q}+\boldsymbol{A}\frac{\partial \boldsymbol{q}}{\partial x}-\boldsymbol{L}\boldsymbol{q}=0,
    \label{eqn:NS_eqs}
\end{equation}
\noindent where the explicit form of operators $\boldsymbol{A}$ and $\boldsymbol{L}$ can be found in appendix~\ref{app:OWNS_op}.
The system of equations~\ref{eqn:NS_eqs} may be marched spatially by discretising the streamwise derivatives using an appropriate method.
As in~\citet{zhu2023recursive}, an implicit midpoint method was chosen as it provides a good balance between accuracy and computational cost.
Since the Navier--Stokes operators contain both downstream- and upstream-propagating modes, the spatial march is generally unstable; this problem can be avoided by either rewriting the formulation so that the resulting system is approximately parabolic~\citep{bertolotti1991analysis,herbert1994}, or by explicitly filtering the upstream modes at each step of the march~\citep{towne2015one,towne2019critical}.
OWNS follows the latter approach, with filtering in the current formulation performed in a recursive manner~\citep{zhu2023recursive} as
\begin{eqnarray}\label{eqn:OWNS_R}
    \boldsymbol{q}^0_n=\frac{1}{h}\boldsymbol{q}_n, \\
    (\boldsymbol{L}+\ii\omega-\ii\boldsymbol{A}\beta^*_j)\boldsymbol{q}^j_n=(\boldsymbol{L}+\ii\omega-\ii\boldsymbol{A}\beta^-_j)\boldsymbol{q}^{j-1}_n, \ j=1,2,...,N_\beta, \\
    \boldsymbol{q}_n=\boldsymbol{q}^{N_\beta}_n,
\end{eqnarray}
\noindent where $\boldsymbol{q}_n$ is the solution at the position $x_n$ in the flow.
From the initial solution $\boldsymbol{q}^0_n$, obtained from the spatial march, the field is filtered $N_\beta$ times to remove the signature of upstream waves in the march.
The definition of the recursion parameters $\beta^*, \beta^-$ and the factor $h$ follows previous works and can be found in~\citet{zhu2023recursive}.

The method requires an initial field, to be marched in the downstream direction.
This initial solution is obtained by solving the locally-parallel problem close to the nozzle plane ($x/D = 0.1$), which can be written as
\begin{equation}
    \ii \alpha \boldsymbol{A}\boldsymbol{q}_0=(\ii \omega \boldsymbol{I} + \boldsymbol{L})\boldsymbol{q}_0.
    \label{eqn:NS_eqs_LSA}
\end{equation}
\noindent where $\alpha$ is the complex eigenvalue whose real and imaginary parts represent the wavenumber and growth rate of the disturbances, respectively.
This value is also used in the definition of recursion parameters to ensure that the desired mode is well captured in the spatial march.
The eigenspectrum for the current problem contains several modes associated with different wavenumbers following the Floquet ansatz~\citep{lajus2019spatial}, and care must be taken to select the most appropriate mode for the march.
Here, the most unstable mode whose eigenfunction is dominated by $m = 0$ disturbances for each frequency were chosen.
This mode is expected to be dominant further downstream, and to be responsible for most of the downstream radiation in these jets~\citep{cavalieri2012axisymmetric}.

The computational domain is discretised using Chebyshev polynomials in the radial direction and Fourier in azimuth~\citep{trefethen2000spectral,weideman2000matlab}, as in previous studies.
A radial mapping following~\citet{lesshafft2007linear} was used in order to provide an appropriate point density in the region of interest, around the shear layer.
The mean flow is defined based on the experimental mean velocity slices, and is further described below.
A mesh of $(N_r,N_\phi) = (100,32)$ was used, considering that only a sector of $\Delta \phi = 2\pi/12$ needed to be discretised due to the application of the Floquet ansatz.
The streamwise step size was also chosen as $\Delta x/D = 0.1$, which was sufficient to converge the main features of the coherent structures evaluated herein.

\subsection{Comparisons with experimental data}
\label{sec:wavepacket_comparisons}

\subsubsection{Definition of base flow}

The current experimental dataset contains time-resolved velocity fields at selected streamwise stations, which allows for the azimuthal characterisation of the dominant coherent structures in the flow.
On the modelling side, the cross-plane slices also allow evaluation of locally supported waves using locally parallel linear stability analysis, but this approach cannot be used to evaluate the streamwise development of the structures.
Here we use OWNS to overcome that limitation.
The method requires a three-dimensional mean flow as input; here we attempt to define an appropriate mean flow based on the limited planar measurements available.
It is important to keep in mind the limitations of this approach, as no information about the exact state of the shear layers at each spatial point is available, and the spacing between streamwise slices is too large for reliable interpolation.
Nevertheless, the mean flow characteristics follow the same trends observed in previous results in the literature, which provides some level of confidence in the approach, despite these limitations. 

\begin{figure}
	\centering
	   \includegraphics[width=0.33\textwidth]{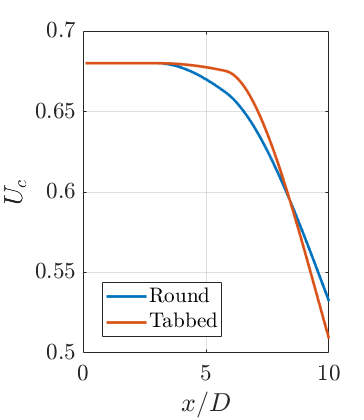}\includegraphics[width=0.33\textwidth]{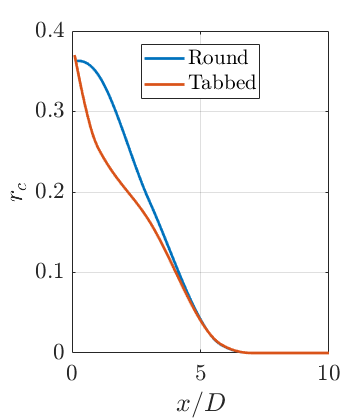}\includegraphics[width=0.33\textwidth]{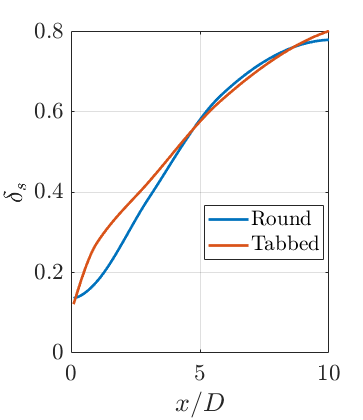}
	\caption{Streamwise variation of the mean flow parameters for both round and tabbed nozzles.}
	\label{fig:OWNS_mean_par}
\end{figure}

\begin{figure}
	\centering
	\begin{subfigure}{0.45\textwidth}
	   \includegraphics[clip=true, trim= 0 60 10 100, width=\textwidth]{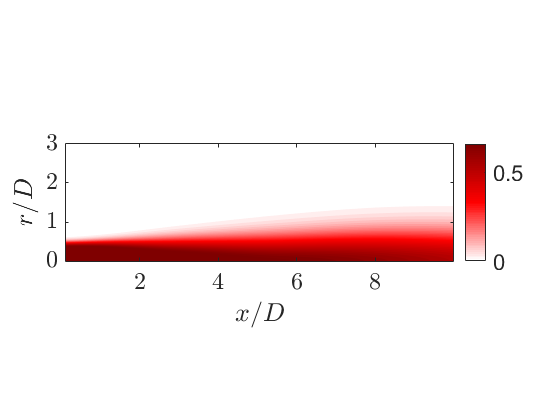}
    \caption{Round, $m=0$}
	\end{subfigure} 
 \begin{subfigure}{0.45\textwidth}
	   \includegraphics[clip=true, trim= 0 60 10 100, width=\textwidth]{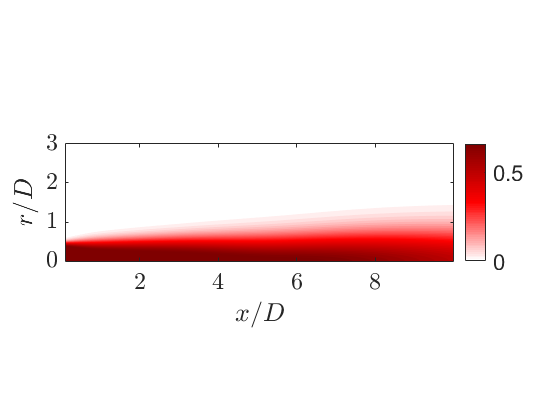}
    \caption{Tabbed, $m=0$}
	\end{subfigure} 
 \begin{subfigure}{0.45\textwidth}
	   \includegraphics[clip=true, trim= 0 60 10 100, width=\textwidth]{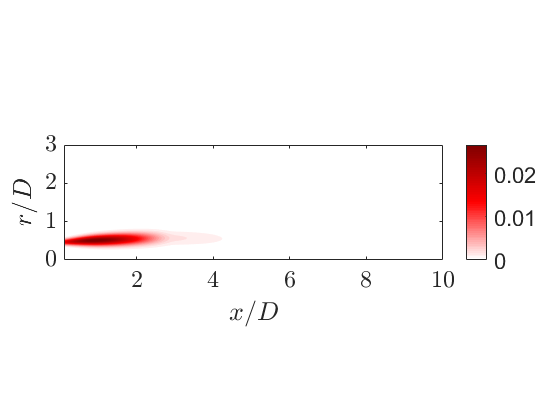}
    \caption{Tabbed, $m=12$}
	\end{subfigure}
	\caption{Azimuthal components of mean flow for both round and tabbed nozzles.}
	\label{fig:OWNS_mean_xr}
\end{figure}

The mean flows at each streamwise station are Fourier transformed in azimuth to separate the streak component from the axisymmetric component of the mean.
The latter is then fitted to the expression
\begin{equation}
    U_0(x,r)= 
     \begin{cases}
       U_c(x) &\quad\text{if } r \leq r_c(x) \\
       U_c(x) \ \ee^{-\left( \frac{r-r_c(x)}{\delta_s(x)} \right)^2} &\quad\text{if } r > r_c(x),\\
     \end{cases}
     \label{eqn:meanmodel}
\end{equation}
\noindent where $U_c$ is the centreline velocity, $r_c$ controls the position of the shear layer and $\delta_s$ controls its thickness.
As shown in~\citet{rodriguez2015study}, all these parameters are slowly varying in the streamwise direction, which will lead to a well-behaved interpolation process.
This interpolation is performed using a modified Akima cubic Hermite method in Matlab~\citep{akima1974method}, with special attention to ensuring the non-negativity of the $r_c$ parameter.
The same process is followed for the $m = 12$ component of the mean flow, where streaks will be present in the tabbed nozzle; for this component, a Gaussian profile is assumed in radius, and the resulting coefficients are smoothly interpolated in the streamwise direction.
The resulting behaviour of all three coefficients in expression \ref{eqn:meanmodel} is shown in figure~\ref{fig:OWNS_mean_par} for both nozzles.
At the initial streamwise station, both nozzles have roughly the same coefficients, but changes in the shear layer development (parameters $r_c$ and $\delta_s$) are observed very close to the nozzle.
These changes also appear to cause a shortening of the potential core, contrasting with previous observations for tabs and chevron nozzles~\citep{wernet2021characterization,zaman2011evolution}.
Such differences may be related to the lack of data near the end of the potential core or the high number of tabs used here~\citep{bridges2004parametric}.
These features are confirmed in the reconstructed mean flows in the $(x,r)$ plane, shown in figure~\ref{fig:OWNS_mean_xr}.
Generally, the shear layer of the tabbed nozzle develops more rapidly and is thicker throughout most of the domain, becoming very similar near the domain's end.
The $m = 12$ component of the mean flow has higher amplitudes at the initial three diameters and decays afterwards, in line with the behaviour observed in~\citet{nogueira2019large}.
The radial and azimuthal velocity components of the mean velocity were not considered, as no clear model exists for non-axisymmetric jets.

\begin{figure}
	\centering
	\begin{subfigure}{\textwidth}
	   \includegraphics[width=\textwidth]{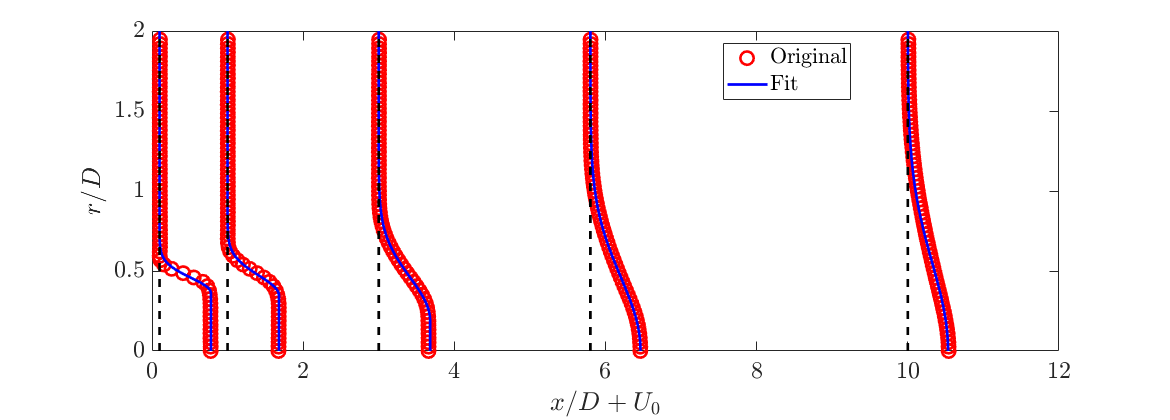}
    \caption{Round}
	\end{subfigure}
 \begin{subfigure}{\textwidth}
	   \includegraphics[width=\textwidth]{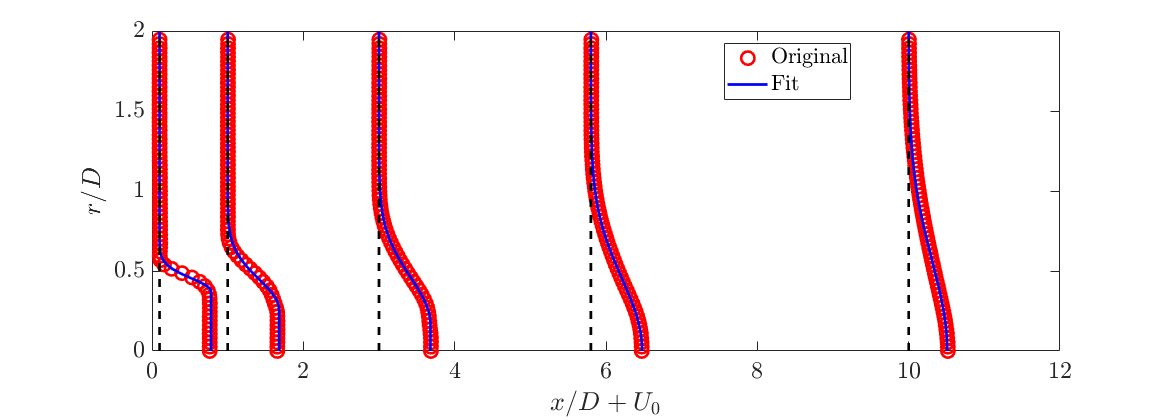}
    \caption{Tabbed}
	\end{subfigure}
	\caption{Comparison between the axisymmetric component of the mean flow from experimental data and the resulting fitted curve at the streamwise stations used in the analysis for both round (a) and tabbed nozzles (b).}
	\label{fig:OWNS_mean_lines}
\end{figure}

\begin{figure}
	\centering
     \includegraphics[clip=true, trim= 30 0 70 0, width=0.33\textwidth]{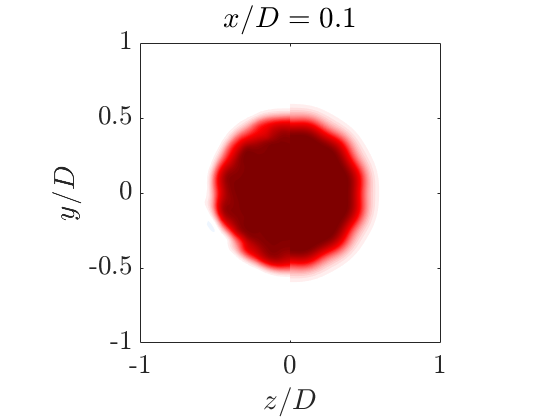}\includegraphics[clip=true, trim= 30 0 70 0, width=0.33\textwidth]{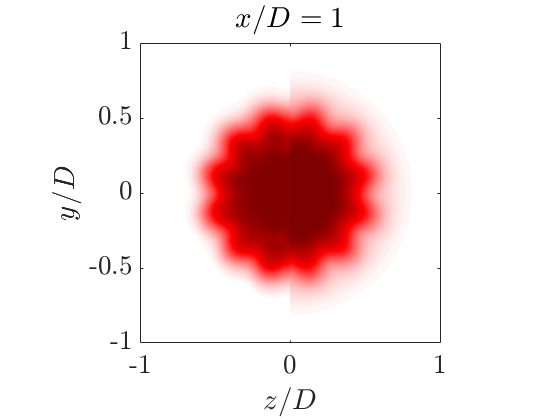}\includegraphics[clip=true, trim= 30 0 70 0, width=0.33\textwidth]{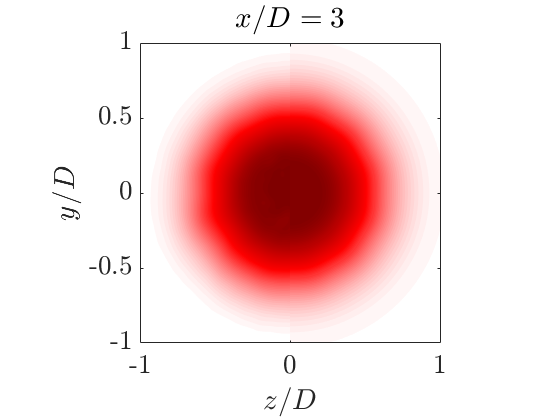}
	\caption{Cross plane reconstructions of the mean flow (right-half of the frames) compared to experimental data (left-half of the frames) for $x/D = 0.1$, 1 and 3.}
	\label{fig:OWNS_mean_yz}
\end{figure}

Thus, the reduced-order mean flow used in this work is based on a reconstruction using only the $m=0$ and 12 components of the mean flow, which are then curve-fitted into analytical functions (known to capture the overall behaviour of these components), and interpolated in the streamwise direction.
The three-dimensional flow reconstruction is therefore composed by only two azimuthal components, which are expected to capture most of the effects of the streaks in the flow.
A comparison between the original profiles extracted from the experimental data and the fitted curves is shown in figure~\ref{fig:OWNS_mean_lines}, confirming that the current approach successfully approximates the mean flow characteristics at the stations where data are available. 
A reconstruction using both the $m=0$ and 12 components of the mean flow at some of the most relevant streamwise stations in the flow is shown in figure~\ref{fig:OWNS_mean_yz}.
These plots confirm that the resulting 3D mean flow retains some of the most important characteristics of the flow, including the shear-layer modulation induced by the streaks, especially in the near-nozzle region; this level of fidelity is essential for faithful predictions using linear models.

\subsubsection{Streak-modulated wavepackets}


The remainder of the analysis focuses on the coherent structures predicted by OWNS using as input the base flows obtained from velocity measurements.
We will focus on three frequencies: $St = 0.3$, $0.5$ and $0.7$, which covers most of the region of interest for jet noise.
It is important to note that the tabs lead to two significant changes in the problem.
The first is the changes in the axisymmetric part of the mean flow induced by the presence of the streaks, as shown in figure~\ref{fig:OWNS_mean_xr}.
The modified $m = 0$ component of the tabbed jet clearly exhibits different shear layer characteristics, which also affects the potential core length and centreline decay; thus, it is expected that coherent structures generated from both nozzles already differ due to the significant changes in the mean flow.
The second effect is the presence of strong steady streaks in the time averaged flow, which cause the azimuthal modulation of the shear layer.
We will look at these two effects separately by performing OWNS calculations using only the $m=0$ component of the mean flow for both nozzles, and the reconstruction using $m=0$ and 12 components for the tabbed nozzle.
Wavepackets are predicted for $\mu = 0$, which is known to be responsible for most of the downstream acoustic radiation~\citep{cavalieri2012axisymmetric}.

\begin{figure}
	\centering
	\begin{subfigure}{0.45\textwidth}
	   \includegraphics[width=\textwidth]{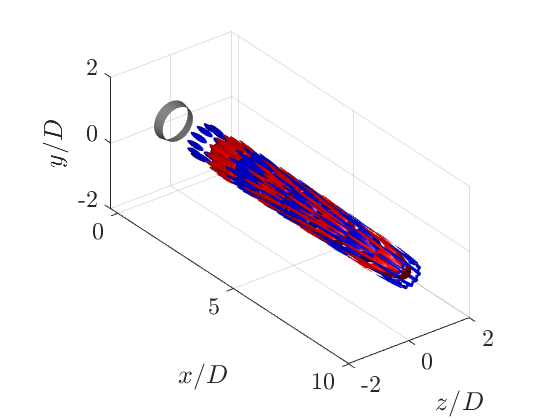}
    \caption{$u$, $St = 0.3$}
	\end{subfigure}
 \begin{subfigure}{0.45\textwidth}
	   \includegraphics[width=\textwidth]{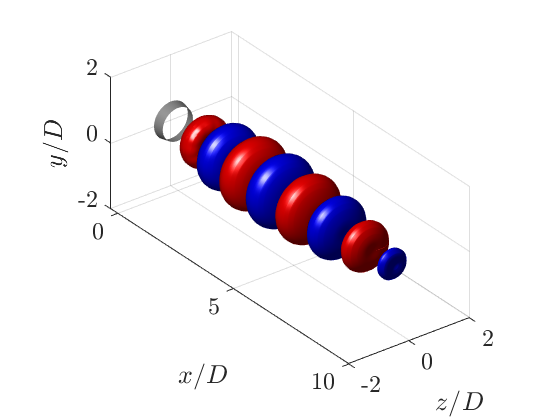}
    \caption{$p$, $St = 0.3$}
	\end{subfigure}
 \begin{subfigure}{0.45\textwidth}
	   \includegraphics[width=\textwidth]{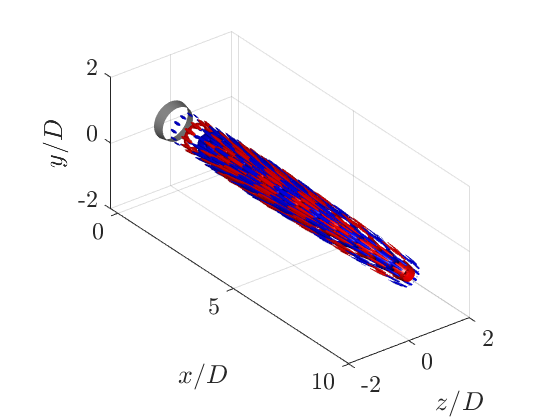}
    \caption{$u$, $St = 0.5$}
	\end{subfigure}
 \begin{subfigure}{0.45\textwidth}
	   \includegraphics[width=\textwidth]{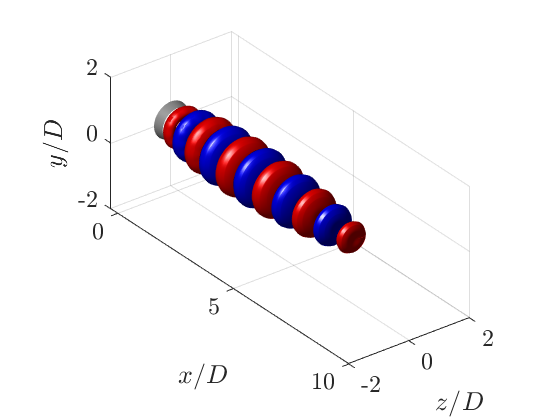}
    \caption{$p$, $St = 0.5$}
	\end{subfigure}
 \begin{subfigure}{0.45\textwidth}
	   \includegraphics[width=\textwidth]{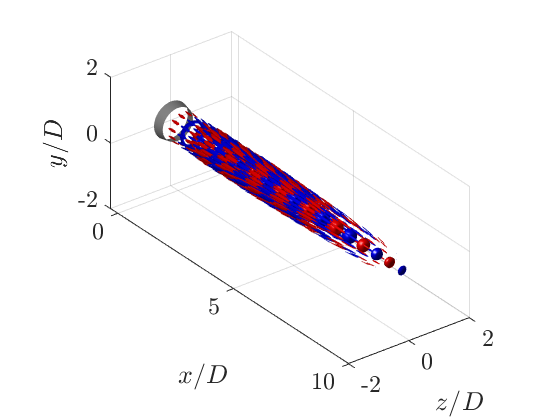}
    \caption{$u$, $St = 0.7$}
	\end{subfigure}
 \begin{subfigure}{0.45\textwidth}
	   \includegraphics[width=\textwidth]{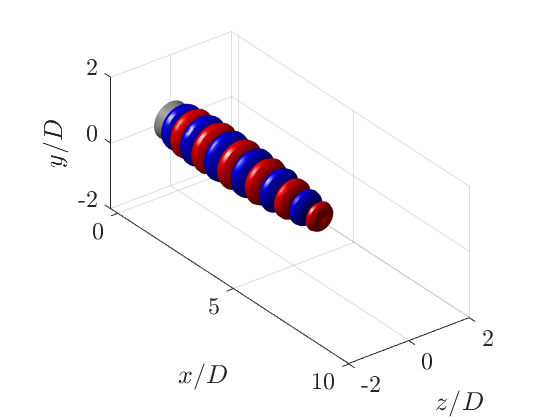}
    \caption{$p$, $St = 0.7$}
	\end{subfigure}
	\caption{Isosurfaces at $30\%$ of maximum amplitude of streamwise velocity (left) and pressure (right) for $St = 0.3$, 0.5 and 0.7 and $\mu = 0$. Only the real part of the tabbed case modes are shown.}
	\label{fig:OWNS_iso}
\end{figure}

An overview of the resulting 3D streamwise velocity and pressure fields obtained from OWNS is shown in figure~\ref{fig:OWNS_iso} for the tabbed case, which highlights the qualitative effect of shear-layer modulation in the coherent structures.
The first interesting feature shown in these plots is the fact that the azimuthal modulation is observed more clearly in the velocity, while the pressure field remains mostly axisymmetric.
All structures have the same overall wavepacket shape, with their envelopes and wavelengths decreasing in size for increasing frequency, in line with the literature~\citep{sasaki2017high}. 
The azimuthal modulation in the streamwise velocity follows the same trend, but its phase compared to the axisymmetric part of the wavepacket seems to change as it evolves in $x$. 
Interestingly, the azimuthal modulation remains strong for the entire extent of the structure, even after the mean flow has become essentially axisymmetric.
This suggests that even high-wavenumber streaks (which have a small streamwise extent, as shown in~\citet{nogueira2019large}), can affect the entire wavepacket structure, especially if they are present in the region where the K--H instability is most sensitive.

\begin{figure}
	\centering
	\begin{subfigure}{0.45\textwidth}
	   \includegraphics[clip=true, trim= 0 0 0 0, width=\textwidth]{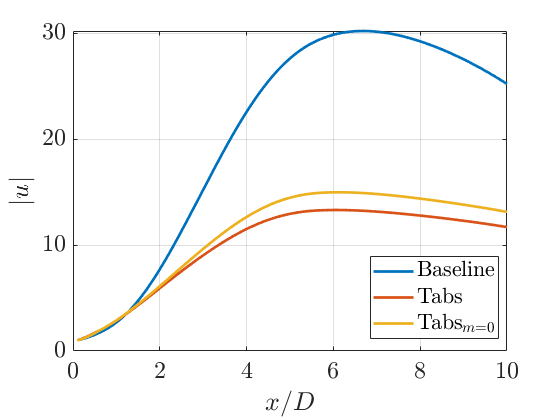}
    \caption{$St = 0.3$}
	\end{subfigure} 
 \begin{subfigure}{0.45\textwidth}
	   \includegraphics[clip=true, trim= 0 0 0 0, width=\textwidth]{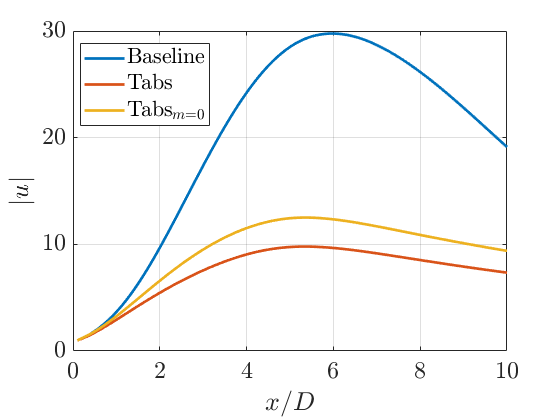}
    \caption{$St = 0.5$}
	\end{subfigure} 
 \begin{subfigure}{0.45\textwidth}
	   \includegraphics[clip=true, trim= 0 0 0 0, width=\textwidth]{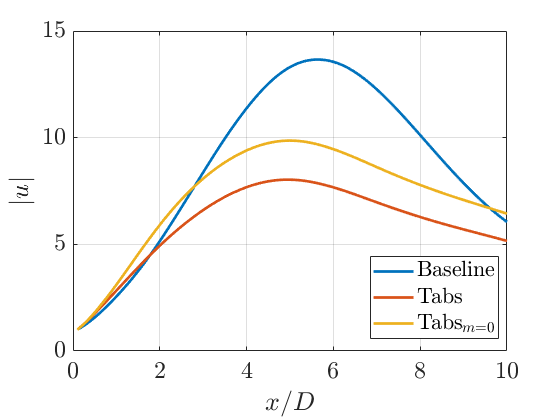}
    \caption{$St = 0.7$}
	\end{subfigure}
	\caption{Centreline signature of $m = 0$ modes.}
	\label{fig:OWNS_centreline}
\end{figure}

To evaluate the effect of the tabs in the actual growth of the wavepacket, the absolute value of the velocity disturbances predicted by OWNS at the centreline is plotted in figure~\ref{fig:OWNS_centreline}.
Here, the initial amplitude of the wavepackets are considered to be the same; this is consistent with the similar initial rms magnitudes observed in both round and tabbed nozzles, shown in figure~\ref{fig:rms_profiles_xD1_3_10}.
It is clear that the K--H mode grows more strongly in the round nozzle for all frequencies, when compared to the tabbed nozzles.
This is in line with the fact that the tabbed nozzles have thicker shear layers at most streamwise positions in the flow, which explains the stabilisation observed in the SPOD results and the far-field noise reductions.
By removing the streaks from the mean flow and recomputing the wavepackets using only the axisymmetric part, a strong damping is still observed at the centreline, confirming that most of the effect of the tabs in the wavepacket development is via changes in the axisymmetric part of the mean flow, which contains the azimuthally-averaged effect of the streaks.
That said, the presence of streaks in the 3D calculations provides a further reduction in amplitude, suggesting that the interaction between the streaks and the K--H instability leads to a further stabilisation not directly related to changes in the $m = 0$ component of the mean flow particularly at higher frequencies.

\begin{figure}
	\centering
	\begin{subfigure}{\textwidth}
	   \includegraphics[width=0.33\textwidth]{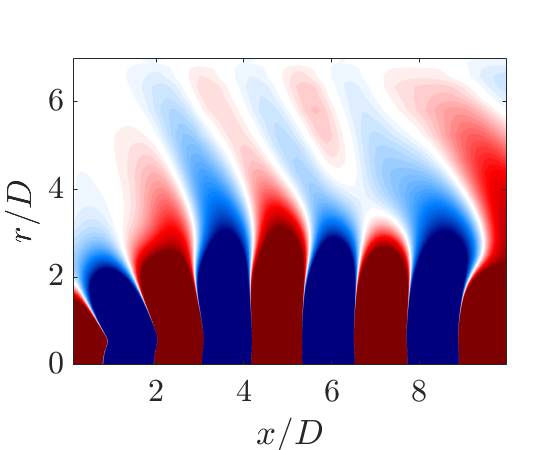}\includegraphics[width=0.33\textwidth]{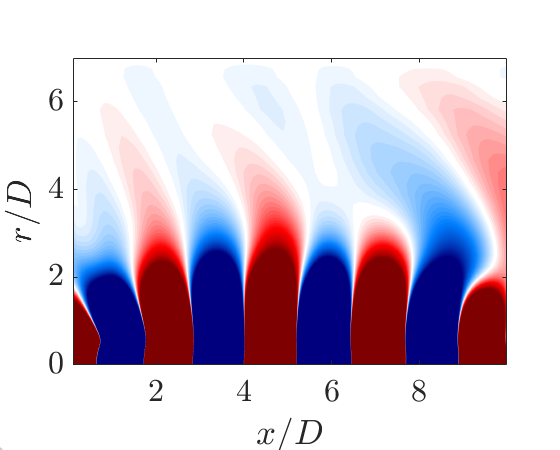}\includegraphics[width=0.33\textwidth]{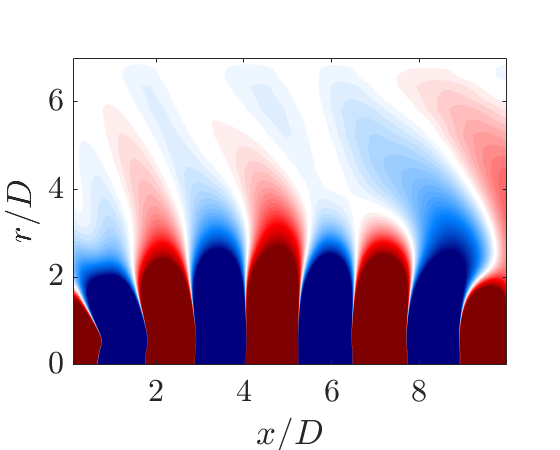}
    \caption{$St = 0.3$}
	\end{subfigure}
 \begin{subfigure}{\textwidth}
	   \includegraphics[width=0.33\textwidth]{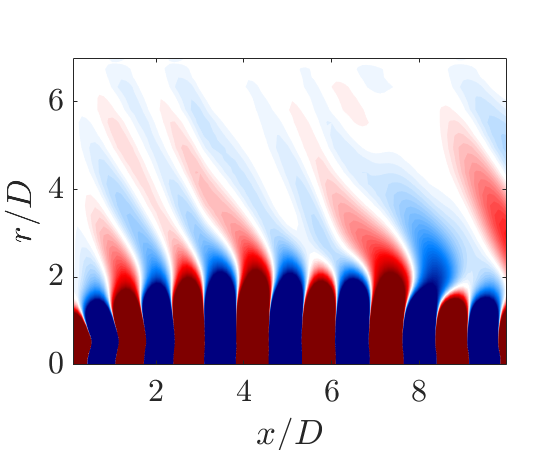}\includegraphics[width=0.33\textwidth]{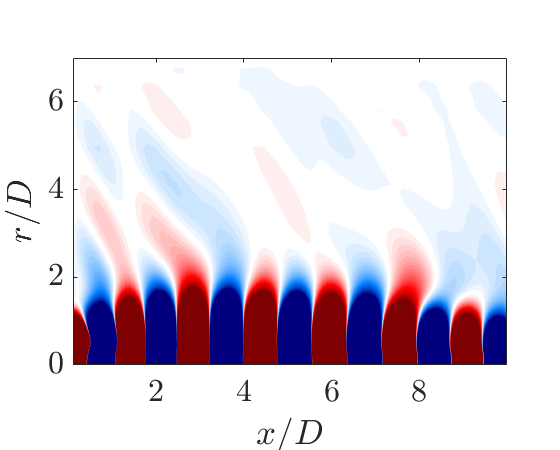}\includegraphics[width=0.33\textwidth]{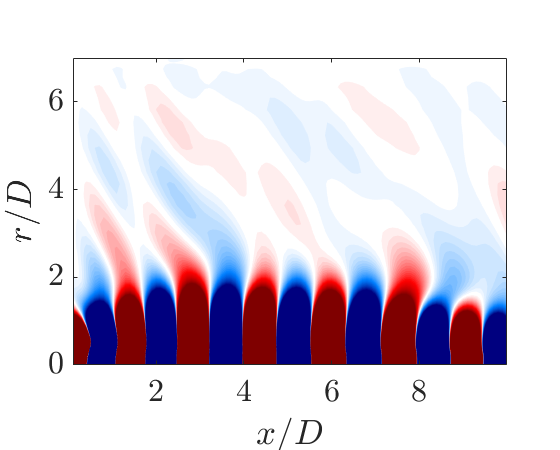}
    \caption{$St = 0.5$}
	\end{subfigure}
\begin{subfigure}{\textwidth}
	   \includegraphics[width=0.33\textwidth]{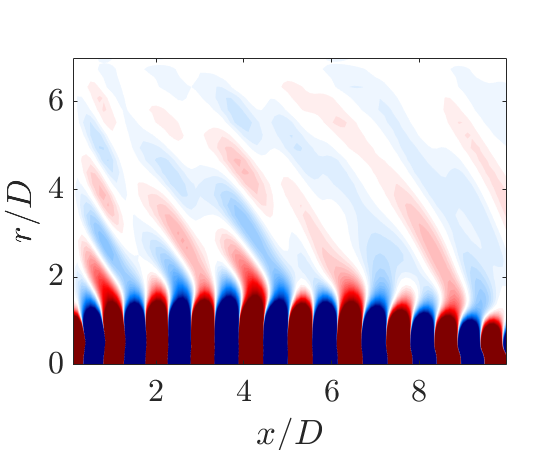}\includegraphics[width=0.33\textwidth]{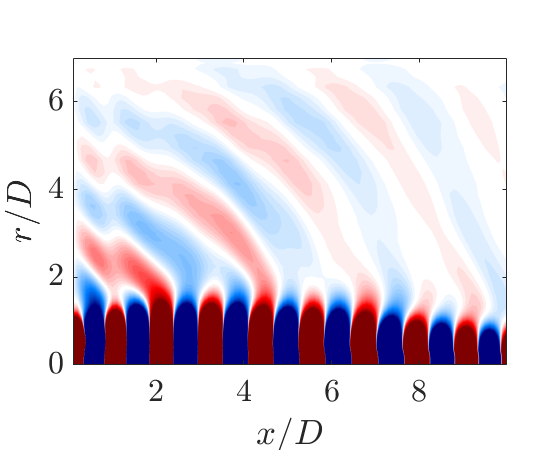}\includegraphics[width=0.33\textwidth]{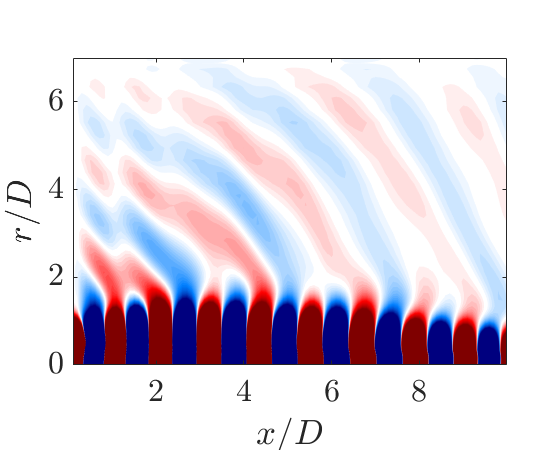}
    \caption{$St = 0.7$}
	\end{subfigure}
	\caption{Saturated pressure fields (real part) associated with the wavepackets predicted by one-way Navier--Stokes for $\mu=0$. Contours shown for $St = 0.3$ (a), 0.5 (b) and 0.7 (c), and for calculations based on the round nozzle (left), tabbed nozzle (middle) and the $m = 0$ component of the tabbed nozzle (right).}
	\label{fig:OWNS_press}
\end{figure}

One of the advantages of OWNS is the fact that the method captures part of the acoustic field associated with the predicted coherent structures~\citep{towne2015one,towne2019critical}.
In figure~\ref{fig:OWNS_press}, saturated contours of pressure are shown for the cases studied herein, with all structures forced to have the same initial amplitude at the centreline, as in figure~\ref{fig:OWNS_centreline}.
One limitation of these plots is the fact that the domain only ranges from $x/D = 0.1$ to $10$, and the characteristic acoustic beam generated from the structures should be outside this range for the lower frequencies.
Even so, some interesting trends are observed.
First, it is clear that the near-field acoustic signature of the wavepackets at lower frequencies ($St = 0.3$ and 0.5) is weaker for the tabbed cases, especially if the streaky component of the mean flow is considered in the calculations.
The beginning of the acoustic beam for these frequencies is observed by the end of the domain, and has a significant lower amplitude for the tabbed cases as well, suggesting that some level of sound reduction can be achieved directly by the damping of the wavepacket using streaks.
For the higher frequency ($St=0.7$), the acoustic signature is generally weaker (which is expected as the wavepacket is less acoustically matched,~\citep{cavalieri2019wave}).
The changes in the mean flow by the tabs leads to some changes in the near field directivity, and the model predicts a shift in the equivalent sound source position, which could lead to more sound radiation at slightly higher polar angles.
It is important to remember that OWNS does not predict the entire acoustic field, but only the downstream-travelling part associated with the linear wavepacket; the sound field is also be determined by other factors, including coherence decay~\citep{cavalieri2011jittering,cavalieri2014coherence,baqui2015coherence}, which is essential to recover the actual far-field sound radiation.
Despite these limitations, the observed trends of sound reduction at lower frequencies are in line with the acoustic measurements presented in \S~\ref{sec:acoustic_results}.

\begin{figure}
	\centering
     \includegraphics[clip=true, trim= 30 0 70 0, width=0.33\textwidth]{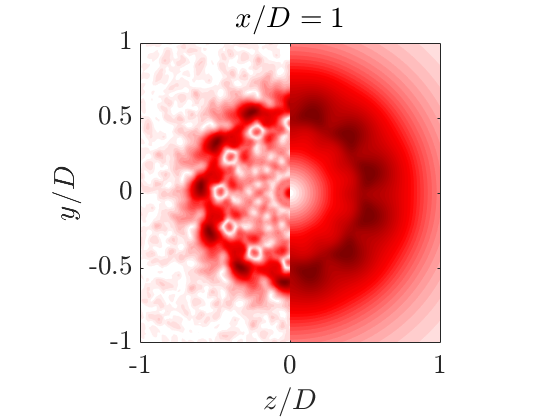}\includegraphics[clip=true, trim= 30 0 70 0, width=0.33\textwidth]{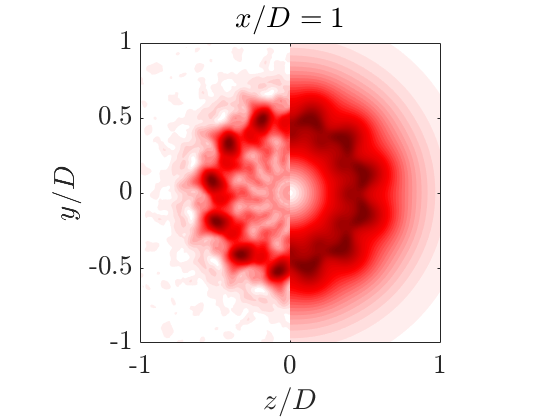}\includegraphics[clip=true, trim= 30 0 70 0, width=0.33\textwidth]{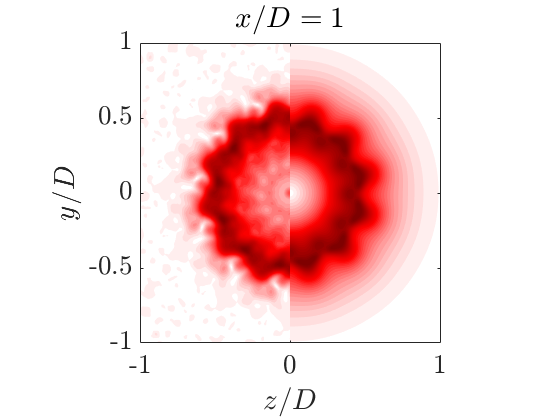} \\
     \includegraphics[clip=true, trim= 30 0 70 0, width=0.33\textwidth]{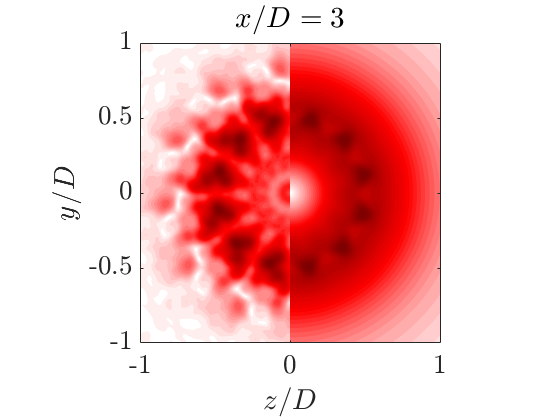}\includegraphics[clip=true, trim= 30 0 70 0, width=0.33\textwidth]{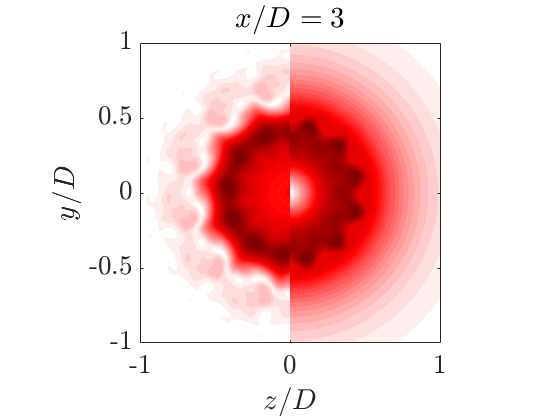}\includegraphics[clip=true, trim= 30 0 70 0, width=0.33\textwidth]{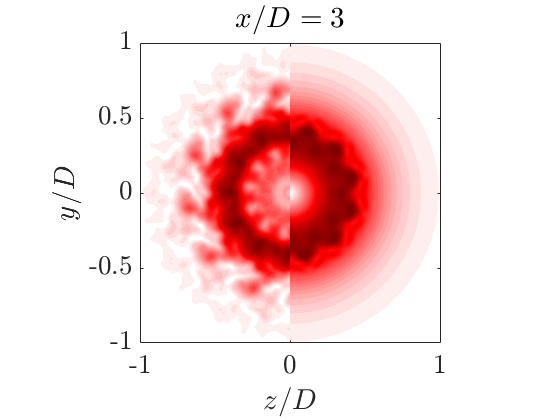}
	\caption{Absolute value of radial velocity from SPOD and OWNS. Results for $x/D = 1$ (top) and 3 (bottom), and $St = 0.3$ (left). 0.5 (middle) and 0.7 (right).}
	\label{fig:OWNS_exp_yz}
\end{figure}

Finally, a comparison between the structures obtained from OWNS and the dominant SPOD mode is shown in figure~\ref{fig:OWNS_exp_yz}.
Depending on the streamwise station, structures extracted using SPOD may be associated with streaks, Orr mechanism, or K--H wavepackets~\citep{pickering2020lift}, and the lack of streamwise coherence complicates direct mode identification.
In an attempt to overcome the issue, comparisons are restricted to positions where the K--H mode is expected to be dominant, namely $x/D=1$ and 3, and comparisons are based on the absolute value of the radial velocity component, as this component offers clearer comparisons with modelling approaches~\citep{edgington2021waves,nogueira2021investigation}.
Good agreement is found at higher frequencies, where the azimuthal modulation predicted by the model agrees well with the experimental data, despite the interpolation process of the mean flow.
Interestingly, the model predicts a significantly broader radial support for the radial velocity, which could also be related to the differences in mean flow profile or due to the fact that the mean radial velocity was not included in the calculations.
The agreement deteriorates for $St = 0.3$ and $x/D = 1$, but at the moment it is unclear if the structure obtained from SPOD at this station is the same as the one observed further downstream; in fact, the shape of the velocity field for this mode resembles other $m = 12$-dominated modes (either K--H or streak-like), which may be responsible for the discrepancy. 

\section{Conclusions}
\label{sec:conclusions}

This work investigated changes in the acoustic field and in the dynamics of noise-generating coherent structures in jets by tab-inducing streaks.
The tabs were designed to stabilize the K--H mechanism, following the guidelines by~\citet{pujals2010forcing} and nozzle boundary-layer characteristics at $M_j = 0.4$~\citep{kaplan2021nozzle}.
Acoustic measurements were carried out using an azimuthal antenna and the data were analysed as a function of azimuthal wavenumber.
Time-resolved particle image velocimetry (TR-PIV) measurements were performed to explore the flow field.
The velocity field was decomposed into Floquet coefficients, corresponding to azimuthal Fourier modes, due to the $L$-fold azimuthal symmetry imposed by the tabs.
Spectral proper orthogonal decomposition (SPOD) was then used to further decompose the Floquet modes in the inhomogeneous radial direction.

The presence of the streak-inducing tabs led to noise reductions of up to 3 dB/St for the tripped and up to 6 dB/St for the untripped configurations.
The noise reduction becomes smaller as the Mach number increases, attaining a maximum reduction of 1.5 dB/St for $M_j = 0.9$.
Overall sound pressure level (OASPL) results over the $0.05 \leq St \leq 3.5$ range confirm reductions of up to 2 dB for the tripped and 3 dB for the untripped configurations.
Such reductions were observed for all Mach numbers in the range $0.4 \leq M_j  \leq 0.9$.
Moreover, the achieved noise reduction levels are comparable to or exceed those reported in the literature on chevrons, micro-jets, and tabs~\citep{saiyed1999tabs,tam2000subsonic,bridges2004parametric,callender2005far,alkislar2007effect,zaman2011evolution}.

The TR-PIV results show that tabs leave a signature on the mean flow data at least up to the end of the potential core.
While streaky structures are present for the baseline configuration (as shown by~\citet{nogueira2019large} and~\citet{pickering2020lift}), mainly for low Strouhal numbers ($St \rightarrow 0$), such structures are strongly enhanced by the tabs.
Near the end of the potential core, the most energetic SPOD modes for the baseline nozzle resemble K--H structures, and in the case of the tabbed nozzle these K--H structures are found to be strongly attenuated, in agreement with previous studies~\citep{sinha2016parabolized,marant2018influence,lajus2019spatial,wang2021effect}.

The attenuation effect is reproduced in a linear model based on the one-way Navier--Stokes (OWNS) equations, where linearisation is performed about the mean flow.
The growth of wavepackets was mitigated for all frequencies considered, with stronger damping at lower frequencies; this is also the region in the spectrum where greatest sound reductions are observed, suggesting a link between streak-induced hydrodynamic effects and far-field radiation.
It was also observed that the modulation by streaks persists beyond the point where the flow becomes axisymmetric and that the associated near-pressure field is generally weaker for the tabbed nozzle.
Finally, comparison between OWNS and SPOD revealed good qualitative agreement between the empirical and modelled coherent structures, showing that the linear model captures the key underlying physics.
Interestingly, it is found that modifications of the axisymmetric part of the mean flow induced by the tabs is responsible for most of the wavepacket attenuation, but stronger reductions are observed if the streaky component of the mean flow is explicitly included in the calculations.
While it is hard to isolate the effect of the streaks on wavepacket development from their effect on the mean flow, these results seem to indicate that this effect may be partially represented in models based solely on the axisymmetric ($m = 0$) mean flow component.

Although the physical mechanisms behind the use of tabs and chevrons are quite similar, modifying the jet shear layer by inducing rolls and streaks, the design of the streak-inducing tabs studied in this paper is physics-based, rather than empirical.
The tab position and geometry were designed to promote large-amplitude streaks in the turbulent boundary layer at the nozzle exit.
This provides an interesting perspective for jet noise reduction, as these or other devices may be designed targeting the induction of streak growth.
Note that the use of chevrons is often associated with a thrust penalty~\citep{krothapalli1993role, saiyed2003acoustics, zaman2011evolution}.
Unfortunately, thrust could not be measured in the present test facility, and this remains an open point to be addressed in future studies.


\backsection[Acknowledgements]{The authors would like to thank the \emph{Prométée plateforme} staff, mainly Dr. Anton Lebedev and Eng. Damien Eysseric. We are also grateful to Dr. Barbara G. Hasparyk for her technical support.}


\backsection[Funding]{A.V.G. Cavalieri was supported by the National Council for Scientific and Technological Development (CNPq/Brazil), grant \#310523/2017-6. I.A. Maia received funding from São Paulo Research Foundation (FAPESP/Brazil), grant \#2022/06824-4. P.A.S. Nogueira was supported by the Australian Research Council through the Discovery Early-Career Research Award: DE240100933.}

\backsection[Declaration of interests]{The authors report no conflict of interest.}


\backsection[Author ORCIDs]{F.R. do Amaral, \url{https://orcid.org/0000-0003-1158-3216}; P.A.S. Nogueira, \url{https://orcid.org/0000-0001-7831-8121}; I.A. Maia, \url{https://orcid.org/0000-0003-2530-0897}; A.V.G. Cavalieri, \url{https://orcid.org/0000-0003-4283-0232}; P. Jordan, \url{https://orcid.org/0000-0001-8576-5587}.}



\appendix

\section{Complementary acoustic results}
\label{app:acoustic_extra}

\begin{figure}
	\centering
	\includegraphics[width=\textwidth]{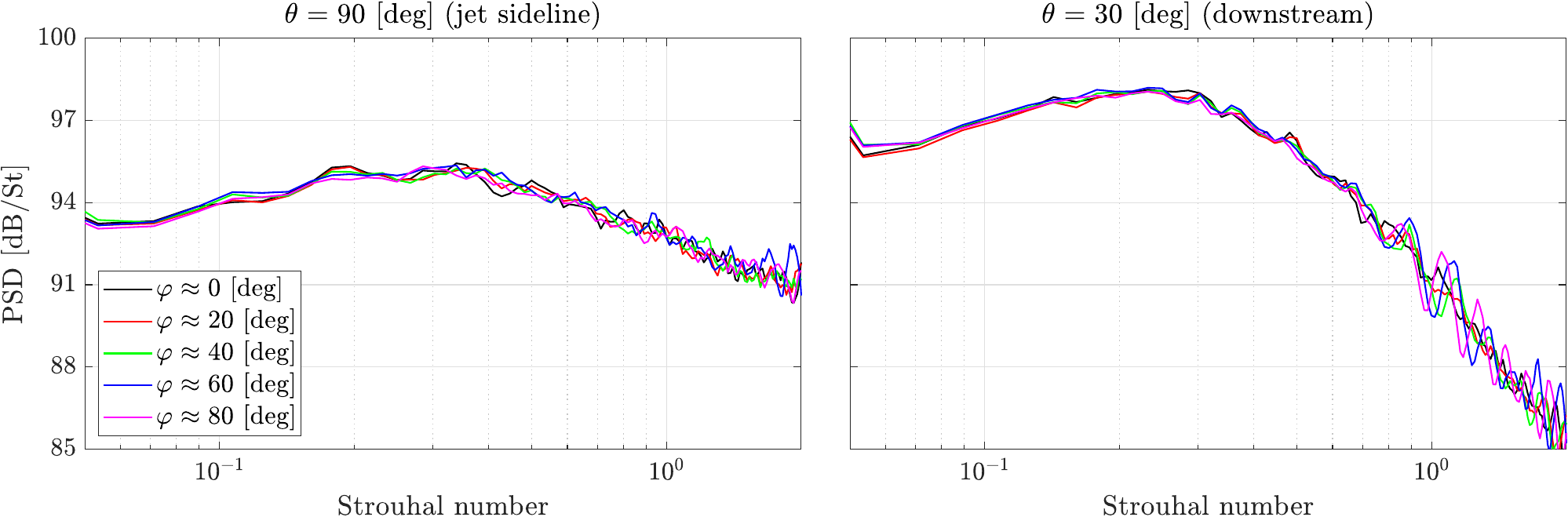}
	\caption{Acoustic spectra for $M_j = 0.4$ for both the tripped tabbed nozzle configuration and several azimuthal angles $\varphi$ in the 0 to 80$^\circ$ range. Left frame: $\theta =$ 90 deg (nozzle exit, sideline). Right frame: $\theta =$ 30 deg.} 
	\label{fig:spectra_theta90-30_phi0-80_extra}
\end{figure}

Figure~\ref{fig:spectra_theta90-30_phi0-80_extra} shows the acoustic spectra for several array microphones in the $0^\circ \geq \varphi \geq 80^\circ$ for polar angles of $\theta = 90^\circ$ and $30^\circ$, respectively (see the sketch in Figure~\ref{fig:sketch_antenna}).
The measurements correspond to a jet flow at $M_j = 0.4$.
Spectra for the tripped tabbed nozzle configuration are presented in the figure and demonstrate the homogeneity of the acoustic field, which justifies the use of azimuthal decomposition in the subsequent analyses.
Similar results were obtained for the untripped configuration.

\begin{figure}
	\centering
	\includegraphics[width=\textwidth]{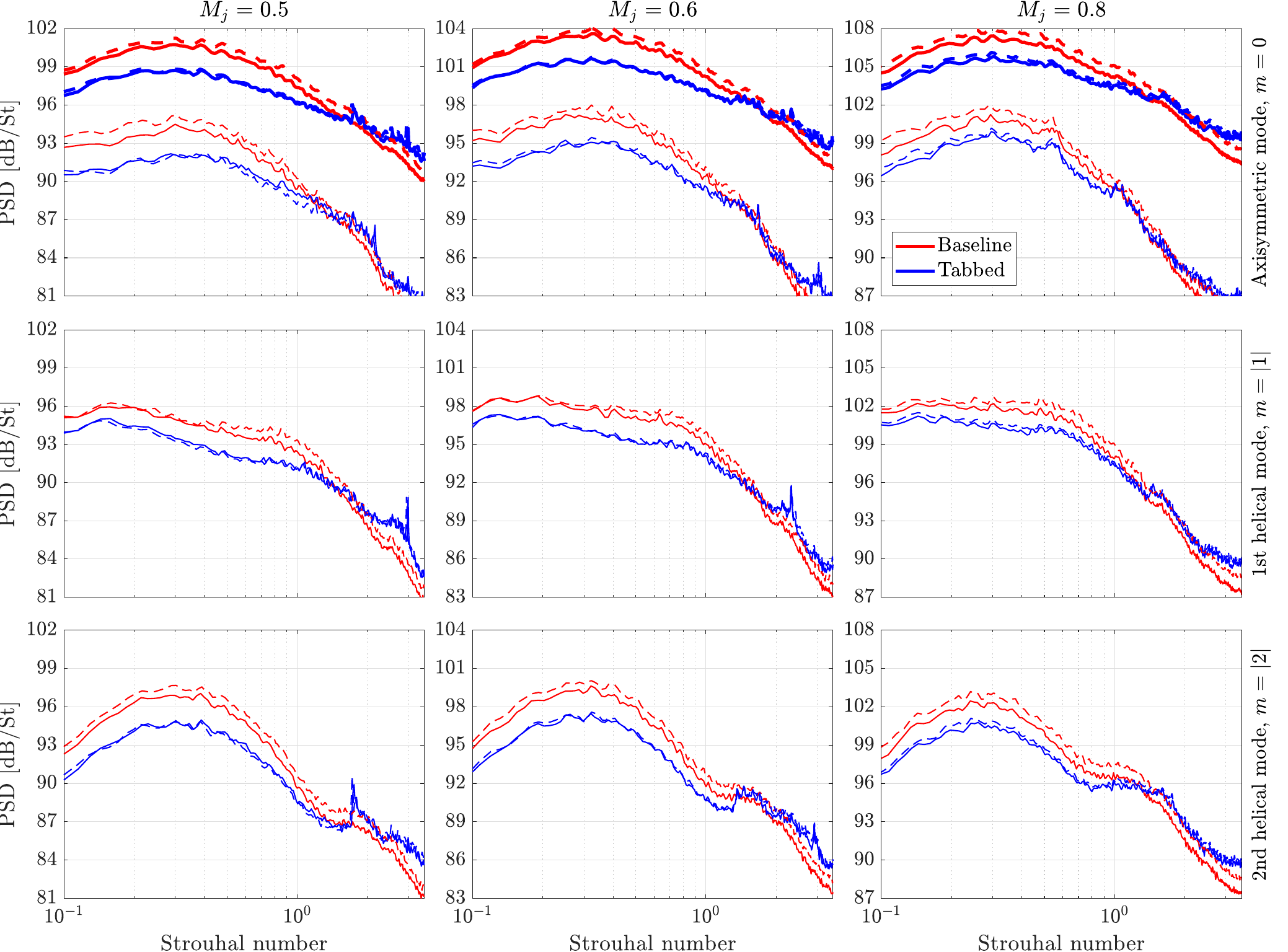}
	\caption{Acoustic spectra for polar angle $\theta =$ 90 deg (nozzle exit, sideline) for both tripped ($-$) and untripped ($--$) configurations. Frames, from top to bottom: axisymmetric mode ($m =$ 0), 1st helical mode ($|m| =$ 1) and 2nd helical mode ($|m| =$ 2). Frames, from left to right: Mach 0.5, 0.6 and 0.8. The top frames display the full sound radiation without azimuthal decomposition, shown as thick curves.} 
	\label{fig:spectra_az_theta90_extra}
\end{figure}

\begin{figure}
	\centering
	\includegraphics[width=\textwidth]{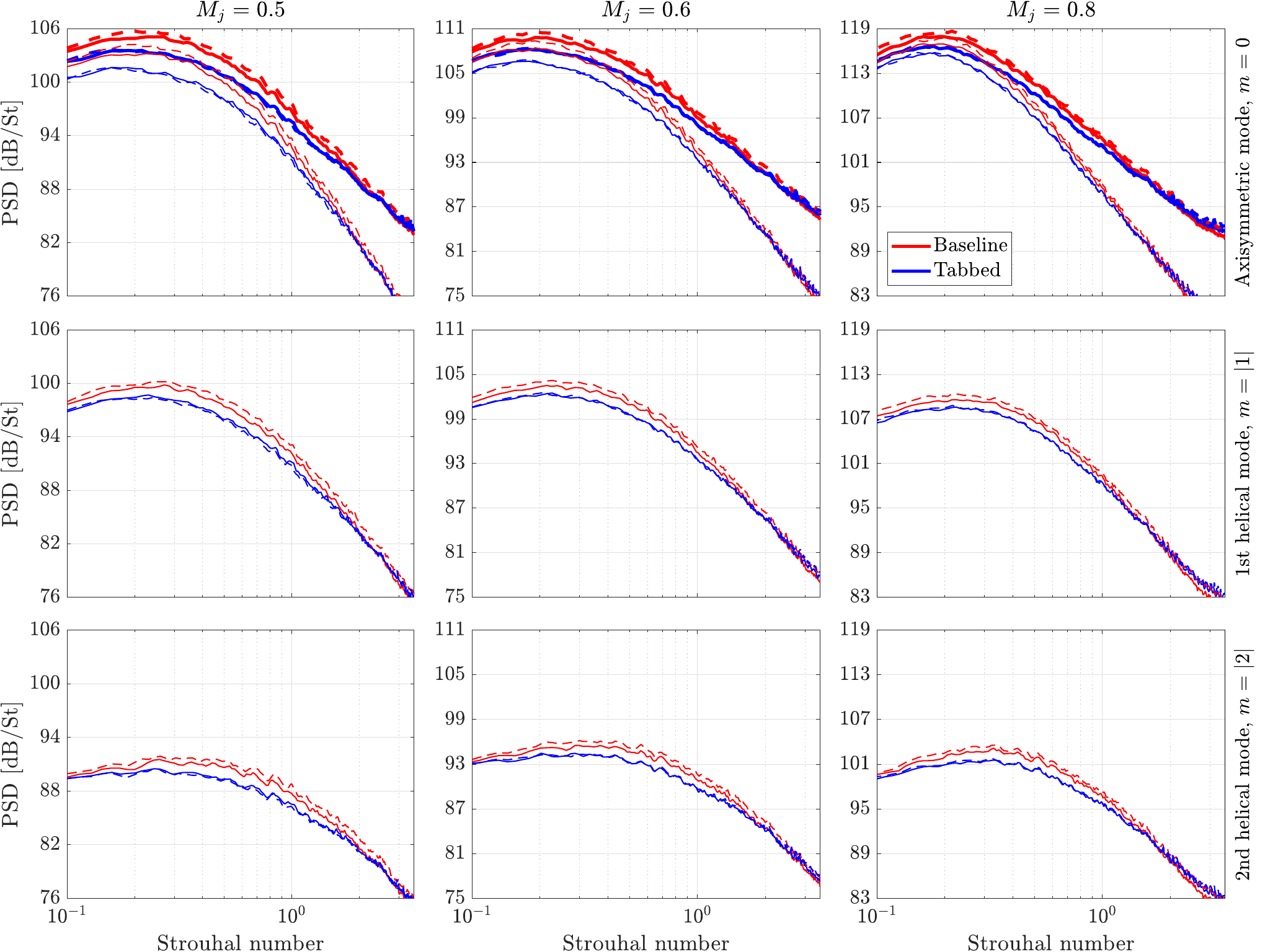}
	\caption{Acoustic spectra for polar angle $\theta =$ 30 deg. See the comments in the caption of figure~\ref{fig:spectra_az_theta90_extra}.} 
	\label{fig:spectra_az_theta30_extra}
\end{figure}

\begin{figure}
	\centering
	\includegraphics[width=\textwidth]{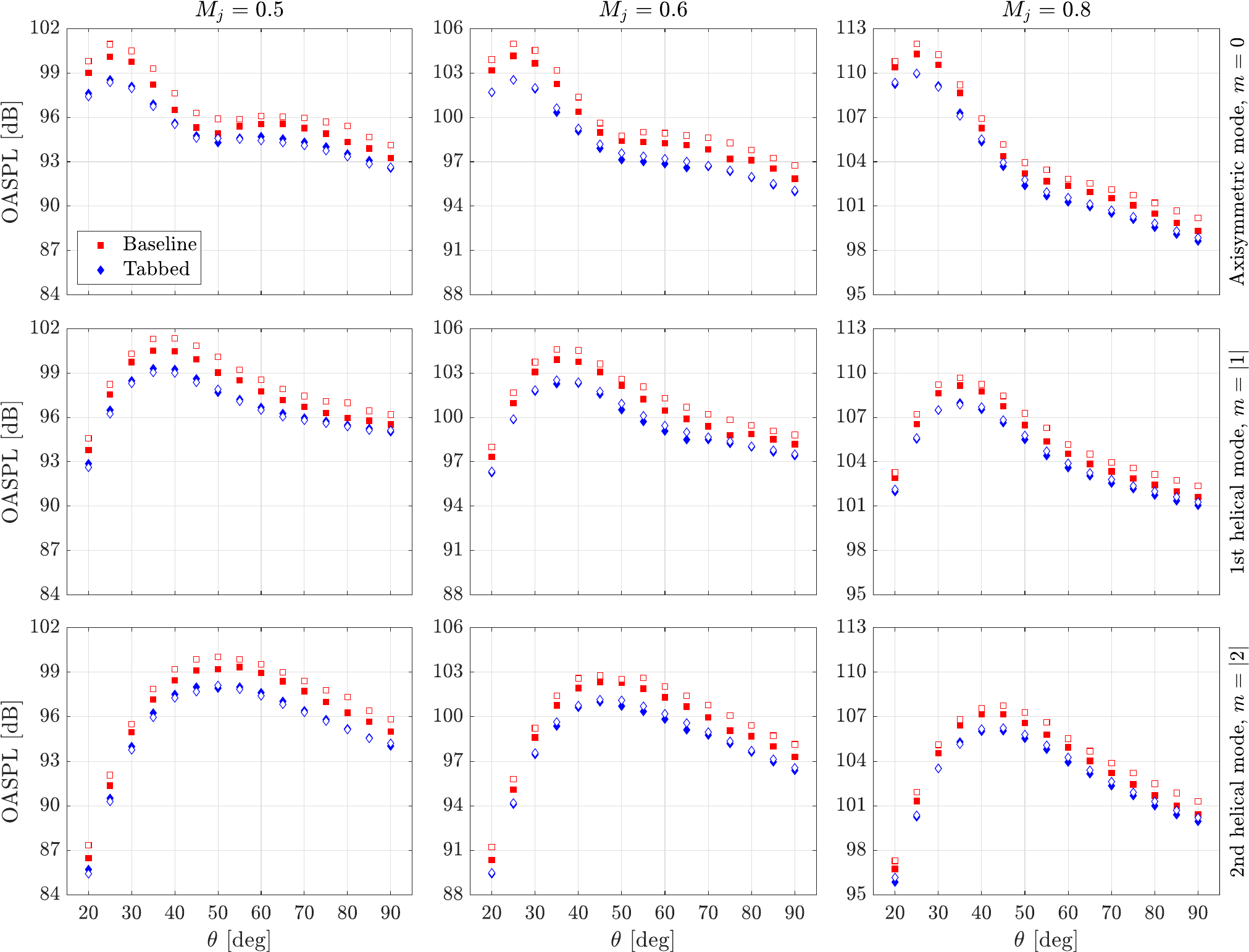}
	\caption{OASPL for polar angles in the $\theta =$ 20 deg to 90 deg interval for tripped (closed symbols) and untripped (open symbols) configurations. Power levels evaluated for 0.1 $\leq St \leq$ 3.5. Frames, from top to bottom: axisymmetric mode ($m =$ 0), 1st helical mode ($|m| =$ 1) and 2nd helical mode ($|m| =$ 2). Frames, from left to right: Mach 0.5, 0.6 and 0.8.} 
	\label{fig:oaspl_az_extra}
\end{figure}

Figures~\ref{fig:spectra_az_theta90_extra}, \ref{fig:spectra_az_theta30_extra} and \ref{fig:oaspl_az_extra} exhibit the acoustic spectra at $\theta =$ 90 and 30 deg and the corresponding OASPL, respectively, for $M_j =$ 0.5, 0.6 and 0.8.
The trends observed are consistent with those shown in figures~\ref{fig:spectra_az_theta90}, \ref{fig:spectra_az_theta30} and \ref{fig:oaspl_az}, corresponding to $M_j =$ 0.4, 0.7 and 0.9.
In terms of spectral levels, the noise reduction achieved by the tabs decreases smoothly from 3 to 1.5 dB/St as the Mach number increases from 0.4 to 0.9 for the tripped nozzles.
For the untripped nozzles, the noise reduction ranges from 6 to 1.5 dB/St over the same Mach numbers.
The OASPL indicate reductions between 1 to 3 dB, depending on the Mach number, polar angle and azimuthal mode.

\begin{figure}
	\centering
	\includegraphics[width=\textwidth]{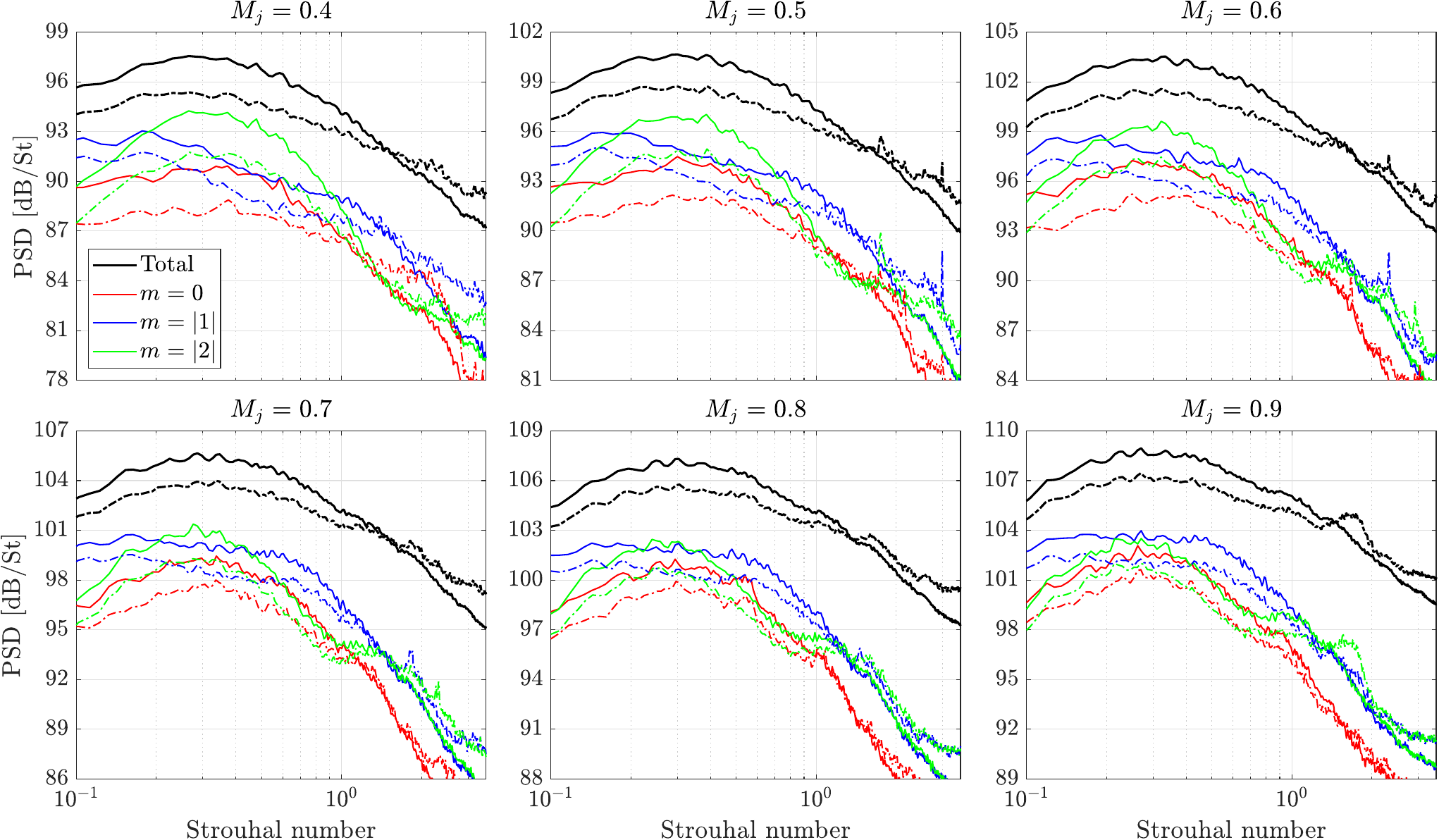}
	\caption{Acoustic spectra for polar angle $\theta =$ 90 deg  (nozzle exit, sideline) and tripped configurations. Continuous and dash-dotted curves regard baseline and tabbed nozzles, respectively. Top frames: Mach 0.4, 0.5 and 0.6 Bottom frames: Mach 0.7, 0.8 and 0.9.} 
	\label{fig:spectra_az_theta90_extra2}
\end{figure}

\begin{figure}
	\centering
	\includegraphics[width=\textwidth]{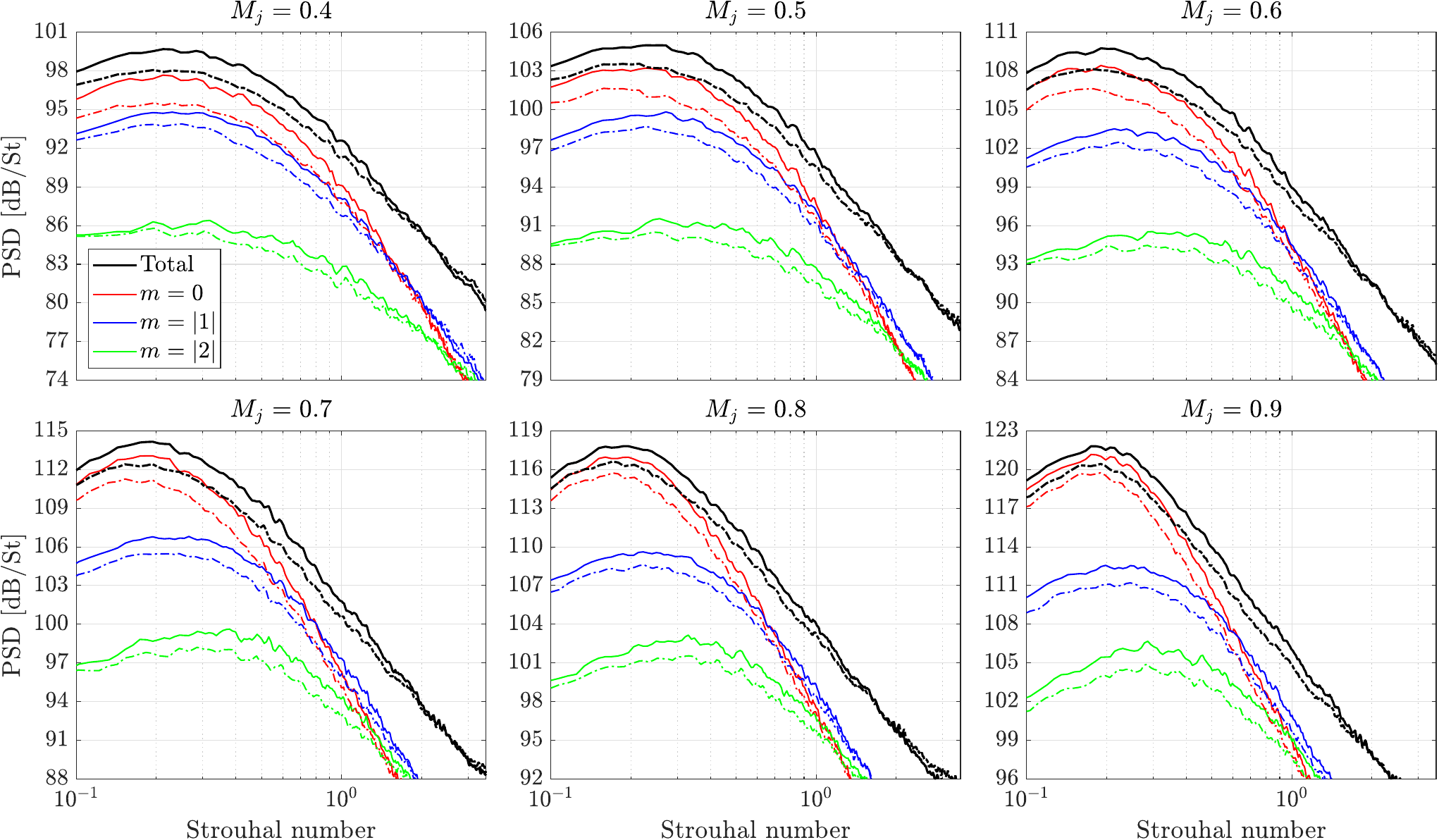}
	\caption{Acoustic spectra for polar angle $\theta =$ 30 deg and tripped nozzles. See the comments in the caption of figure~\ref{fig:spectra_az_theta90_extra2}.} 
	\label{fig:spectra_az_theta30_extra2}
\end{figure}

\begin{figure}
	\centering
	\includegraphics[width=\textwidth]{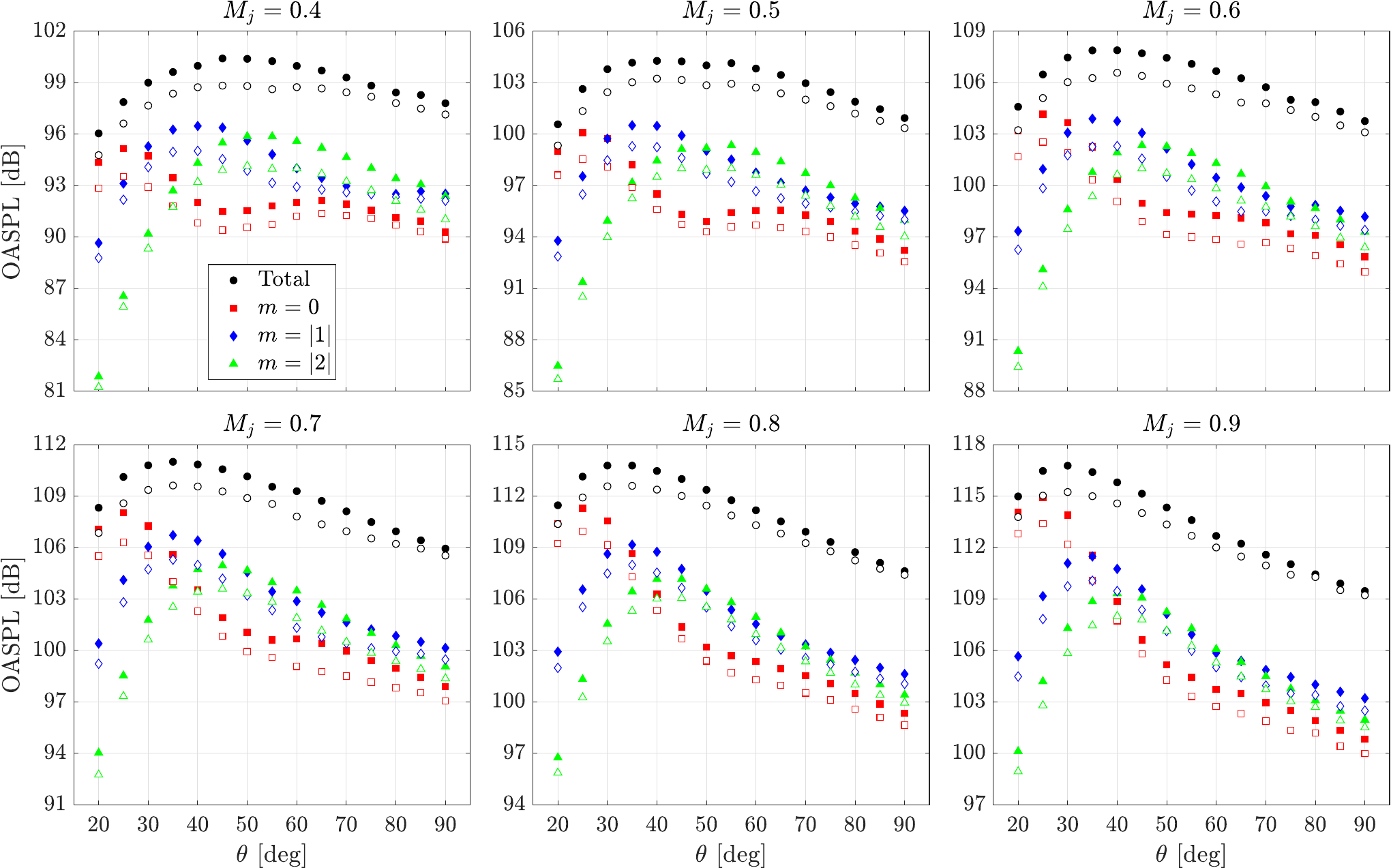}
	\caption{OASPL for polar angles in the $\theta =$ 20 deg to 90 deg interval and tripped configurations. Power levels evaluated for 0.1 $\leq St \leq$ 3.5. Closed and open symbols regard baseline and tabbed nozzles, respectively. See the comments in the caption of figure~\ref{fig:spectra_az_theta90_extra2}.} 
	\label{fig:oaspl_az_extra2}
\end{figure}

Figures~\ref{fig:spectra_az_theta90_extra2} to \ref{fig:oaspl_az_extra2} show the acoustic spectra for $\theta =$ 90 and 30 deg and the OASPL for the baseline and tabbed nozzles, respectively.
Only the tripped configurations are considered, with the Mach numbers in the 0.4 to 0.9 range and azimuthal modes 0, 1(-1) and 2(-2).
These plots were made in the same fashion as~\citet{cavalieri2012axisymmetric} and enable a direct comparison among the azimuthal modes spectral levels.

\section{Operators for one-way Navier--Stokes}
\label{app:OWNS_op}

The explicit form of the operators used in \S~\ref{sec:wavepacket_owns} are given by 
\begin{equation}
    \boldsymbol{L}=\boldsymbol{L_0}-\boldsymbol{L_{visc}},
\end{equation}
\noindent with
\begin{eqnarray}
\boldsymbol{L_0}=
    \setlength{\arraycolsep}{4pt}
    \renewcommand{\arraystretch}{1.3}
    \left[
    \begin{array}{ccccc}
    - \partial_x \bar{u}  &  \partial_x \bar{\nu} &  \partial_r \bar{\nu}-\bar{\nu}(D_r+\frac{1}{r})  &  \frac{\partial_\theta \bar{\nu}}{r}-\frac{\bar{\nu}}{r}\left(\ii \mu + D_\theta  \right)  &  0  \\
    \partial_x \bar{p} & \partial_x \bar{u}  &  \partial_r \bar{u}  &  \frac{\partial_\theta \bar{u}}{r}  &  0  \\
    \partial_r \bar{p}  &  0  & 0 &  0  &  \bar{\nu} D_r   \\
    \frac{\partial_\theta \bar{p}}{r}  &  0  &  0 &  0 & \frac{\bar{\nu}}{r}\left( \ii\mu + D_\theta \right)   \\
    0  &  \partial_x \bar{p}  &  \partial_r \bar{p} + \gamma \bar{p} D_r + \gamma \frac{\bar{p}}{r}  &  \frac{\partial_\theta \bar{p}}{r}+\gamma \frac{\bar{p}}{r}\left( \ii\mu + D_\theta \right) & \gamma \partial_x \bar{u}  \\
    \end{array}  \right] \nonumber \\
    \
\end{eqnarray}  
\begin{eqnarray}
\boldsymbol{L_{visc}}=\frac{1}{Re}
    \setlength{\arraycolsep}{4pt}
    \renewcommand{\arraystretch}{1.3}
    \left[
    \begin{array}{ccccc}
    0  &  0 &  0  &  0  &  0  \\
    \partial_r^2\bar{u}+\frac{\partial_r\bar{u}}{r}+\frac{\partial_\theta^2\bar{u}}{r^2} & \bar{\nu} \Delta  &  0  &  0  &  0  \\
    0  &  0  & \bar{\nu} (\Delta - \frac{1}{r^2}) &  -\frac{2 \bar{\nu}}{r^2} \left( \ii \mu + D_\theta \right)  &  0   \\
    0  &  0  &  \frac{2 \bar{\nu}}{r^2} \left( \ii \mu + D_\theta \right) &  \bar{\nu} (\Delta - \frac{1}{r^2}) & 0   \\
    \frac{\gamma}{Pr}\Delta_\nu   &  0  &  0  &  0 & \frac{\gamma}{Pr}\Delta_p  \\
    \end{array}  \right], \nonumber \\
    \
    \label{eqn:ap.Lv}
\end{eqnarray}
\noindent and
\begin{eqnarray}
\boldsymbol{A}=
    \setlength{\arraycolsep}{5pt}
    \renewcommand{\arraystretch}{1.3}
    \left[
    \begin{array}{ccccc}
			\bar{u} & -\bar{\nu}  		& 0  			& 0  			& 0  				\\
			0  			& \bar{u}  				& 0  			& 0  			& \bar{\nu}	\\
			0  			& 0  							& \bar{u}	& 0  			& 0   			\\
			0  			& 0  							& 0 			& \bar{u}	& 0   			\\
			0  			& \gamma \bar{p}	& 0  			& 0  			& \bar{u}		\\
    \end{array}  \right],
    \label{eqn:ap.B0}
\end{eqnarray}
\noindent where $\bar{u}$, $\bar{\nu}$ and $\bar{p}$ are the mean streamwise velocity, specific volume and pressure respectively. Note that, in the above operators, the streamwise pressure and specific volume gradients were neglected as the jet is isothermal and shock-free. The operators $\Delta$, $\Delta_\nu$ and $\Delta_p$ are given by
\begin{equation}
    \Delta=D_r^2+\frac{1}{r}D_r-\frac{1}{r^2}\left( -\mu^2 + 2\ii\mu D_\theta + D_\theta^2 \right),
\end{equation}
\begin{equation}
    \Delta_\nu=\bar{p} \Delta + \partial_r^2\bar{p}+\frac{\partial_r\bar{p}}{r}+\frac{\partial_\theta^2\bar{p}}{r^2} + 2\partial_r\bar{p}D_r+\frac{\partial_t\bar{p}}{r^2}(\ii\mu+D_\theta),
\end{equation}
\begin{equation}
    \Delta_p=\bar{\nu} \Delta + \partial_r^2\bar{\nu}+\frac{\partial_r\bar{\nu}}{r}+2\partial_r\bar{\nu}D_r+\frac{\partial_t\bar{\nu}}{r^2}(\ii\mu+D_\theta).
\end{equation}

\section{SPOD convergence}
\label{app:SPOD_convergence}

In this appendix, a convergence analysis of the SPOD computations is carried out using two subsets of the dataset: the full time series ($N_s = 27,000$ snapshots) and its first and second halves ($N_s = 13,500$ snapshots).
These subsets correspond to $N_b = 421$ and 210 blocks, respectively, using a fixed block size of $N_{\mathit{fft}} = 128$ with 50\% overlap, as described in \S~\ref{sec:piv_processing}.

Following~\citet{lesshafft2019resolvent}, the alignment between SPOD modes computed using the full time series ($\boldsymbol{\Phi}_\mathit{full}$) and those obtained from either the first or second halves ($\boldsymbol{\Phi}i$, with $i = 1, 2$) is quantified as
\begin{equation}
    \beta(\mathit{St,\mu}) = \frac{\left\lvert\ \langle \boldsymbol{\Phi}_\mathit{full}(\mathit{St},\mu),~{\boldsymbol{\Phi}_i}(\mathit{St},\mu) \rangle \right\rvert}{ \vert\vert \boldsymbol{\Phi}_\mathit{full}(\mathit{St},\mu)\vert\vert \hspace{2pt} \vert\vert {\boldsymbol{\Phi}_i}(\mathit{St},\mu)\vert\vert} \mbox{.}
\label{eq:beta}
\end{equation}
A perfectly converged mode yields $\beta = 1$, while complete misalignment results in $\beta = 0$.

\begin{figure}
	\centering
	\includegraphics[width=\textwidth]{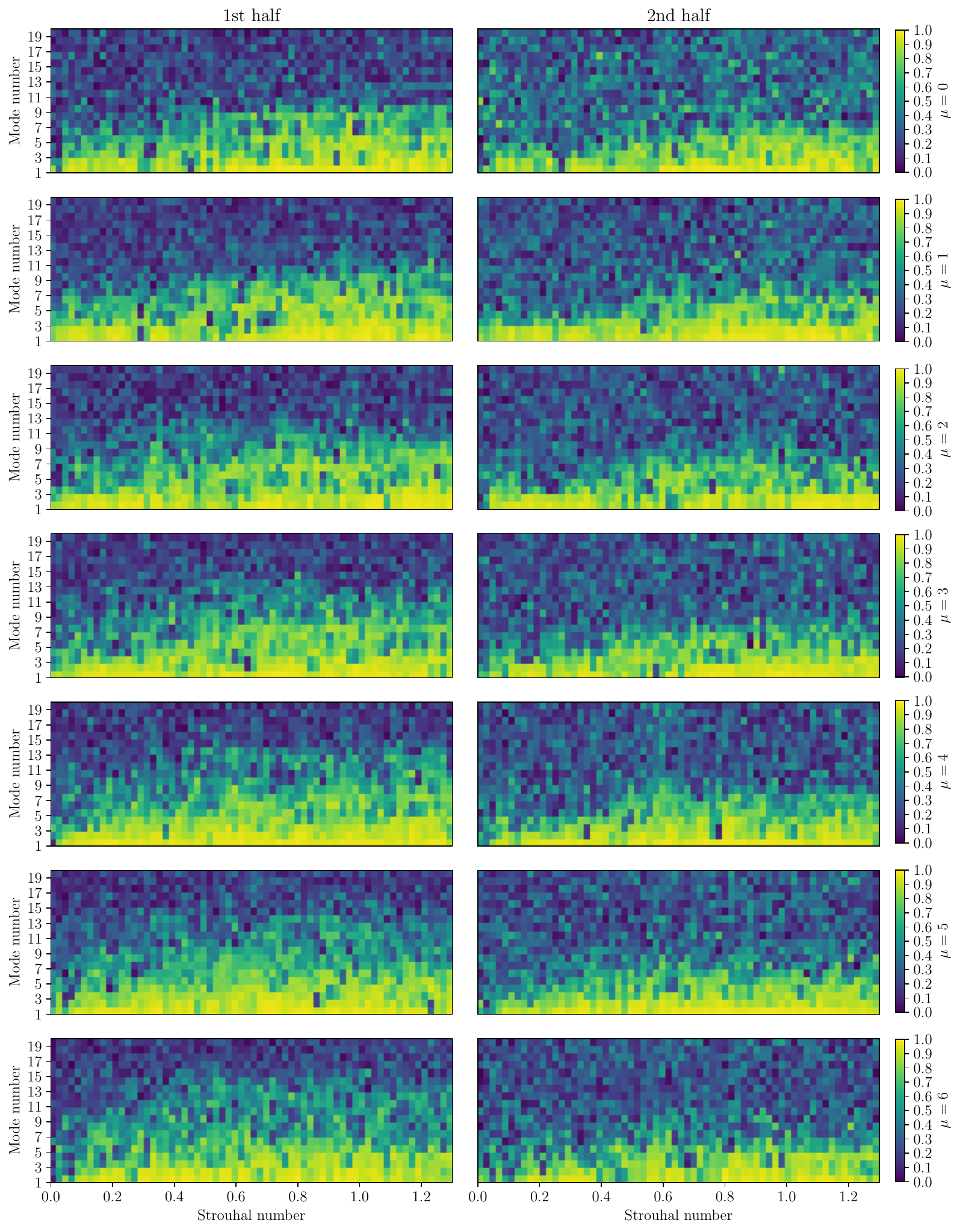}
	\caption{SPOD convergence levels for the baseline configuration and $x/D$ = 1.}
	\label{fig:spod_convergence_baseline_xD1}
\end{figure}

\begin{figure}
	\centering
	\includegraphics[width=\textwidth]{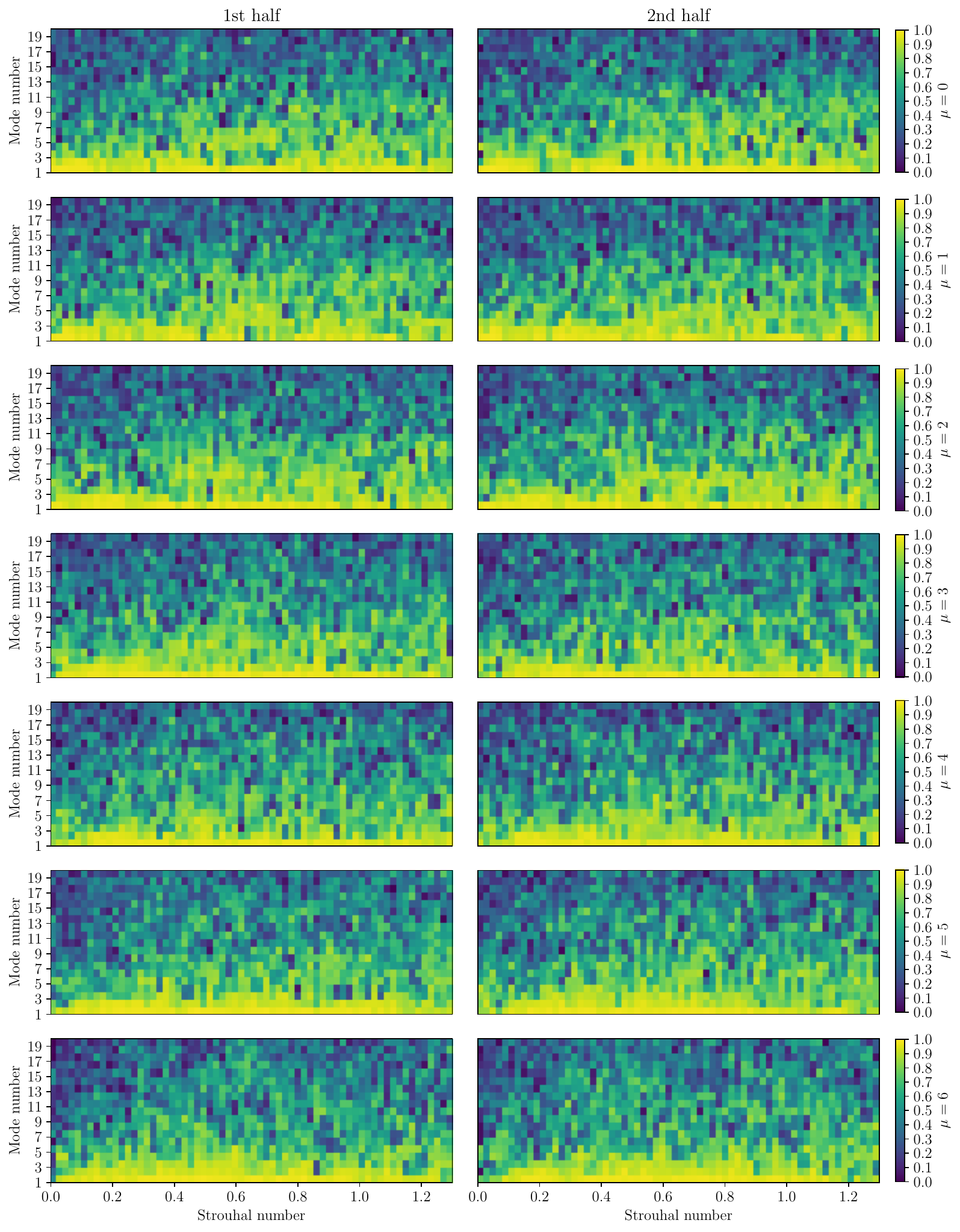}
	\caption{SPOD convergence levels for the tabbed configuration and $x/D$ = 1.}
	\label{fig:spod_convergence_tabbed_xD1}
\end{figure}

Figures~\ref{fig:spod_convergence_baseline_xD1} and \ref{fig:spod_convergence_tabbed_xD1} show the convergence levels of the SPOD modes for the first 20 azimuthal orders, as a function of Strouhal number and Floquet coefficient, at $x/D = 1$.
The leading modes exhibit good convergence, with $\beta$ values approaching unity.
Further downstream (not shown), convergence improves even for higher-order modes.
It is expected that even higher levels of convergence could be achieved with longer time series.

\section{Momentum flux estimates}
\label{app:momentum_estimates}

Following~\citet{hussein1994velocity}, the total momentum flux can be evaluated for each TR-PIV plane as
\begin{equation}
	\dot{P} = \rho \int_0^{2\pi} \int_0^\infty \left[\bar{u}^2 + \langle {u}^2 \rangle - \frac{1}{2}\left( \langle {v}^2 \rangle + \langle {w}^2 \rangle \right)\right] r\, dr\, d\varphi \mbox{,}
	\label{eq:momentum}
\end{equation}
\noindent where $r$ and $\varphi$ denote the radial and azimuthal directions, respectively, $\rho$ indicates the flow density $\bar{u}$ is the mean streamwise velocity, and $\langle {u}^2 \rangle$, $\langle {v}^2 \rangle$ and $\langle {w}^2 \rangle$ are the central momentum components in the streamwise and cross-flow directions.
In the streamwise direction, the central momentum is given as
\begin{equation}
	\langle {u}^2 \rangle = \frac{\sum_{i=1}^{N_s} (u - \bar{u}^2) {\Delta t_i}}{\sum_{i=1}^{N_s} {\Delta t_i}} \mbox{,}
	\label{eq:central_momentum}
\end{equation}
\noindent where $u$ is a streamwise velocity component snapshot, $\Delta t_i$ is the time between snapshots and ${N_s}$ is the number of snapshots.
Similar expressions are obtained for the cross-flow velocity components $\langle {v}^2 \rangle$ and $\langle {w}^2 \rangle$.

The total momentum flux $\dot{P}$ was computed for both tabbed and baseline nozzles at each streamwise station, using the velocity fields obtained from the TR-PIV measurements and assuming constant density $\rho$.
In order to obtain a statistical quantity, the mean momentum flux ($\overline{\dot{P}}$) and the standard deviation ($\sigma_{\dot{P}}$) were computed based on the six streamwise station measurements.

\begin{figure}
	\centering
	\includegraphics[width=0.8\textwidth]{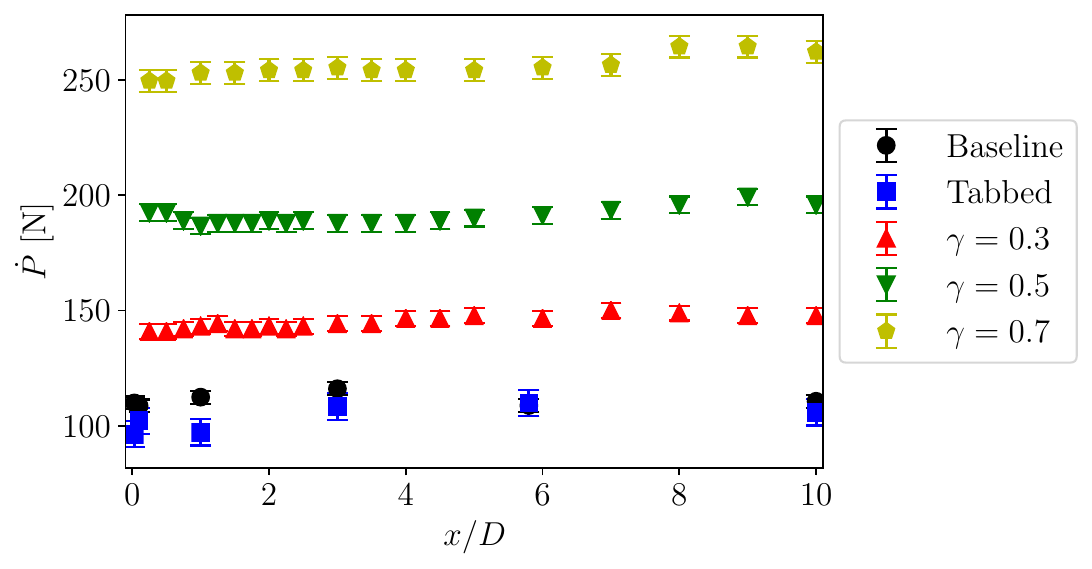}
	\caption{Total momentum flux estimates. Error bars denote one standard deviation. $\gamma$ values extracted from~\citet{guitton2007measurements}. Here, $\gamma = U_s/U_p$, i.e. the exit velocity ratio (secondary $s$ and primary $p$ jet flows).}
	\label{fig:total_momentum_estimates}
\end{figure}

Figure~\ref{fig:total_momentum_estimates} presents the total momentum integral (defined by~\eqref{eq:central_momentum}), which includes both mean and fluctuating components (error bars given as one standard deviation) and should be conserved as the flow evolves downstream.
Data from the present study are compared with the experimental results of \citet{guitton2007measurements} for coaxial jets (using radial laser Doppler anemometry (LDA) profiles, which explains why at downstream stations, it becomes difficult to ensure that measurement exactly follows a radial profile).
While the total momentum is approximately conserved, all data sets exhibit comparable uncertainty levels, underlining the inherent difficulty of such measurements.

\begin{figure}
	\centering
	\includegraphics[width=0.65\textwidth]{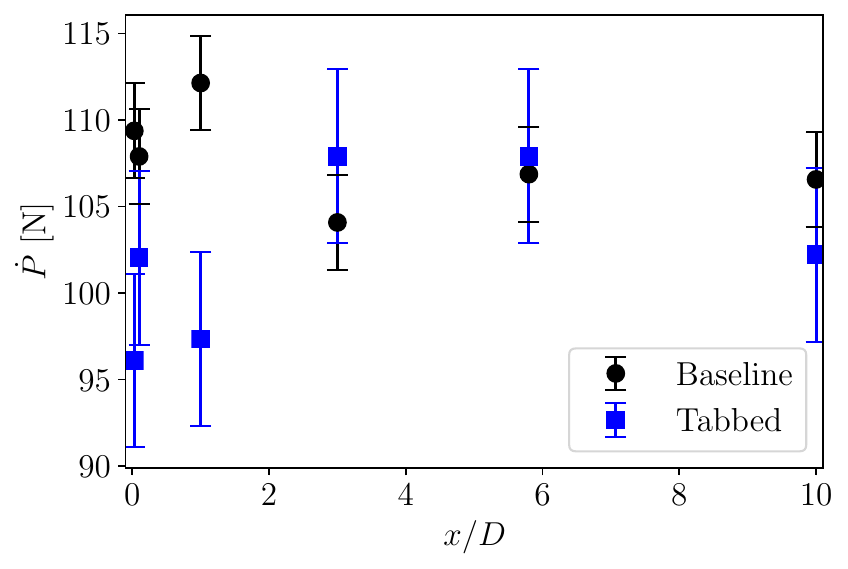}
	\caption{Streamwise momentum flux estimates. Error bars denote one standard deviation.}
	\label{fig:streamwise_momentum_estimates}
\end{figure}

Figure~\ref{fig:streamwise_momentum_estimates} shows the integrated mean streamwise momentum, an indicator of thrust, only for the baseline and tabbed nozzle.
The terms $\langle {u}^2 \rangle$, $\langle {v}^2 \rangle$ and $\langle {w}^2 \rangle$ were zeroed in equation~\eqref{eq:momentum}.
Two observations emerge: 1)	measurement uncertainty is highest near the nozzle, where velocity gradients are steepest; and 2)	in the tabbed flow, uncertainty is noticeably greater than in the baseline case, due to the additional gradients and flow distortion introduced by the tabs.

In the downstream region ($x/D$ > 3), where uncertainties are lower, the data do not offer a conclusive assessment of thrust change.
However, a cautious interpretation suggests that the thrust may not be significantly affected by the tabs.
Nonetheless, given the limitations of this indirect method, we refrain from drawing strong conclusions.
A dedicated thrust measurement using an aerodynamic balance is planned for future work to address this question definitively.

\bibliographystyle{abbrvjfm}
\bibliography{references}

\end{document}